\documentclass[preprint2]{pp7}

\input{pp7.h}

\usepackage{amsmath}
\usepackage{multicol}
\usepackage{multirow}
\usepackage{hyperref}
\usepackage[switch,columnwise]{lineno}
\usepackage{savesym}
\begin{document}

\title{\textbf{\LARGE Structured Distributions of Gas and Solids in Protoplanetary Disks\index{Protoplanetary disks}}} 

\author {\textbf{\large Jaehan Bae}}
\affil{\small\em University of Florida}
\affil{\small\em Carnegie Institution for Science}
\author {\textbf{\large Andrea Isella}}
\affil{\small\em Rice University}
\author {\textbf{\large Zhaohuan Zhu}}
\affil{\small\em University of Nevada, Las Vegas}
\author {\textbf{\large Rebecca Martin}}
\affil{\small\em University of Nevada, Las Vegas}
\author {\textbf{\large Satoshi Okuzumi}}
\affil{\small\em Tokyo Institute of Technology}
\author {\textbf{\large Scott Suriano}}
\affil{\small\em Union College}
\affil{\small\em SUNY Corning Community College}
\affil{\small\em The University of Tokyo}

\begin{abstract}
\baselineskip = 11pt
\leftskip = 0.65in 
\rightskip = 0.65in
\parindent=1pc
{\small 
Recent spatially-resolved observations of protoplanetary disks\index{Protoplanetary disks} revealed a plethora of substructures\index{Protoplanetary disks! substructures}, including concentric rings and gaps\index{Protoplanetary disk substructure!rings and gaps}, inner cavities, misalignments\index{Disk breaking and misalignment}, spiral arms, and azimuthal asymmetries. This is the major breakthrough in studies of protoplanetary disks since {\it Protostars and Planets VI} and is reshaping the field of planet formation. However, while the capability of imaging substructures in protoplanetary disks has been steadily improving, the origin of many substructures are still largely debated. The structured distributions of gas and solids in protoplanetary disks likely reflect the outcome of physical processes at work, including the formation of planets. Yet, the diverse properties among the observed protoplanetary disk population, for example, the number and radial location of rings and gaps in the dust distribution, suggest that the controlling process may differ between disks and/or the outcome may be sensitive to stellar or disk properties. In this review, we (1) summarize the existing observations of protoplanetary disk substructures collected from the literature; (2) provide a comprehensive theoretical review of various processes proposed to explain observed protoplanetary disk substructures; (3) compare current theoretical predictions with existing observations and highlight future research directions to distinguish between different origins; and (4) discuss implications of state-of-the-art protoplanetary disk observations to protoplanetary disk and planet formation theory. \\~\\~\\~}
 %leave this in to get the correct vertical space after the abstract
\end{abstract}  

\section{Introduction}

Since the modern planet formation theory was established \citep[e.g.,][]{safronov1972}, many physical processes associated with planet formation (including protoplanetary disk turbulence, dust dynamics, planetesimal formation, assembly and evolution of planetary cores) evolved to active research fields by their own and have been reviewed through previous {\it Protostars and Planets} series \citep[e.g.,][in {\it Protostars and Planets VI}]{helled2014,johansen2014,raymond2014,testi2014,turner2014}. However, despite the large amount of theoretical efforts, the planet-forming environments and processes have been largely unconstrained by observations until recently. 

Over the last few decades, more than 5,000 exoplanets have been discovered\footnote{\url{https://exoplanetarchive.ipac.caltech.edu/}}, and we now know that the solar system planets are not the only planets in our galaxy. Interestingly, a large fraction of the observed exoplanet population has very distinct properties compared with those of solar system planets, potentially suggesting that the typical planet forming environment may be different from what we had 4.6 billion years ago in the solar nebula. In order to better understand the diversity in the exoplanet population and the potential uniqueness of the solar system, it is crucial to study protoplanetary disks\index{Protoplanetary disks} -- the birthplace of planets.

With the advent of the state-of-the-art observing facilities, including the Atacama Large Millimeter/submillimeter Array (ALMA\index{ALMA}) and ground-based telescopes equipped with extreme adaptive optics, we now have the capability of spatially resolving protoplanetary disks at as fine as a $0\farcs01$ angular resolution, corresponding to a few astronomical unit (au) in linear scale for the disks in nearby star-forming regions. At the time {\it Protostars and Planets VI} was written, radio interferometric observations could achieve a $\gtrsim0\farcs2$ angular resolution with the Submillimeter Array, Combined Array for Research in Millimeter-wave Astronomy, and Plateau de Bure Interferometer (see e.g., \citealt{dutrey2014,espaillat2014,testi2014} in {\it Protostars and Planets VI}), so we now have about an order of magnitude better resolving power than back then. Notably, the observations from the 2014 ALMA\index{ALMA} Long Baseline Campaign revealed multiple sets of concentric bright rings and dark gaps in the protoplanetary disk surrounding the young star HL Tau \citep{alma2015}\index[obj]{HL Tau}. It has long been suggested that protoplanetary disks\index{Protoplanetary disks} have structures, inferred based on the spectral energy distribution (SED; \citealt{strom1989}, see also review by \citealt{espaillat2014} in {\it Protostars and Planets VI}) and images taken with ground-based telescopes, the Hubble Space Telescope \citep[e.g.,][]{grady1999}, and pre-ALMA\index{ALMA} radio interferometers \citep[e.g.,][]{andrews2009}. However, the ALMA\index{ALMA} observation of the HL Tau disk was the first time that au-scale fine structures were imaged in protoplanetary disks\index{Protoplanetary disks}. Since then, high-resolution observations have revealed a myriad of disk structures, including rings and gaps, spirals, and crescents which, collectively, are commonly referred to as {\it protoplanetary disk substructures}\index{Protoplanetary disks!substructures}. 

One of the reasons why protoplanetary disk substructures\index{Protoplanetary disks! substructures} have been drawing  large attention from the community is that their presence could be related to planet formation, although whether substructures are the cause or effect of planet formation is unclear at the time of writing this Chapter (we further discuss this point in Section \ref{sec:implications}). In either case, the increasingly powerful observational facilities and techniques have led us to the point where planet formation theory can be finally  tested by observations. Several Chapters in this book are dedicated to recent high-resolution observations of protoplanetary disks ({\it Benisty et al.}, {\it Pinte et al.}) and theoretical studies of protoplanetary disks and planet formation therein ({\it Drazkowska et al.}, {\it Lesur et al.}, {\it Paardekooper et al.}). In this Chapter, we aim to connect observations and theory, by comparing existing observational data with current theoretical predictions for disk substructures.

This Chapter is organized as follows. In Section \ref{sec:observation_overview}, we provide an overview of recent high-resolution observations of protoplanetary disks\index{Protoplanetary disks}. We provide basic statistics and discuss whether or not we find any correlation between substructure properties and stellar/disk properties. As we will show in Section \ref{sec:observation_overview}, a large fraction of substructures observed to date have been detected in radio continuum and/or optical/near-infrared (NIR) scattered light observations, both of which probe dust grains in the disks. When interpreting disk substructures, it is important to keep in mind that the spatial distribution of the gas, which contains about 99\% of the total disk mass, and that of the dust can differ due to the aerodynamic drag\index{Aerodynamic drag} that exerts on the dust. We discuss the coupling between the gas and the dust in Section \ref{sec:coupling}. We then review various substructure-forming mechanisms from Section \ref{sec:hydro} to Section \ref{sec:dust}. There are more than 20 mechanisms that we review in this Chapter, and we group them into hydrodynamic processes (Section \ref{sec:hydro}), magnetohydrodynamic processes (Section \ref{sec:MHD}), tidal interactions with perturbers (Section \ref{sec:companion}), disk self-gravity and the effects it has on other substructure-forming processes (Section \ref{sec:self-gravity}), and processes induced by dust particles (Section \ref{sec:dust}). In each of these sections, we follow a homogenized structure (when applicable) such that we first introduce how each mechanism operates, then describe in which regions of protoplanetary disks it is expected to operate, and finally summarize the properties of substructures it creates. After we have discussed the substructure-forming mechanisms, in Section \ref{sec:obs_theory_connection} we compare the properties of the substructures predicted from theories and numerical simulations with those from observations, discuss potential origins of observed disk substructures, and provide future directions to distinguish between different origins. Finally, we discuss the implications of disk substructures to protoplanetary disk\index{Protoplanetary disks} and planet formation theory in Section \ref{sec:implications} and conclude the Chapter in Section \ref{sec:conclusion}.

\section{Overview of Observed Protoplanetary Disk Substructures\index{Protoplanetary disks! substructures}}
\label{sec:observation_overview}

\subsection{Observational Primer} 
\label{sec:obs_primer}

Assuming that protoplanetary disks\index{Protoplanetary disks} inherit their composition from the parent molecular clouds, about 99\% of their mass is composed of gas, while the remaining 1\% is in the form of refractory material, i.e., solid particles. Of the gas mass, about 70\% is in the form of hydrogen (both atomic and molecular), about 28\% is composed of helium, and only about 2\% is composed of heavier elements. However, at densities and temperatures that are typical for protoplanetary disks, hydrogen and helium are hardly observable and our knowledge of protoplanetary disks must therefore rely on observations of much rarer components such as dust particles and tracer molecules. 

For dust particles, thermal and scattered continuum emission at a wavelength of $\lambda$ are dominated by grains with a size $a \sim \lambda/2\pi$. Consequently, optical/near-infrared observations mainly probe sub-micrometer particles, while mm and cm-wave observations probe sand-sized and pebble-size solids. This characteristic, coupled to the fact that the absorption and scattering opacity of dust grains strongly depends on the grain size \citep[see, e.g.,][]{draine2016}, results in a strong wavelength dependence of the optical depth of the dust continuum emission. For example, assuming a typical dust composition and grain size distribution \citep[see, e.g.,][]{birnstiel2018}, the dust opacity $\kappa$ scales roughly as $\kappa \sim 10^3~(\lambda/ 1\mu m)^{-2/3}$~cm$^2$ g$^{-1}$. Therefore, the continuum emission at 1 $\mu$m becomes optically thick for dust column densities larger than $\sim10^{-3}$ g cm$^{-2}$, while a much larger dust column density of 0.1 g cm$^{-2}$ is required to have an optical depth of 1 at 1 mm. Assuming a standard gas-to-dust ratio of 100, the transition from optically thin to optically thick continuum emission would happen at a gas column density of 0.1 g cm$^{-2}$ at 1 $\mu$m, and 10 g cm$^{-2}$ at 1 mm.

To understand the importance of the optical depth in detecting substructures in the dust continuum emission, it is useful to put these numbers in the context of the Minimum Mass Solar Nebula (MMSN) model \citep{weidenschilling1977_MMSN,hayashi1981}.  Following \cite{hayashi1981}, the gas surface density of the MMSN is given by $\Sigma_g(R) = 1700~(R/1~{\rm au})^{-3/2}~{\rm g~cm^{-2}}$, where $\Sigma_g$ is the gas surface density and $R$ is the radial distance from the central star. At a wavelength of 1 $\mu$m, the emission of the MMSN would be optically thick out to a radius of about 660~au (assuming that the MMSN extends that far), while the emission at 1~mm would be optically thick within the inner 30~au only. Due to the difference in optical depth, along with the different vertical distributions for dust particles with different sizes (see Section \ref{sec:settling}), observations at optical/NIR wavelengths typically constrain the scattering properties of small particles floating in the uppermost layers of protoplanetary disks\index{Protoplanetary disks}, while mm and cm-wave observations typically probe the column density of larger dust grains located near the disk midplane. 

Similar considerations can be done for atomic and molecular line emission. Depending on the abundance and excitation properties, the optical depth of spectral lines typically observed in protoplanetary disks\index{Protoplanetary disks} may vary by multiple orders of magnitude. For example, rotational and vibrational transitions of abundant molecules like $^{12}$CO are optically thick, and are used to constrain the temperature of the emitting gas, while transitions of rarer compounds are a better probe of the molecular gas density. However, the conversion between the density of a specific gas tracer and that of hydrogen is hampered by the limited knowledge of molecular abundances which are set by complex chemical networks controlled by the spectrum of the incident radiation as well as by the gas density itself (see {\it Miotello et al.} of this book). Furthermore, because wide-band continuum observations achieve a better sensitivity than narrow-band spectral line observations, most disk substructures have been observed in the mm continuum, although there have been an increasing number of substructures observed in the line observations \citep[e.g.,]{law2021a}. In this review, we mostly focus on disk substructures observed in scattered/thermal emission from the dust. 

\subsection{Substructure Occurrence\index{Protoplanetary disk substructure!occurrence}}
\label{sec:substructure_occurrence}

\begin{figure*}[!t]
\centering 
\includegraphics[width=0.8\textwidth]{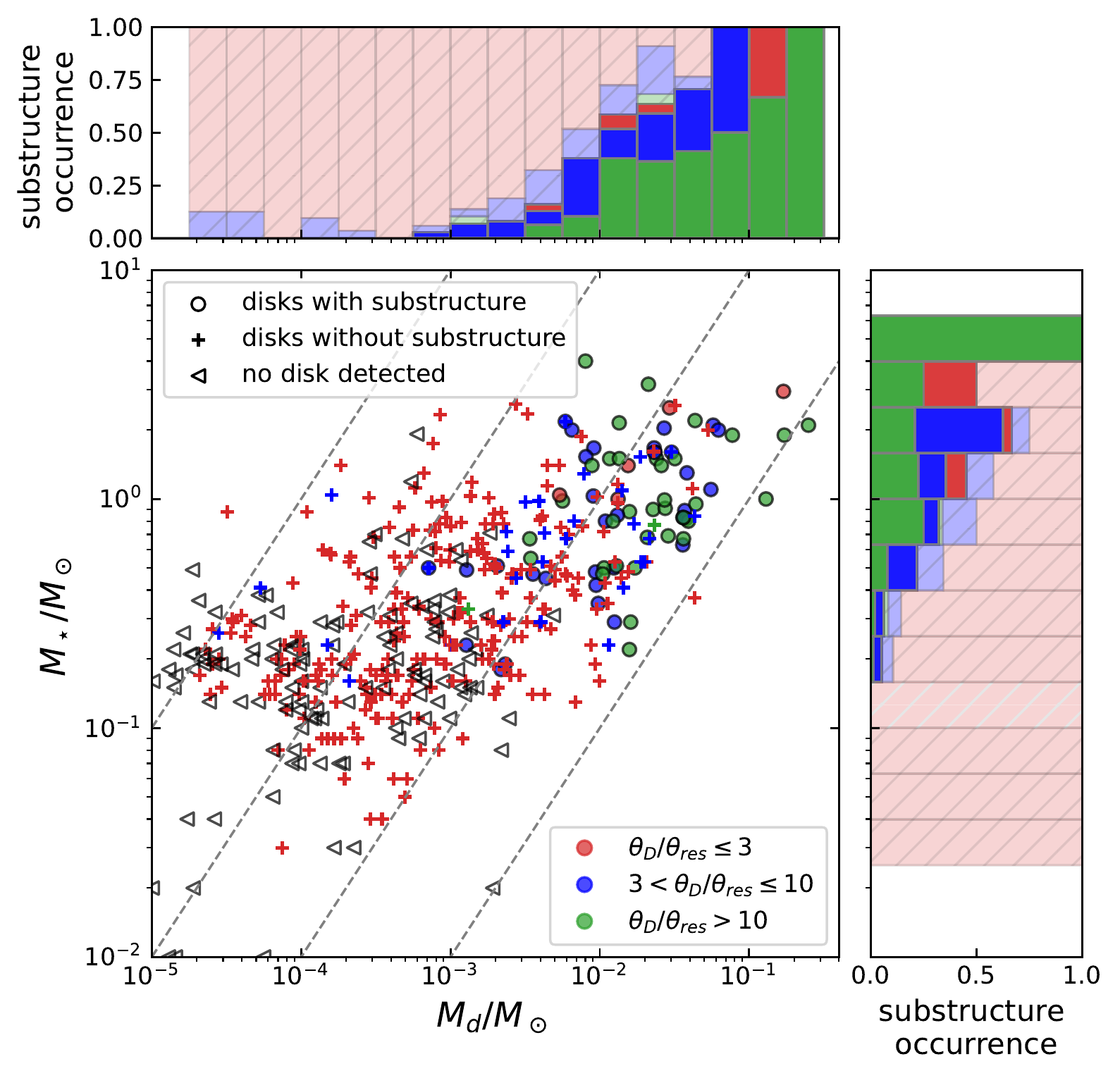}
\caption{Mass of the disk inferred from mm continuum flux vs. mass of the host star for a sample of 479 young stars (see Section \ref{sec:substructure_occurrence}). Solid circles indicate disks with substructures, crosses indicate disks where no substructures were observed, and triangles indicate disks with an upper limit for the continuum flux (i.e., non-detection). Data points are color-coded by the ratio between the angular diameter containing 90\% of the continuum emission ($\theta_D$) and the angular resolution of the observations ($\theta_{res}$). See the legend in the lower-right corner. The four diagonal dashed lines show constant disk-to-star mass ratios of $M_{\rm d}/M_* = 0.0001, 0.001, 0.01$, and 0.1 (from left to right).  The histograms presented in the upper and right panels show the fraction of disks with (solid) and without (hatched) substructures in each (upper panel) disk mass and (right panel) stellar mass bin. The histograms are color-coded by $\theta_D/\theta_{res}$ in the same way to the scatter plot. The histograms suggest that the fraction of disks with substructures may increase with $M_*$ and $M_d$; however, note that the histograms clearly depict that the majority of disks without substructures have been observed at low effective angular resolution of $\theta_D/\theta_{res} \leq 3$ (red hatched regions of the histograms). Future high angular resolution observations are crucial to properly assess whether or not substructure occurrence\index{Protoplanetary disk substructure!occurrence} depends on $M_*$ and/or $M_d$.}
\label{fig:disk_fluxes1}
\end{figure*}

Before we dive into a discussion of disk substructures, it is important to consider any observational biases that might affect the occurrence of disk substructures\index{Protoplanetary disk substructure!occurrence}. More specifically, we must keep in mind that disk substructures can be observed only if the disk emission is spatially resolved and recorded at a sufficiently high signal-to-noise ratio. To demonstrate this point, we compiled from the literature a sample of 479 young stars that have been observed at NIR and/or mm wavelengths, most of which are located in the Taurus, Ophiuchus, Upper Scorpius, Lupus, and Chameleon I star-forming regions. Figure~\ref{fig:disk_fluxes1} shows the disk masses and the stellar masses for this sample. The disk mass, $M_d$, is calculated from measured integrated 1.3~mm fluxes as $M_d = F d^2/\kappa B(T)$, where $F$ is the flux, $d$ is the distance to the object, $\kappa=0.023$~cm$^2$~g$^{-1}$ is the opacity, and $B(T)$ is the Planck function for which we adopt $T=20$~K as in \cite{pascucci2016} and \cite{testi2022}. When 1.3~mm fluxes are not available, as in the case of Upper Scorpius, we convert fluxes measured at other mm wavelengths, e.g., 0.87~mm, assuming a spectral index of 3.6.  Finally, we convert the dust mass to the total disk mass by multiplying by a gas-to-dust mass ratio of 100. Millimeter data and stellar masses were obtained from a number of sources including \cite{akeson2014, andrews2011, andrews2013, Ansdell2018, barenfeld2017, cieza2019, cieza2021, isella2009, long2018, long2019, pascucci2016, testi2022} (see also the references in Table \ref{tab:disks_with_rings}, \ref{tab:disks_with_spirals}, and \ref{tab:disks_with_crescents}).   

Out of the 479 young stars, disks have been detected around 355 objects (circles and crosses in Figure \ref{fig:disk_fluxes1}). For 124 non-detections (triangles in Figure \ref{fig:disk_fluxes1}), we put an upper limit to the disk mass. Of the 355 disks, substructures have been found in 73 disks (circles in Figure \ref{fig:disk_fluxes1}), either in the NIR scattered light emission, mm-wave continuum, or mm-wave molecular line emission, while substructures have not been found in the rest 282 disks (crosses in Figure \ref{fig:disk_fluxes1}) at the angular resolution and sensitivity of current observations. The data points use different colors to show the different ranges of the {\it effective} angular resolution $\theta_D/\theta_{res}$, where $\theta_D$ is the angular diameter containing 90\% of the continuum emission \cite[see, e.g.,][]{hendler2020} and $\theta_{res}$ is the angular resolution of the observation.

In addition to the scatter plot, in Figure \ref{fig:disk_fluxes1} we present histograms showing the substructure occurrence\index{Protoplanetary disk substructure!occurrence} as a function of the disk and stellar masses. The histograms show that the fraction of disks with substructures increases steeply with the disk mass: the substructure occurrence rate is below 50\% for disks with $M_d \lesssim 0.01 M_\odot$ (corresponding to 1.3 mm flux of $\lesssim 50$~mJy), while the substructure occurrence rate exceeds 80\% for disks with $M_d \gtrsim 0.01 M_\odot$. In agreement with previous analyses, the figure shows that the dust disk mass positively correlates with the stellar mass (see e.g., \citealt{andrews2013,pascucci2016} and {\it Manara et al.} of this book). As a result, the larger substructure occurrence rate for high-mass disks translates to large substructure occurrence rate for high-mass stars. Taken at face value, the correlation between substructure occurrence\index{Protoplanetary disk substructure!occurrence} and the disk mass (or stellar mass) might offer valuable information for understanding the origin of the substructures. However, caution must be taken because this correlation might instead result from observational biases. In fact, for the 256 disks observed with a low effective angular resolution of $\theta_D/\theta_{res} \leq 3$, substructures have been detected in only $2\%$ of the sample (5 disks). When disks are observed with an intermediate effective angular resolution of $3 < \theta_D/\theta_{res} \leq 10$, substructures are found in about half of the sample (32 disks with substructures out of total 62 disks). Finally, when disks are observed with a high effective angular resolution of $\theta_D/\theta_{res} > 10$, substructures are found in $95\%$ of the sample (35 disks with substructures out of total 37 disks). As we mentioned at the beginning of this subsection, sufficiently high angular resolution (and sensitivity) is required to detect disk substructures, and it is not surprising that most of the disks with substructures have been observed with high effective angular resolution. The apparent lack of substructures in less massive disks could perhaps be explained with the fact that current observations have not yet achieved the angular resolution necessary to discover their substructures.

As we will discuss in the subsequent Sections, certain substructure-forming scenarios, such as gravitational instability\index{Gravitational instability} and companions, favor a large disk mass (or large $M_d/M_*$). On the other hand, other scenarios favor a small disk mass (e.g., vertical shear instability\index{Vertical shear instability} since it requires rapid cooling) or are expected to be relatively insensitive to the disk mass (e.g., icelines). Future high resolution observations across a range of disk mass bins are highly desirable to obtain more complete statistics on disk substructures, but also to start to answer whether the increasing substructure occurrence rate\index{Protoplanetary disk substructure!occurrence} with disk mass is simply due to the fact that low-mass disks have not been observed at sufficiently high angular resolution or the trend reflects the underlying substructure-forming processes.

\begin{figure*}[!t]
\centering
\includegraphics[width=0.85\textwidth]{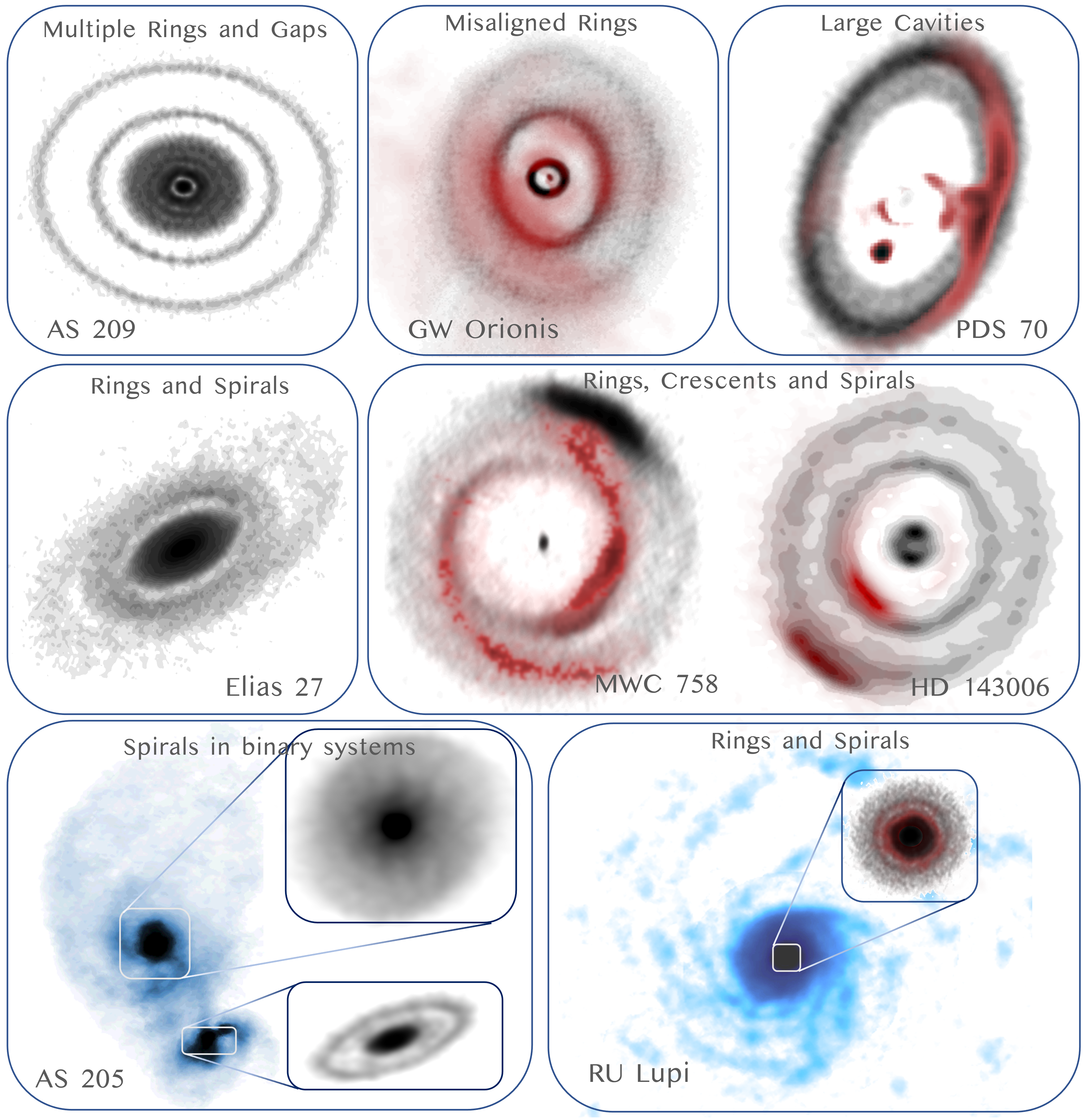}
\caption{
Some of the morphological features observed in protoplanetary disks, including concentric rings and gaps\index{Protoplanetary disk substructure!rings and gaps}, misaligned rings\index{Disk breaking and misalignment}, large cavities, spirals, and crescents. Note that there are some disks which have more than one substructure: Elias~2-27, MWC~758, HD~143006, AS~205, RU~Lup. Black and white contours show mm continuum observations, blue contours show mm line observations, and red contours show NIR observations, respectively. The data presented in this figure are originally from: AS~209 -- \citet{andrews_dsharp}, GW~Ori -- \citet{bi2020,kraus2020}, PDS~70 -- \citet{keppler2018,benisty2021}, Elias~2-27 -- \citet{andrews_dsharp}, MWC~758 -- \citet{Benisty2015,dong2018b}, HD~143006 -- \citet{andrews_dsharp,benisty2018}, AS~205 -- \citet{andrews_dsharp,Kurtovic2018}, RU~Lup -- \citet{andrews_dsharp,Huang2018a}.}
\label{fig:substructures}
\end{figure*}

\subsection{Substructure Morphology}
\label{sec:disk_substructures}

Figure~\ref{fig:substructures} presents some examples of observed disk substructures. It has been customary to group disk substructures into three main classes: rings and gaps\index{Protoplanetary disk substructure!rings and gaps}, spirals\index{Protoplanetary disk substructure!spirals}, and crescents\index{Protoplanetary disk substructure!crescents}. However, it is worth noting that a large variety of properties have been observed and thus any rigid classification is challenging. Also, it turned out that a single disk can have multiple types of substructures, and it is possible that a disk shows different classes of substructures when observed at different wavelengths. For examples, rings, gaps, spirals, and crescents are observed in the mm continuum image of the MWC 758 disk, while only spirals are observed in the IR observations (Figure~\ref{fig:substructures}; \citealt{dong2018b, Benisty2015}). 

Keeping in mind these nuances, in Table \ref{tab:disks_with_rings}, \ref{tab:disks_with_spirals}, and \ref{tab:disks_with_crescents}\footnote{The data presented in the tables are available online at \url{http://ppvii.org/chapter/12/}.} we group the 73 disks with substructures introduced above based on the presence of rings, spirals\index{Protoplanetary disk substructure!spirals}, and crescents\index{Protoplanetary disk substructure!crescents}. 
For each object, we list the distance from Earth ($d$), the stellar mass ($M_*$), the stellar luminosity ($L_*$), the infrared SED classification, the disk mass ($M_d$), the number of substructures, the radial location or extent of the substructures, the main properties of the substructures, and the wavelength (mm or IR) at which the substructures were observed. For binary systems, we also provide the angular separation between the stars. References from which the information is adopted are listed in the captions of the Tables. 

Substructures were observed at both infrared and mm wavelengths in 21 disks, while substructures were observed at only mm or infrared wavelengths in the other 49 and 3 objects, respectively. Rings and gaps\index{Protoplanetary disk substructure!rings and gaps} are observed in 62 disks and are the most common type of substructure, while spirals and crescents are observed in 22 and 13 disks, respectively. Finally, 19 disks have more than one type of substructure. In the following subsections, we discuss the occurrence and properties of each type of substructure. Note that in this section, we mostly focus on reporting the findings from the observational data and defer the discussion as to what those findings may indicate in terms of their origin to Section \ref{sec:obs_theory_connection}, after we introduce substructure-forming processes in Sections \ref{sec:hydro} -- \ref{sec:dust}.

\begin{figure*}[t!]
\centering
\includegraphics[width=0.9\textwidth]{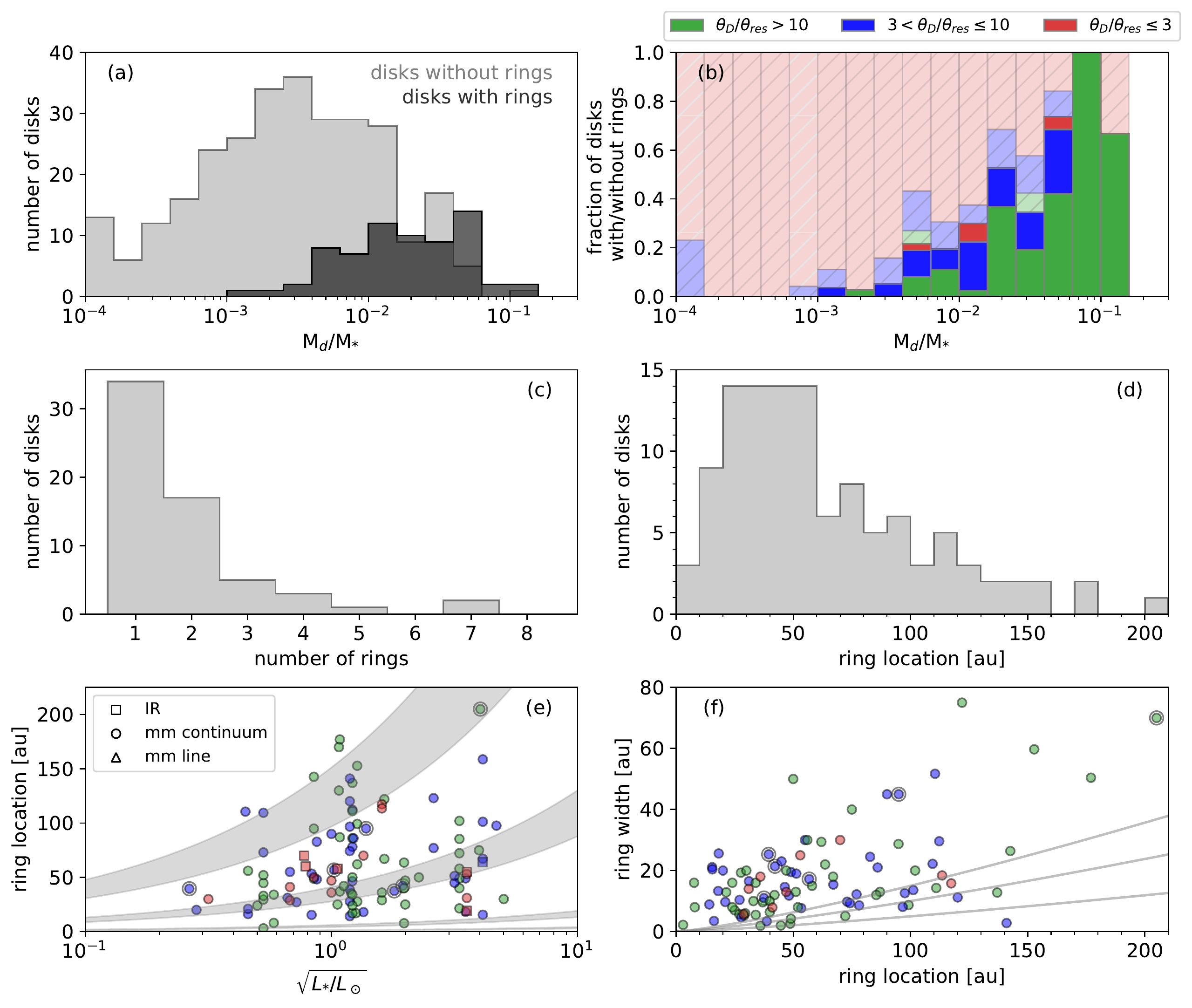}
\caption{(a) Histograms showing the number of disks  (gray) without rings and (black) with rings, as a function of $M_{\rm d}/M_*$. The bin size is 0.2 dex in the logarithmic scale. (b) A histogram showing the fraction of disks with and without rings in each $M_{\rm d}/M_*$ bin. The same colors as in Figure \ref{fig:disk_fluxes1} are used to present the effective angular resolution used for observations. (c) A histogram showing the number of rings in each disk that is known to have rings. (d) A histogram showing the radial location of individual rings. (e) A scatter plot presenting the radial location of rings vs. $\sqrt{L_*/L_\odot}$. Squares show the rings detected in IR observations, circles show the rings detected in mm continuum observations, and triangles show the rings detected in mm line observations, respectively. Double symbols (e.g., two concentric circles) indicate binary systems. The shaded gray regions represent the parameter space where N$_2$, CO, CO$_2$, and H$_2$O icelines (from top to bottom) would be located.  (f) A scatter plot showing the radial width of the rings vs. the radial location of the rings. The three gray curves show the gas scale height $H = (H/R)_{100~{\rm au}} (R/100~{\rm au})^{1/4}$ adopting $(H/R)_{100~{\rm au}} = $ 0.15, 0.1, and 0.05 (from top to bottom).}
\label{fig:rings}
\end{figure*}

\subsubsection{Rings and gaps\index{Protoplanetary disk substructure!rings and gaps}} 
\label{sec:disk_substructures:rings}

{\it Occurrence:\index{Protoplanetary disk substructure!occurrence}}
Rings are generally defined as being intrinsically circular and azimuthally symmetric. They are the most common type of substructure observed in protoplanetary disks\index{Protoplanetary disks} thus far. A total of 120 unique rings (113 at mm and 7 at NIR) have been detected in 62 disks (Table~\ref{tab:disks_with_rings}) among the 73 disks with substructures compiled in this review. Rings and gaps\index{Protoplanetary disk substructure!rings and gaps} are far more common in mm observations (61 disks) compared to NIR observations (6 disks, see also {\it Benisty et al.} in this book). One possible explanation to the apparent discrepancy between mm and NIR observations is that the observed rings coincide with pressure maxima, efficiently trapping large particles and facilitating the detection in mm observations (see Section \ref{sec:coupling}). Disks with large ($>20$ au) inner cavities are a noteworthy subset of disks with rings and gaps. At least 15 objects could be included in this class. About half of these objects show an imaged inner disk and/or infrared excess in the SED indicating the presence of a small ($< 3-5$ au) dusty disk. An example is MWC~758, where the inner disk is detected but not spatially resolved by existing ALMA\index{ALMA} observations with angular resolution down to $0\farcs04$ (corresponding to 6~au linear resolution; \citealt{dong2018b}). The other half show no or little infrared excess in the SED implying that dust is depleted in the innermost disk regions. These disks are classified as transitional disks (TD; Table \ref{tab:disks_with_rings}). We find that the fraction of disks with rings increases with $M_{d}/M_{\star}$ (Figure~\ref{fig:rings} (a) and (b)) although we caution that this correlation may be due to the observational bias we discussed in Section~\ref{sec:substructure_occurrence}. 

{\it Multiplicity:} 
About half of systems with rings show a single ring, but multi-ring systems have been commonly observed (see Figure~\ref{fig:rings} (c)) particularly at the highest spatial resolution provided by ALMA\index{ALMA}. Outstanding examples of multi-ring systems include HL~Tau and AS~209 with 7 rings each \citep{alma2015,andrews2018a,Huang2018a}, TW~Hya with 5 rings \citep{andrews2016,huang2018c}, and HD~163296, CI Tau, and RU Lup with 4 rings each \citep{clarke2018, isella2018, Huang2020b}. Some of the single-ring systems have been observed at a relatively coarse resolution, and it is possible that future high-resolution observations may detect additional rings and gaps. As an example, this has been the case for LkCa~15 for which observations at $0.\farcs15 - 0.\farcs25$ angular resolution (corresponding to $\simeq 24 - 40$~au linear resolution) initially revealed a single ring \citep{andrews2011b, Isella2014, Jin2019}, but later ALMA\index{ALMA} observations at $0.\farcs05$ angular resolution (corresponding to $\simeq 8$~au linear resolution) resolved the previously observed single ring to have at least two, and perhaps three, components \citep{facchini2020}.

{\it Radial location:}
Rings can be described using a 1-D Gaussian function $I(R) \propto I_0 e^{-(R-R_0)^2/2\sigma_R^2}$, where $R_0$ is the radius of the ring and $\sigma_R$ is its radial width. For mm observations, $R_0$ and $\sigma_R$ are often measured by fitting the Gaussian model to visibilities. Rings are observed at essentially all radii from as close as 3~au to the host star (TW~Hya) to as far as 205~au (HD~142527) from the host star, with a maximum frequency occurring at 20 -- 50~au (Figure \ref{fig:rings}(d)). The steep drop in frequency at small radii ($< 20$~au) is due to the limited angular resolution of the observations, while the gradual drop at large radii beyond 50~au likely represents the radial extent of the continuum disks.
Besides that, we do not find any correlation between the radial location of the rings and stellar/disk properties (e.g., Figure \ref{fig:rings}(e)). 

{\it Radial width:} Measurements of the radial width of the continuum rings have been obtained for 88 rings out of a total of 113. Ring widths range from a few au, set by the best resolution achieved with ALMA\index{ALMA}, to about 80 au (Figure~\ref{fig:rings} (f)). However, some of the widest rings have only been marginally resolved, and it is therefore possible that they might be composed of multiple narrow rings, as in the case of LkCa~15 mentioned above. Similar to the radial location of rings, we do not find any correlation between the radial width of the rings and the stellar or disk properties.

{\it Misalignment:\index{Disk breaking and misalignment}} 
If mm-wave dust rings are optically thin and trace the disk midplane, their aspect ratio provides a direct measurement of the disk inclination relative to the line-of-sight. Although most of the multi-ring systems appear to be co-planar, there are systems where the disk inclination varies with the distance from the central star, implying a misalignment or warp in the disk. Such examples include HD~143006, DoAr44, AA~Tau\index[obj]{AA Tau}, HD~142527, HD~100453, HD~100546, J1604-2130, HD~139614, and GW~Ori \citep{benisty2018,perez2018,casassus2018,loomis2017,marino2015,benisty2017,walsh2017,mayama2018,bi2020,kraus2020}. For these disks, the degree of misalignment appears to range from $30^\circ$ (HD~143006, DoAr~44) to $80^\circ$ (HD~100546). In NIR observations, misalignments can manifest themselves as a shadow cast on the misaligned outer disk \citep[e.g.,][]{min2017,bohn2021}. For a more complete review of misalignments and shadows, we refer readers to {\it Benisty et al.} in this book.

\subsubsection{Spirals\index{Protoplanetary disk substructure!spirals}}
\label{sec:spirals_obs}

\begin{figure*}[t!]
\centering
\includegraphics[width=0.9\textwidth]{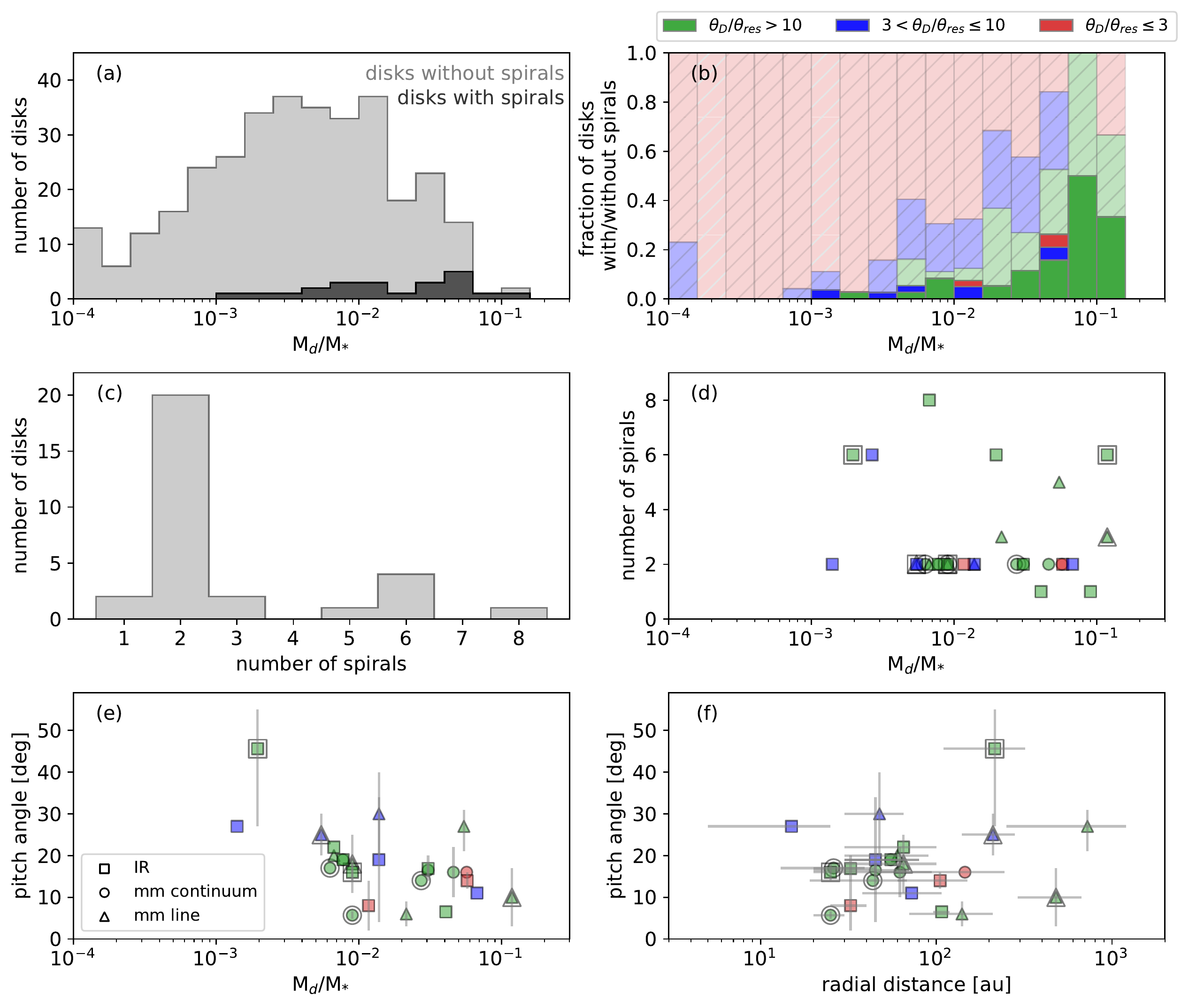}
\caption{(a) Histograms showing the number of disks  (gray) without spirals and (black) with spirals, as a function of $M_{\rm d}/M_*$. The bin size is 0.2 dex in the logarithmic scale. (b) A histogram showing the fraction of disks with and without spirals in each $M_{\rm d}/M_*$ bin. The same colors as in Figure \ref{fig:disk_fluxes1} are used to present the effective angular resolution used for observations. (c) A histogram showing the number of disks as a function of the number of spirals. (d) A scatter plot presenting the number of spirals vs. $M_{\rm d}/M_*$. Squares show the spirals detected in IR observations, circles show the spirals detected in mm continuum observations, and triangles show the spirals detected in mm line observations, respectively. Double symbols (e.g., two concentric circles) indicate binary systems.  (e) A scatter plot showing the pitch angle vs. $M_{\rm d}/M_*$. The vertical bars show the range of pitch angle of spirals in multi-spiral systems. (f) A scatter plot showing the pitch angle vs. radial extent of the spirals (shown with horizontal bars).}
\label{fig:spirals}
\end{figure*}

{\it Occurrence:\index{Protoplanetary disk substructure!occurrence}} Spirals\index{Protoplanetary disk substructure!spirals} are less common than rings and have been observed in only 22 disks so far (Table~\ref{tab:disks_with_spirals}). Among the 22 disks, 17 disks show spirals in NIR observations and 12 disks show spirals in the mm continuum or line emission. In 7 disks, spirals are detected at both NIR and mm wavelengths although the properties of the spirals inferred at different wavelengths do not necessarily match (see Table \ref{tab:disks_with_spirals}). So far, spirals have been observed in 6 multiple stellar systems, either in the disk around the primary star as in AS~205N \citep[Figure 2,][]{Kurtovic2018}, or in the circumbinary disk as in HD~142527 \citep{Fukagawa2006}. Similar to rings and gaps, the fraction of disks with spirals appears to increase with the $M_d/M_\star$ ratio (Figure~\ref{fig:spirals} (a) and (b)), although we caution again that this may be a result of observational biases (see Section~\ref{sec:substructure_occurrence}).

{\it Multiplicity:}
The number of spirals in the 22 disks ranges from 1 to 8, but there is a significant peak at 2 (Figure~\ref{fig:spirals} (c)).  We find no obvious correlation between the number of spirals and the stellar or disk properties including $M_d/M_\star$ (Figure~\ref{fig:spirals} (d)). However, we find one noticeable difference when spirals in mm and NIR observations are compared. Spirals in NIR observations tend to reveal a larger number of spirals: all five systems with 6 or more spirals are observed in NIR, whereas mm observations have so far revealed only two- or three-armed spirals. 

{\it Pitch angle:}
Pitch angle defines how tightly a spiral is wound. Mathematically, the pitch angle $\psi$ is given by $\tan \psi = -{\rm d}R/(R {\rm d}\phi)$. Pitch angles are generally smaller than 30\arcdeg, with the exception of HD34700A and CQ~Tau which have pitch angles as large as $40-55$\arcdeg\ \citep{Uyama2020,Wolfer2021}. We find a weak trend that the pitch angle decreases as a function of $M_d/M_{\star}$ (Figure~\ref{fig:spirals} (e)). At face value, this appears to be consistent with the finding by \citet{yu2019}. However, we note that this trend may be exclusive to the spirals detected in IR and is weak or missing for the spirals detected in mm continuum observations or mm line observations.  The pitch angle does not appear to be correlated with the radial location of the spirals (Figure~\ref{fig:spirals} (f)). There are seven disks where spirals are probed at more than one wavelength. For MWC~758, WaOph6, and HD~100453, spirals are detected in NIR and mm continuum observations which must probe significantly different heights in the disks (see Section \ref{sec:obs_primer}). Among the three disks, in HD~100453 the pitch angle of the spirals measured in NIR and mm molecular line observations ($14 - 18^\circ$ and $11 - 25^\circ$, respectively) are larger than the pitch angle measured in mm continuum ($5 - 7 ^\circ$; \citealt{rosotti2020}). For the other two disks, the pitch angles measured in NIR and mm continuum observations are comparable.

{\it Radial extent:}
Spirals are observed across a wide range of radii from 5 to $> 1000$~au (Figure~\ref{fig:spirals} (f)). The innermost radius at which spirals are observed depends either on the angular resolution of the observations, or, as in the case of NIR observations, on the size of the coronograph used to occult the central star.

{\it Pattern speed and time variation:} Pattern speed and time variation of spirals are not shown in Table \ref{tab:disks_with_spirals} or Figure \ref{fig:spirals} because such studies require long-term monitoring observations which are still scarce. However, they can potentially be used to determine the origin of the spirals. One example is presented in \citet{ren2020}, where the pattern speed of the spirals in the MWC 758 disk was inferred to be $0^\circ.22\pm0^\circ.03~{\rm yr}^{-1}$ over a 5 year time baseline. The pattern speed of the spirals is comparable to the orbital frequency at $172^{+18}_{-14}$~au assuming a $1.9~M_\odot$ central stellar mass. Because the spirals extend from about 40 to 80~au and the local Keplerian frequency over that radial extent ($0.7-2^\circ.0~{\rm yr}^{-1}$) is significantly larger than the observed spiral pattern speed, \citet{ren2020} concluded that the spirals are not locally excited, but launched at larger orbital radii and propagated inward. Another example for which the motion of spirals is monitored is SAO 206462. \citet{xie2021} explains the spiral motion with a single planet in a circular orbit at 86 au,  although there is tentative evidence that the two spirals have different pattern speed, which may require the presence of two planets or one planet in an eccentric orbit \citep{Zhu2022}.

\subsubsection{Crescents\index{Protoplanetary disk substructure!crescents}}
\label{sec:crescents_obs}

\begin{figure*}[t!]
\centering
\includegraphics[width=0.9\textwidth]{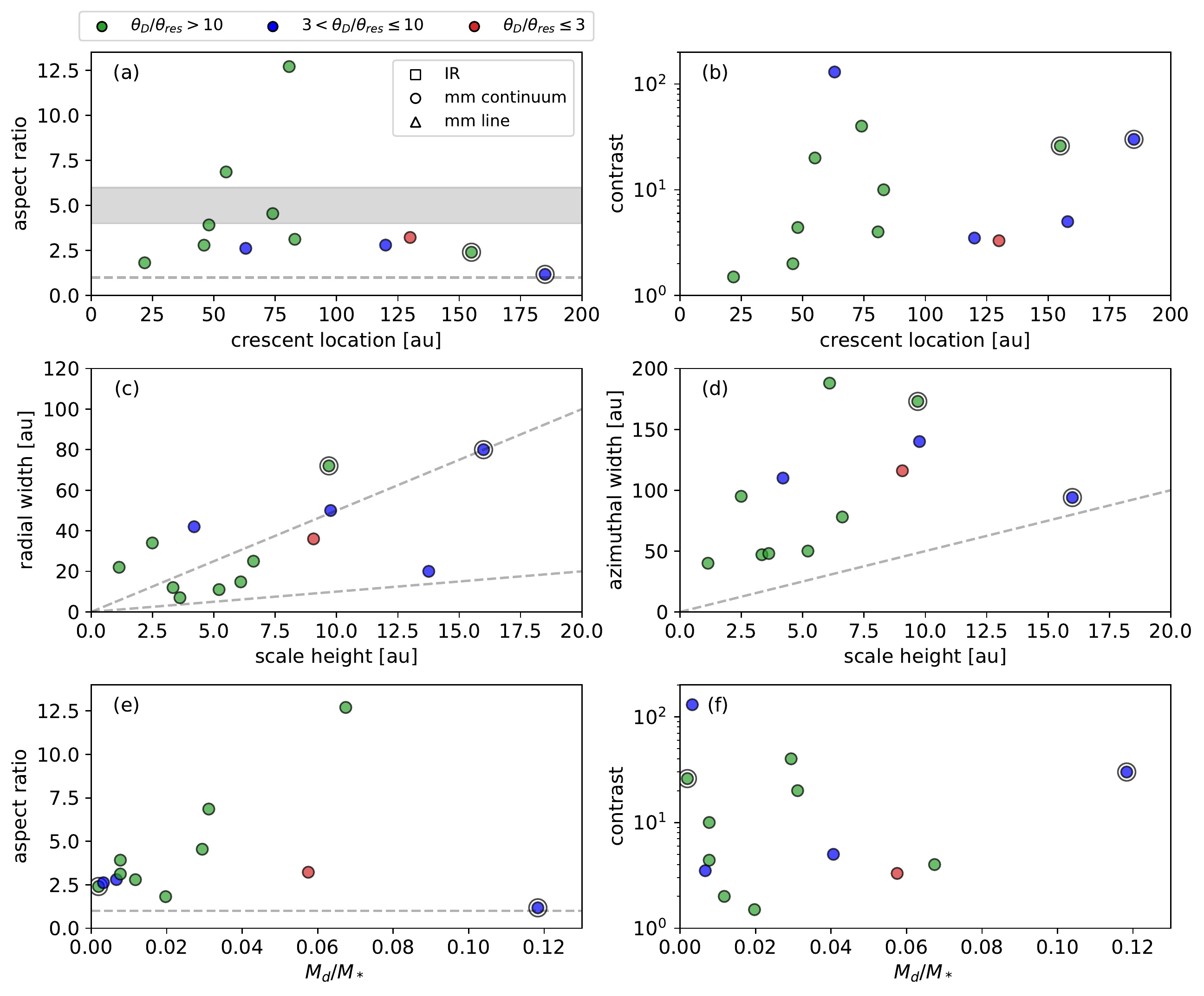}
\caption{(a) Aspect ratio of crescents, defined as the ratio between the azimuthal FWHM and the radial FWHM, as a function of the radial location of the crescents. Squares show the crescents detected in IR observations, circles show the crescents detected in mm continuum observations, and triangles show the crescents detected in mm line observations, respectively. Double symbols (e.g., two concentric circles) indicate binary systems. The same colors as in Figure \ref{fig:disk_fluxes1} are used to present the effective angular resolution used for observations. The gray shaded region shows aspect ratio between 4 and 6, a range with which a vortex can survive from elliptical instability (see Section \ref{sec:crescents}). (b) Contrast of crescents, defined as the ratio between the peak brightness and minimum brightness at the radius of the crescent, as a function of the radial location of the crescents. (c) The radial width of crescent $w_R$ as a function of the disk scale height $H$. The gas scale height $H = c_s/\Omega_K$, where $c_s = \sqrt{\mathcal{R}T_{\rm irr}/\mu}$, is computed using the stellar irradiation-dominated temperature $T_{\rm irr} = [\phi L_*/(4\pi r^2 \sigma)]^{1/4}$ \citep[e.g.,][]{chiang1997,dalessio1998,dullemond2001}. Here, $\phi$ accounts for the non-normal flaring angle. We adopt $\phi=0.02$, the value adopted in DSHARP papers \citep{Huang2018a,dullemondbirnstiel2018} for $H/R \approx 0.07$ at 100~au for a star of solar mass and luminosity; however, we find that adopting a different value does not change the trend that $w_R > H$. The two dotted lines show (top) $w_R = 5H$ and (bottom) $w_R = H$. (d) The azimuthal width of crescent $w_\phi$ as a function of the disk scale height $H$. The dotted line shows $w_\phi = 5H$. (e) The aspect ratio as a function of $M_{\rm d}/M_*$. The dotted lines shows aspect ratio of 1. (f) The contrast as a function of $M_{\rm d}/M_*$.}
\label{fig:crescent}
\end{figure*}

Crescents can be described as rings that have an azimuthal variation in intensity. Their identification is straightforward when the azimuthal intensity variation is apparent, as in the case of IRS~48 \citep{vandermarel2013}, HD~142527 \citep{Boehler2018}, and HD~143006 \citep[][see also Figure \ref{fig:substructures}]{andrews_dsharp,perez2018}. However, the distinction between rings and crescents is not straightforward when the azimuthal intensity contrast is close to unity. In addition, the apparent asymmetry can be due to dust's scattering phase function and/or the disk's geometry, instead of the intrinsic dust distribution in the disk. Disentangling all these factors is fruitful, but challenging. As such, in this review we will mainly focus on the crescents in the (sub-)millimeter continuum observations, because they are believed to largely trace the dust distribution at the midplane and less sensitive to scattering phase function or disk geometry. Also, we will narrow our discussion on highly asymmetric disks with a large brightness contrast, which is a clear indication that dust has been azimuthally trapped in some disk features. With this in mind, in  Table~\ref{tab:disks_with_crescents}, we compile a list of disks with crescents with the ratio between the maximum and minimum intensities along the azimuth equal to or greater than 1.5. 

{\it Occurrence:\index{Protoplanetary disk substructure!occurrence}}
Crescents are rarer than rings and spirals, and only a total of 19 crescents have been observed so far in 13 different disks. Due to the small number, it is not possible to draw a conclusion as to if the occurrence of crescents depends on the disk mass. However, it is worth noting that the occurrence rate of crescents is about 30\% for the disks with $M_d / M_\star > 0.05$, while the rate drops significantly to $< 5\%$ for the disks with $M_d / M_\star < 0.05$. 

{\it Multiplicity:}
Millimeter observations show one crescent per disk, except in MWC~758 for which two crescents are observed at different radial locations, 48 and 83~au.  There are two disks for which crescents are observed in IR observations: HD~143006 has 2 crescents \citep{benisty2018} and HD~139614 has 4 crescents \citep{muroarena2020}. We note that the crescent at 74~au in the HD~143006 disk is the only crescent observed at both IR and mm wavelengths thus far \citep{benisty2018,andrews_dsharp,perez2018}. For HD~139614, \citet{muroarena2020} suggested that the observed azimuthally asymmetry in the brightness can be explained with multiple misaligned/warped inner disks. In general, NIR scattered light observations are sensitive to shadows (see {\it Benisty et al.} in this book), so caution is needed when interpreting azimuthal intensity variations especially when a co-locating mm counterpart is not observed. Among all 13 disks that we compiled, there is no case where multiple crescents are observed at the same radial location within a single annular structure. 

{\it Radial location:} Crescents are discovered at various distances from 22~au to 185~au from the star, and they do not appear to be clustered at any particular radial locations (Figure \ref{fig:crescent} (a)). A potential pattern we see is that crescents are located at large distance in circumbinary disks (HD~34700A, HD142527). This may suggest that the presence of the companion star can have an influence on the location of crescents, although small number of the sample limits us to make any firm conclusions.

{\it Width and aspect ratio:}
Radial and azimuthal widths of a crescent are usually calculated by fitting 2D-Gaussian functions to the intensity. The ratio between the azimuthal and radial width defines the aspect ratio, which, as we will discuss in Section~\ref{sec:crescents}, is important to constrain the formation process.  Radial widths are measured for 13 out of a total of 19 crescents and vary between about 7 and 80~au (Figure \ref{fig:crescent} (c)). Azimuthal widths are measured for 12 crescents and vary between about 40 and 190~au (Figure \ref{fig:crescent} (d)). Combined together, the resulting aspect ratio ranges from $\sim1$ (i.e., a nearly circular crescent) to $\sim13$ (i.e., a crescent very elongated in azimuthal direction). However, most of the aspect ratios lie between 2 and 5 (Figure \ref{fig:crescent} (a) and (e)). The aspect ratio of crescents does not appear to be related to their radial locations (Figure \ref{fig:crescent} (a)) but there seems to be a tentative trend that crescents with larger aspect ratios are found in more massive disks (Figure \ref{fig:crescent} (e)). 

{\it Contrast:} The contrast of the observed crescents varies from a minimum of 1.5, the threshold used to distinguish between rings and crescents, to a maximum of $> 130$ in the case of IRS~48. In IRS~48 and few other disks, the observations only provide a lower limit for the contrast because no continuum emission is measured on the opposite side of crescent. There is a very tentative trend that the contrast of crescents decreases with $M_d/M_\star$ (Figure \ref{fig:crescent} (f)). Beside that, we find no correlation between the contrast of the crescents and the disk or stellar properties. 

\subsubsection{Other (sub)structures}

Although we focus on rings/gaps, spirals, and crescents in this review, we note that there are an increasing number of other types of (sub)structures. These include, but not limited to, kinematic substructure \citep[][see also {\it Pinte et al.} in this book]{pinte2018,pinte2019}, and tails or streamers \citep[e.g.,][see also {\it Pineda et al.} and {\it Pinte et al.} in this book]{akiyama2019,pineda2020,Ginski2021}.

\section{\textbf{Coupling Between Gas and Dust}} 
\label{sec:coupling}

When interpreting substructures probed by emission from the dust, it is important to keep in mind that the spatial distribution of the dust may be different to that of the gas. This is because the gas is supported against the central star's gravity by pressure gradients (both radially and vertically), whereas the dust is not. The resulting relative velocity between the gas and the dust causes aerodynamic drag\index{Aerodynamic drag}. In general, the mobility of dust particles relative to the gas can lead to concentrations of dust particles in narrow regions, amplifying the contrast of their emission and facilitating the detection of disk substructures in the dust emission. 

\subsection{Aerodynamic drag\index{Aerodynamic drag}} 
\label{sec:drag}

For a particle with radius, $a$, that is smaller than the mean free path of the ambient gas molecules, $\lambda_{\rm mfp}$ (more precisely $a < 9\lambda_{\rm mfp}/4$), the drag occurs as the particle collides with individual gas molecules. This is called Epstein drag \citep{epstein1924}. $\lambda_{\rm mfp}$ is of order of $\sim 1$~cm in the midplane at 1~au in the MMSN, and increases with the inverse of the gas density $1/\rho_g$. For the disk regions ($\gtrsim 10$~au) and particle sizes ($\lesssim 1$~cm) that this Chapter focuses on, it is safe to assume that particles experience Epstein drag. 
In the Epstein regime, the drag force $F_D$ is given by
\begin{equation}
\label{eqn:drag_force2}
    F_D = - {4 \over 3} \pi a^2 \rho_g \Delta v v_{\rm th},
\end{equation}
where $\rho_g$ is the gas density, $\Delta v$ is the relative velocity between the particle and the gas, $v_{\rm th} \equiv \sqrt{8/\pi} c_s$ is the thermal speed of the gas molecules and $c_s$ is the sound speed.

The drag force acts in the opposite direction to $\Delta v$ such that the solid particle loses (angular) momentum. The timescale over which this happens is called stopping time $t_{\rm stop} \equiv m \Delta v/|F_D|$, where $m$ is the mass of the solid particle of interest. The stopping time is the key parameter describing the coupling between the gas and the dust. 
Using Equation (\ref{eqn:drag_force2}), the stopping time is given by
\begin{equation}
\label{eqn:tstop}
    t_{\rm stop} = {\rho_s \over \rho_g} {a \over v_{\rm th}},
\end{equation}
where $\rho_s$ is the mean internal density of the particle. In the limit of short stopping time $t_{\rm stop} \ll t_{\rm dyn}$, where $t_{\rm dyn}$ is the dynamical timescale, $\Delta v$ vanishes (because otherwise the drag force would diverge) and the particle follows the gas motion. In the opposite limit of long stopping time $t_{\rm stop} \gg t_{\rm dyn}$, the aerodynamic drag\index{Aerodynamic drag} is negligible and the particle decouples from the gas. In the intermediate regime where the stopping time is comparable to the dynamical timescale $t_{\rm stop} \simeq t_{\rm dyn}$, the particle can move relative to the gas but its motion is still largely affected by aerodynamic drag\index{Aerodynamic drag}. This partial coupling drives secular drift motion and plays a crucial role in shaping substructures. 

It is often useful to define a dimensionless stopping time, which is also called Stokes number, 
\begin{equation}
\label{eqn:stokes_number}
    {\rm St} = \Omega_K t_{\rm stop},
\end{equation}
where $\Omega_K$ is the local Keplerian frequency. In a vertically isothermal disk where the gas density $\rho_g$ relates to the vertically integrated surface density $\Sigma_g$ as $\rho_g = \Sigma_g / (\sqrt{2 \pi} H) \exp [-Z^2/(2H^2)]$, where $Z$ denotes the height in the disk and $H$ denotes the scale height of the gas, the Stokes number is given by
\begin{eqnarray}
    {\rm St} & = & {\pi \rho_s a \over 2\Sigma_g} \exp\left( {Z^2 \over 2 H^2} \right) \nonumber \\
    & = & 0.16 \exp\left( {Z^2 \over 2 H^2} \right) \left( {\rho_s \over 1~{\rm g~cm}^{-3}} \right) \left( {a \over 1~{\rm mm}} \right)
     \nonumber \\
    & & \times \left( {\Sigma_g \over 1~{\rm g~cm}^{-2}} \right)^{-1}.
\end{eqnarray}
For the outer disk regions where $\Sigma_g \simeq 1~{\rm g~cm}^{-2}$, we thus expect (sub)mm-sized particles to have a Stokes number close to unity (i.e., $t_{\rm stop} \simeq t_{\rm dyn}$) and to experience strong aerodynamic drag\index{Aerodynamic drag}, whereas $\mu$m-sized particles have a Stokes number orders of magnitude smaller than unity (i.e., $t_{\rm stop} \ll t_{\rm dyn}$) and couple well to the gas.

\subsection{Horizontal Drift\index{Dust!horizontal drift}} 
\label{sec:drift}
The most important consequence of aerodynamic drag\index{Aerodynamic drag} in protoplanetary disks is the radial drift of solid particles.
The orbital velocity of the gas $v_{\rm g, \phi}$ is determined by the balance between the centrifugal force, stellar gravity, and the radial gas pressure gradient, ${\rm d}P/{\rm d}R$, with
\begin{equation}
    {v_{\rm g, \phi}^2 \over R} = {GM_*R \over (R^2+Z^2)^{3/2}} + {1 \over \rho_g} {{\rm d}P \over {\rm d}R}.
\label{eqn:radial_balance}
\end{equation}
On global scales, the density and temperature of the gas decrease as a function of radius, so the pressure gradient ${\rm d}P/{\rm d}R$ is negative. This means that the gas orbits at a sub-Keplerian speed. For solid particles, on the other hand, the pressure gradient is negligible and so particles orbit at the Keplerian velocity in the absence of aerodynamic drag\index{Aerodynamic drag}. As a result, the sub-Keplerian gas motion acts on solid particles as a headwind, removing angular momentum from solid particles and causing particles to drift inward. 

The radial drift velocity of solid particles arising from the aerodynamic drag\index{Aerodynamic drag} can be written as
\begin{equation}
\label{eqn:vrad_dust}
    v_{\rm d,R} = {{\rm St}^{-1} v_{\rm g,R} - \eta v_{\rm K} \over {\rm St}+{\rm St}^{-1}},
\end{equation}
where 
\begin{equation}
\eta \equiv -(\rho_g R\Omega_K^2)^{-1} \frac{{\rm d}P}{{\rm d}R} = -(c_s/v_{\rm K})^2 \frac{{\rm d}\ln P}{{\rm d}\ln R}
\end{equation}
\citep{whipple1972,adachi1976,weidenschilling1977,nakagawa1986,takeuchi02}.
For small particles with ${\rm St} \lesssim \alpha \ll 1$, Equation (\ref{eqn:vrad_dust}) reduces to $v_{\rm d,R} \simeq v_{\rm g,R} - {\rm St} \eta v_{\rm K} \simeq v_{\rm g,R}$\footnote{
Here, we have used that 
% These are the particles satisfying 
${\rm St}|\eta v_K| \lesssim |v_{\rm g,R}|$ for ${\rm St} \lesssim \alpha$. 
For a disk where gas accretion can be described with the viscous stress adopting the dimensionless parameter $\alpha$ introduced by \citet{shakura1973}, the radial velocity of the gas is given by $v_{\rm g,R} = -(3/2) \alpha c_s (H/R)$. With this, the condition for 
${\rm St}|\eta v_K| \lesssim |v_{\rm g,R}|$
becomes  ${\rm St} \lesssim (3/2) (\alpha/|\eta|) (H/R)^2 = (3/2) (\alpha/|{\rm d}\ln P / {\rm d} \ln R|) \sim \alpha$.}, so particles follow the gas motion. For larger particles with ${\rm St} \gtrsim \alpha$, where $\alpha$ is the dimensionless parameter describing the viscous stress \citep{shakura1973}, the radial drift velocity approximates to $v_{\rm d,R} \simeq -\eta v_{\rm K}/({\rm St} + {\rm St}^{-1})$ so they decouple from the gas. Note that in the large particle regime, the radial drift velocity maximizes when ${\rm St} = 1$. The timescale for the radial drift is given by $t_{\rm drift} = R/|v_{\rm d,R}|$. For particles with ${\rm St} = 1$, $t_{\rm drift} \simeq 2/(\eta \Omega_{\rm K})$. Since $\eta \simeq 10^{-3} - 10^{-1}$ under typical protoplanetary disk conditions, particles with appropriate sizes can drift (and be lost from the disk) within about $10^3$ orbital periods, corresponding to 1,000~years at 1~au and 1 million years at 100~au around a solar-mass star. Such a short drift timescale can prevent particles from growing in size. As such, this is often called a ``radial-drift barrier'' or a ``meter-size barrier'' in planet/planetesimal formation, where the latter is named after the fact that ${\rm St} = 1$ particles at 1~au in the MMSN have a size of about one meter \citep{weidenschilling1977}.

So far, we have assumed that the pressure gradient ${\rm d}P/{\rm d}R$ is negative. However, when there is an inversion in the pressure gradient (${\rm d}P/{\rm d}R > 0$), the gas orbits at a super-Keplerian speed. In this case, particles experience a tailwind and thus drift outward. Although a global inversion of ${\rm d}P/{\rm d}R$ is unlikely, protoplanetary disks can have a local pressure gradient inversion around pressure bumps formed, for example, at the outer edge of a gap carved by a planet. At the peak of a pressure bump, the pressure gradient is zero and the aerodynamic drag\index{Aerodynamic drag} vanishes. Thus, pressure bumps offer prime sites to trap particles \citep{whipple1972,pinilla2012}. Using ALMA\index{ALMA} molecular line observations, \citet{teague2018} measured the rotational velocity of the gas in the disk around HD~163296 and showed that the pressure peaks co-locate with the continuum rings in the disk (see also \citealt{teague2018b,rosotti2020b}). This provides direct evidence that particles experience radial drift and are trapped in gas pressure bumps.

In addition to the radial drift, one can consider azimuthal drift. Although azimuthal drift is  negligible in axisymmetric disks (${\rm d}P/{\rm d}\phi = 0$), azimuthal drift can become important when a disk has asymmetric structures, such as crescents\index{Protoplanetary disk substructure!crescents} and spirals\index{Protoplanetary disk substructure!spirals}. When there exists a high-pressure anticyclonic vortex in the disk, particles drift toward the pressure peak in both the radial and azimuthal directions and can be trapped therein  \citep{barge1995,adams1995,tanga1996}. Spirals offer another implication, although the situation can be different from vortices. While vortices orbit at the local Keplerian speed, spirals can orbit at a speed that is significantly different from the local Keplerian speed. For example, at a finite distance from a planet, a spiral launched by the planet orbits at the orbital frequency of the planet $\Omega_p$, not at the local Keplerian speed $\Omega_{\rm K}$. As such, particles with a stopping time shorter than the spiral crossing time, $t_{\rm stop} \lesssim \Delta \phi_s / |\Omega_p - \Omega_{\rm K}|$ where $\Delta\phi_s$ is the azimuthal width of the spiral, can be trapped by the spiral,  while particles with a longer stopping time than the spiral crossing time do not have sufficient time to respond to the perturbation and thus would be poorly coupled to the spiral \citep[see e.g.,][]{isella_turner2018,Sturm2020}. 

\subsection{Vertical Settling\index{Dust!vertical settling}} 
\label{sec:settling}

Aerodynamic drag\index{Aerodynamic drag} also changes the vertical distribution of particles relative to the gas. A particle located at height $Z~(\ll R)$ above the midplane feels the vertical component of stellar gravity, $F_{\rm grav} = -m \Omega_K^2 Z$. Equating the gravitational force and the drag force in Equation (\ref{eqn:drag_force2}), the vertical settling velocity is given by $v_{d,Z} = {\rm St}\Omega_K Z$. The vertical settling timescale can be then written as $t_{\rm sett} = |Z/v_{d,Z}| = 1/({\rm St}\Omega_K)$. At one scale height above the midplane in the MMSN, the settling timescale for a $\mu$m-sized particle is about one million years, independent of $R$\footnote{Note that this is because $\Sigma_g \propto R^{-3/2}$ for the MMSN and $t_{\rm sett} \propto \Sigma_g R^{3/2}$. The radial dependency of the vertical settling timescale changes for different gas surface density profiles.}. For a mm-sized particle, the settling timescale is only about 1,000 years since the settling timescale is inversely proportional to the particle size. This suggests that in the absence of turbulent stirring, particles should settle on timescales shorter than the typical lifetime of protoplanetary disks\index{Protoplanetary disks} ($\simeq$ a few to 10~Myr; see {\it Manara et al.} in this book). On the other hand, optical and infrared observations clearly show that $\mu$m-sized particles scatter stellar photons from a few gas scale heights above the midplane (\citealt{avenhaus2018,rich2021}; see also {\it Benisty et al.} in this book), indicating that turbulence may drive vertical diffusion of dust. This is the topic of the following subsection.

\subsection{Turbulent Diffusion and Dust Concentration \index{Dust!turbulent diffusion}}\label{sec:dust_concentration}

As we mentioned in Section~\ref{sec:drift}, particles can drift radially and azimuthally toward local pressure maxima and be trapped therein. However, particle rings and crescents\index{Protoplanetary disk substructure!crescents} cannot be infinitely thin when turbulence is present; instead, their radial/azimuthal widths are determined by the balance between radial/azimuthal drift and turbulent diffusion \citep{dullemondbirnstiel2018}. In the vicinity of a pressure maximum at radius $R_0$ with radial width $w$, whose radial profile is described by a Gaussian function $P(R) = P(R_0)\exp[-(R-R_0)^2/(2w^2)]$, the pressure gradient is ${\rm d} \ln P/{\rm d} R = -(R-R_0)/w^2$. Inserting this into Equation (\ref{eqn:vrad_dust}), at the radial distance $\Delta R$ from the pressure maximum, the radial drift velocity is $v_{{\rm d},R} \simeq ({\rm St}/({\rm St}^2 + 1))(H/w)^2 \Omega \Delta R$ and the drift timescale is $t_{\rm drift} = |\Delta R /v_{{\rm d},R}| \simeq (w/H)^2({\rm St}^2 + 1)/({\rm St}\Omega)$. As explained in Section \ref{sec:drift},  this radial velocity and drift timescale are applicable to the particles with ${\rm St} \gtrsim \alpha$. The radial diffusion timescale is given by $t_{{\rm diff},R} = w_d^2/D_{d,R}$, where $w_d$ is the radial width of the dust ring and $D_{d,R}$ is the radial diffusion coefficient for dust, which relates to the radial diffusion coefficient for gas $D_{g,R}$ as $D_{d,R} = D_{g,R}/(1+{\rm St}^2)$ \citep{youdin2007}. From the balance of the two timescales, the width of the dust ring in drift--diffusion equilibrium is given by
\begin{equation}
\label{eqn:w_d}
    w_d \simeq \left(\frac{\alpha_R}{\rm St}\right)^{1/2}w,
\end{equation}
where $\alpha_R \equiv D_{g,R}/(H^2\Omega_K)$ is the dimensionless radial diffusion coefficient defined in a similar way to the Shakura \& Sunyaev $\alpha$ parameter.
Similarly, the azimuthal extent of a dust clump that is in drift--diffusion equilibrium, $\Delta \phi_d$, is given by $\Delta \phi_d \simeq (\alpha_\phi/{\rm St})^{1/2} \Delta \phi$, where $\alpha_\phi$ is the dimensionless azimuthal diffusion coefficient and $\Delta \phi$ is the azimuthal extent of the associated gas clump \citep{birnstiel2013,lyra2013}. As seen in the derived relations between the dust and gas widths,  the dust substructure has a width comparable to or smaller than the width of the associated gas substructure. 

Equation (\ref{eqn:w_d}) tells us that if we infer $w_d$ from continuum observations and $w$ from molecular line observations, which is possible by measuring the modulation in the rotational velocity of the gas due to pressure gradients (see Equation \ref{eqn:radial_balance}; see also \citealt{teague2018}), we can directly obtain the ratio $\alpha_R/{\rm St}$. \citet{rosotti2020b} used this method and found that $\alpha_R/{\rm St} \sim 0.1$ for three continuum rings in the HD~163296 disk and two continuum rings in the AS~209 disk. Although the Stokes number has uncertainties that arise from the uncertainties in the gas density and the size of particles that the continuum observation is most sensitive to, the estimated $\alpha_R/{\rm St}$ can put constraints on the degree of radial diffusion. Assuming that the continuum observations probe particles of 1~mm in size, \citet{rosotti2020b} found $\alpha_R$ to be $(1 - {\rm few}) \times 10^{-4}$ for the HD~163296 disk.

One can also consider vertical diffusion of solid particles. For a particle layer with a vertical height $H_d$, the timescale of vertical diffusion is given by $t_{{\rm diff},Z} \sim H_d^2/D_{Z}$, where $D_{Z}$ is the vertical turbulent diffusion coefficient. In the equilibrium state where vertical diffusion balances settling, we have $t_{\rm sett} \simeq t_{{\rm diff},Z}$ which gives
\begin{equation}
    \label{eq:H_d}
    H_d \simeq \left(\frac{\alpha_Z}{{\rm St}}\right)^{1/2} H,
\end{equation}
where $\alpha_Z \equiv D_Z/(H^2\Omega_K)$ is the dimensionless vertical diffusion coefficient. \citet{pinte2016} and \citet{isella2016} measured the vertical thicknesses of the dust rings in the HL~Tau and HD~163296 disks from the sharpness of the emission rings and gaps in the ALMA\index{ALMA} millimeter continuum images. They found that (sub-)mm-sized dust particles are settled substantially, with $H_d \la 1$~au at 100~au, corresponding to $\alpha_Z$ of a few $\times 10^{-4}$.  \citet{doi2021} carried out a similar experiment and estimated the vertical thicknesses of the dust rings in the HD~163296 disk (see also \citealt{ohashi2019}). They found  $\alpha_Z/{\rm St} > 2.4$ at the 67~au ring and $\alpha_Z/{\rm St} < 1.1\times10^{-2}$ at the 100~au ring. Assuming that the continuum observations probe particles of 1~mm in size, these numbers correspond to $\alpha_Z > 1.8 \times 10^{-2}$ and $\alpha_Z < 1.8  \times 10^{-4}$, respectively, which might suggest radially varying turbulent diffusion in the disk.

When both the radial width ratio $w_d/w$ and the vertical thickness ratio $H_d/H$ of a ring are measured, we can use them to infer the (an)isotropy of the underlying turbulence: $(H_d/H)/(w_d/w) \simeq (\alpha_Z/\alpha_R)^{1/2}$. Note that the Stokes number, which is often poorly defined as mentioned earlier, cancels out in this relation. Combining the results from \citet{rosotti2020b} and \citet{doi2021} for the continuum rings at 67 and 100~au in the HD 163296 disk, we can infer $\alpha_Z/\alpha_R \gtrsim 10$ for the 67~au ring and $\alpha_Z/\alpha_R \lesssim 0.3$ for the 100~au ring. Such an anisotropy, combined with numerical simulations, may help to determine the origin of the turbulence. 

\begin{figure*}[ht!]
\centering
\includegraphics[width=0.9\textwidth]{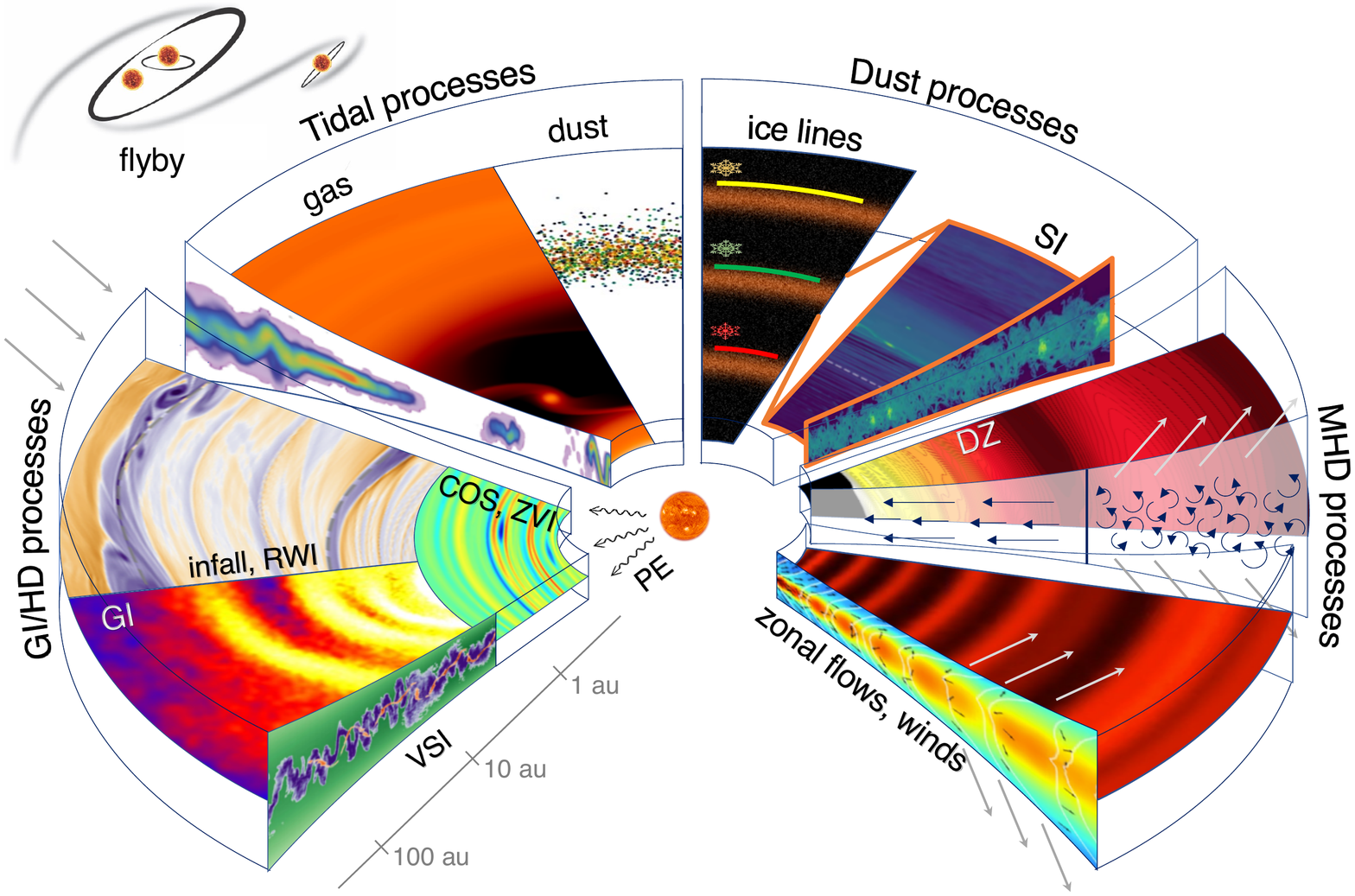}
\vspace{-1.5cm}
\caption{A cartoon showing various substructure-forming processes discussed in this Chapter. In the clockwise direction from the left, we present four main categories of substructure-forming mechanisms: hydrodynamic processes (HD, Section \ref{sec:hydro}) including gravitational instability\index{Gravitational instability} (GI, Section \ref{sec:self-gravity}; figure from \citealt{rice2004}), infall\index{Infall} and Rossby wave instability\index{Rossby wave instability} (RWI; figure from \citealt{kuznetsova2022}), convective overstability\index{Convective overstability} (COS\index{Convective overstability}), zombie vortex instability\index{Zombie vortex instability} (ZVI; figure from \citealt{marcus2015}), photoevaporation\index{Photoevaporation} (PE), and vertical shear instability\index{Vertical shear instability} (VSI; figure from \citealt{flock2017}); tidal processes (Section \ref{sec:companion}) including planet-disk interaction\index{Planet-disk interaction} (figure from \citealt{bae2019}), stellar companions, and stellar flyby\index{Stellar flyby}; dust processes (Section \ref{sec:dust}) including ice lines and streaming instability\index{Streaming instability} (SI; figure from \citealt{li_youdin2021}); magneto-hydrodynamic processes (MHD, Section \ref{sec:MHD}) including dead-zones\index{Dead-zone} (DZ; figure from \citealt{flock2015}), zonal flows\index{Zonal flows} (figure from \citealt{suriano2018}), and winds\index{Magnetized winds}. Figures are reproduced by permission of the AAS, A\&A, and MNRAS.}
\label{fig:cartoon}
\end{figure*}

\section{Hydrodynamic Processes}\label{sec:hydro}

Let us now move onto a discussion of substructure-forming mechanisms. To provide a broad visual overview, in Figure \ref{fig:cartoon} we present a three-dimensional pie chart where we illustrate various substructure-forming processes that we will review in this Chapter. Following the way we structure this Chapter, we group different processes into several categories: hydrodynamic processes and gravitational instability (GI), magnetohydrodynamic processes, tidal processes, and processes induced by dust particles.

In this section, we start by introducing hydrodynamic processes that are capable of creating disk substructures. Assuming a smooth disk structure and simple thermodynamics, Keplarian disks should be linearly stable according to the Rayleigh stability criterion ($\partial (R^4 \Omega^2)/\partial R>0$; \citealt{rayleigh}), in the absence of magnetic fields. However, the disk can be unstable when inhomogeneous density structure, more complicated thermodynamics and/or three-dimensionality are considered. Examples include Rossby wave instability\index{Rossby wave instability} (RWI), vertical shear instability\index{Vertical shear instability} (VSI), convective overstability\index{Convective overstability} (COS\index{Convective overstability}), and zombie vortex instability\index{Zombie vortex instability} (ZVI), which we will discuss in the subsections below. In addition to the aforementioned internally-operating processes, there are externally-driven processes that can produce disk substructures such as infall\index{Infall} and photoevaporation\index{Photoevaporation}. In each of the subsections below, we will introduce how the process operates, describe under which conditions the process operates, and summarize the properties of the substructures formed by the process. For in-depth discussion on the numerical studies of VSI\index{Vertical shear instability}, COS\index{Convective overstability}, and ZVI, we refer readers to the Chapter by {\it Lesur et al.} in this book.

\subsection{Rossby Wave Instability\index{Rossby wave instability}}
\label{sec:RWI}

The RWI arises from a local minimum in the radial profile of the potential vortensity $\eta = \kappa^2 S^{-2/\gamma}/(2\Omega \Sigma_g)$ \citep{lovelace1999,li2000}, where $\kappa  = R^{-3}\partial(R^4 \Omega^2)$ $/\partial R$ is the epicyclic frequency, $S = P/\Sigma_g^\gamma$ is the entropy, and $\gamma$ is the adiabatic index. From its definition, one can see that the radial profile of the potential vortensity is determined by the density, temperature, and orbital frequency of the disk gas, which are related with each other through the radial force balance in Equation (\ref{eqn:radial_balance}). Thus, perturbations given to the gas density and/or temperature transition to shear in the rotation velocity in order to maintain the radial force balance. When the velocity shear is sufficiently large, a Kelvin-Helmholtz type instability can arise from the shear, creating anticyclonic vortices. 

Where can such strong radial potential vortensity variations develop in protoplanetary disks? Previous studies showed that they can develop (1) at the edges of MRI-dead zone\index{Dead-zone} where the accretion rate can rapidly change \citep[][see also Section \ref{sec:deadzone}]{varniere2006,inaba2006,lyra2008,lyra2012,meheut2010,miranda2017,pierens2018}, (2) at the edge of planet-induced gaps where pressure bumps are sustained  \citep{deval-borro2007,lyra2009,fu2014a,fu2014b,lin2014,bae2016a}, (3) at the edge of the region subject to mass addition via infalling flows onto the disk (\citealt{bae2015,kuznetsova2022}, see also Section \ref{sec:infall}), and (4) between corrugated flows induced by the VSI\index{Vertical shear instability} \citep[][see also Section \ref{sec:vsi}]{richard2016}. 

Properties of RWI-induced vortices are often studied by adopting a Gaussian density perturbation to an otherwise smooth disk as an initial condition, with the width and amplitude of the perturbation as variables. \citet{ono2016,ono2018} carried out parameter studies using linear stability analyses and two-dimensional simulations, under a barotropic setup. They showed that disks with a density perturbation of the order of $\Delta \Sigma_g /\Sigma_g \gtrsim 1$ over a radial length scale of the order of the local scale height can become unstable to the RWI. For such perturbations, the RWI grows rapidly with a typical growth timescale of $\sim10~\Omega^{-1}$. During the linear phase of the instability, the disk region subject to a steep potential vortensity variation breaks into multiple anticyclonic vortices, with the most unstable azimuthal mode number increasing with the strength of the perturbation. As the instability saturates, vortices eventually merge into a single vortex over a typical timescale of $\sim100~\Omega^{-1}$. For the vortices formed after the final merger, the radial half-width of the vortices has a maximum value of about $2-2.5$ scale heights \citep{surville2015,ono2018}. These simulations showed that the aspect ratio ranges from 2 to $\sim 20$, although the majority have an aspect ratio in the range $4-8$. Vortices can survive as long as $\gtrsim 1000$ orbital periods, although their lifetime depends strongly on the level of viscous dissipation \citep[e.g.,][]{lin2014} or disk turbulence \citep{zhu2014b}.

As vortices form via the RWI\index{Rossby wave instability}, they can excite spiral waves. While vortices extend within the region that is subsonic with respect to the vortex center, they can perturb the sonic line and launch free traveling waves to the supersonic region \citep{paardekooper2010,ono2018}. After such free waves are launched, their shape (i.e., pitch angle) can be described in the same way as the spirals launched by a planet, which we will discuss in Section \ref{sec:companion}. However, although the spirals driven by vortices have similar shapes as the ones driven by companions, they are much weaker than planet-driven spirals: the surface density contrast of vortex-driven spirals are comparable to those driven by a sub-thermal mass planet (typically a few to a few tens of Earth masses; \citealt{huang2019}). Thus, spirals driven by vortices are less likely to be observable compared with planetary spirals although whether three-dimensional effects can enhance the strength of vortex spirals at the disk surface needs to be studied in the future. Alternatively, it is possible that a vortex itself may appear as an one-arm spiral in scattered light when viewed at a finite inclination \citep{metea2022}.

\subsection{Vertical Shear Instability\index{Vertical shear instability}}
\label{sec:vsi}

The VSI\index{Vertical shear instability} is a linear instability originating from vertical gradients in the disk's rotational velocity \citep{urpin1998,arlt2004,nelson2013}. By solving Equation (\ref{eqn:radial_balance}) along with a similar equation of force balance in the vertical direction, one can show that vertical differential rotation will always arise unless the disk is globally isothermal, a condition that is very unlikely to be found in protoplanetary disks. Protoplanetary disks thus offer favorable conditions for the VSI to develop. However, the VSI requires rapid gas cooling so that the vertical differential rotation is sustained. When cooling is inefficient, vertical buoyancy stabilizes the vertical gas motion and the VSI is suppressed \citep{linyoudin2015}. For typical protoplanetary disk conditions, this cooling requirement is met in the outer regions of protoplanetary disks at $\sim10-100$~au \citep{malygin2017,pfeil2019}. In addition to buoyancy, dust settling and the subsequent back-reaction of the dust on the gas \citep{lin2019}, dust growth \citep{fukuhara2021}, and strong magnetic fields \citep{latter2018,cui2021b} are known to stabilize the VSI.  

When the VSI\index{Vertical shear instability} is fully grown to the non-linear state, the most prominent outcome is nearly axisymmetric vertical corrugation modes (i.e., radially alternating upward and downward gas flows) which have much smaller horizontal wavelengths than vertical wavelengths ($\lambda_R/\lambda_Z \ll 1$; \citealt{nelson2013}). Numerical simulations showed that the VSI induces vertical corrugation modes that are as strong as $\sim10 - 100~{\rm m~s}^{-1}$ \citep{flock2017,flock2020}, which are observable using molecular line observations with ALMA\index{ALMA} \citep{barraza2021}. Because the VSI predominantly drives vertical gas flows, disks with a low inclination offer favorable conditions for the detection of VSI-driven velocity perturbations. Vertical corrugation modes can cause strong vertical diffusion of dust particles, characterized by $\alpha_Z \simeq 0.01$ \citep{stoll2017,flock2020}. For the disk models considered in \citet{flock2017,flock2020}, mm-sized particles with a Stokes number of about 0.1 are lofted up to $Z/R \sim 0.1$ with a scale height $H_d \simeq 0.03~R$. The perturbed dust distribution manifests as concentric bright rings and dark gaps in radio continuum observations \citep{flock2017,blanco2021}.  However, the perturbations driven by the VSI are intrinsically transient, and particles do not experience long-term trapping or net radial concentration \citep{flock2017} unless they are trapped in VSI-induced vortices. 

The velocity perturbations driven by the VSI can give rise to potential vorticity perturbations that break up into  vorticies \citep{richard2016}, in a manner that is similar to the RWI introduced in the previous subsection.  Recent high-resolution numerical simulations showed that the VSI creates long-lived vortices with a lifetime of $\gtrsim 100$ orbits. The VSI-induced vortices typically have a radial size of about a local gas scale height with an azimuthal size of about 10 local gas scale heights, resulting in an aspect ratio of $\simeq 10$ \citep{flock2020,manger2020}. Note that these aspect ratio values are measured from the potential vorticity of the gas, not from the distribution of solid particles trapped therein, so caution is needed when connecting these simulations to the crescents observed in continuum observations\index{Protoplanetary disk substructure!crescents}.

\subsection{Convective Overstability\index{Convective overstability}}
\label{sec:cos}

The COS\index{Convective overstability} is a linear, axisymmetric instability in which radial buoyancy leads to exponential amplification of epicyclic oscillations \citep{klahr2014,lyra2014}. The COS\index{Convective overstability} operates when the cooling time of the gas is comparable to the epicyclic oscillation period ($\simeq$ Keplerian timescale), while it is suppressed in the limits of short and long cooling times: too rapid cooling removes buoyancy, whereas too slow cooling makes the work done on the oscillating gas by the buoyant force over one epicycle vanishingly small. Due to this requirement, the COS\index{Convective overstability} can be active around the midplane at $\sim  1-10$~au and above the midplane at $r \sim 10-100$~au \citep{malygin2017,lyra2019,pfeil2019}. 

Three-dimensional hydrodynamic simulations show that the COS\index{Convective overstability} produces a self-sustained, large-scale vortex in its nonlinear saturated state \citep{lyra2014,raettig2021}. Although the vortices initially created by the instability are small, they merge together and form larger vortices. Three-dimensional simulations using a local shearing-box show that eventually a vortex of radial extent $\approx H$ forms \citep{raettig2021}, although the final vortex size may be affected by the box size. Given that the COS\index{Convective overstability} is expected to operate at $\sim 1-10$~au, the radial extent of COS\index{Convective overstability}-driven vortices would generally be less than 1~au, making it very challenging to observe them using ALMA\index{ALMA}. As we mentioned earlier, the COS\index{Convective overstability} may operate in the surface layers in outer disk regions of $\sim 10 - 100$~au; however, whether the COS\index{Convective overstability} triggered at high altitudes can produce vortex columns extending down to the midplane and collect large particles there to produce observable signatures in radio continuum observations will have to be tested in the future.

\subsection{Zombie Vortex Instability\index{Zombie vortex instability}}
\label{sec:zvi}
The ZVI is another purely hydrodynamic instability that can operate in protoplanetary disks \citep{barranco2005,marcus2013,marcus2015,marcus2016}. The instability is triggered when the so-called baroclinic critical layers are excited. These baroclinic critical layers are where mathematical singularities exist, and as they are excited the singularities generate vortex layers. The vortices formed in the vortex layers excite new critical layers where additional vortices can form and turbulence arises. This self-replicating nature is why the instability is named the ``zombie'' vortex instability. Under typical protoplanetary disk conditions, the radial width of the vortex and the spacing between the critical layers are expected to be of order the vertical pressure scale height, with an aspect ratio of $4-5$ \citep{marcus2015}.

The ZVI\index{Zombie vortex instability} requires strong vertical buoyancy and cooling times much longer than the Keplerian timescale \citep{lesur2016,barranco2018}. As such, the ZVI likely operates in the innermost region of protoplanetary disks within $\lesssim 1$~au \citep{malygin2017}. It is thus unlikely that the crescents observed at tens to hundreds au are associated with the ZVI.

\subsection{Eccentric Modes\index{Eccentric modes}}
\label{sec:eccentric_modes}

An eccentric mode refers to the coherent precession of gas on eccentric orbits. Historically, it has been invoked to explain spirals in spiral galaxies (see review by \citealt{shu2016}). In a locally isothermal disk, \cite{lin2015} has shown that eccentric modes can be excited as the background temperature gradient can exert negative angular momentum to the mode to destabilize it. One of the reasons why eccentric modes are of interest is that they are less susceptible to dissipation process (e.g., viscous dissipation) and can be long-lived \citep{lubow1991,tremaine2001}. The eccentric modes form a single spiral arm in the disk and precess at rates that are much lower than the local orbital frequency \cite{lee2019}. Due to the slow orbital motion, it is expected that the spiral associated with eccentric modes traps small particles only (${\rm St} \ll 1$; see Section \ref{sec:coupling}). \cite{lin2015} showed that spirals associated with eccentric modes satisfy $k R \sim \pi G \Sigma R / c_{s}^2 \sim 1/(HQ)$ where $k$ is the wavenumber.  With $Q\sim1$ and $H/R \ll 1$, the spirals are tightly wound with pitch angle $\lesssim 5^\circ$. At this moment, it is unclear if more open spirals can form with the eccentric mode. Eccentric modes can also evolve with time. Simulations by \cite{li2021} have shown that the spirals become multiple rings over the timescale of several thousands orbits.

\subsection{Infall\index{Infall}}
\label{sec:infall}

So far, we have focused on hydrodynamic processes that originate within protoplantary disks. However, there are important substructure-forming processes that are  externally driven. The first example we introduce in this subsection is infalling flows onto the disk. In the traditional rotating singular isothermal sphere model of star and disk formation  \citep{ulrich1976,shu1977,cassen1981}, the rotational velocity of the infalling flow when it reaches the disk midplane at the centrifugal radius (i.e., outer edge of the mass landing on the disk) is the same as the local Keplerian velocity. As a result, the infalling material tends to pile up near the centrifugal radius and create a pressure bump. Using two-dimensional hydrodynamic simulations implementing infall from a rotating singular isothermal cloud, \citet{bae2015} showed that a strong pressure bump can develop and trigger the RWI near the outer edge of the infall, which, in turn, creates vortices. \cite{kuznetsova2022} further investigated this scenario by adopting various infall models in which infalling materials have different specific angular momentum. The vortices induced by infall generally share similar properties to other RWI-induced vortices which are summarized in Section \ref{sec:RWI}. 

When the specific angular momentum of the infalling material significantly differs from that of the disk gas, the mismatch can lead to a rapid redistribution of the infalling gas. Using two-dimensional hydrodynamic simulations, \citet{lesur2015} showed that such a redistribution can occur via large-amplitude spiral density waves (see also \citealt{kuznetsova2022}). These spirals can propagate inward to radii ten times smaller than the radius at which infalling material generates shock to the gas. By numerically solving the ordinary equations in locally isothermal disks under the constraint that the flow must satisfy the Rankine-Hugoniot conditions through the shock, \citet{hennebelle2016} showed that the pitch angle of infall-driven spirals are generally small, $3-9^\circ$ for $H/R = 0.05 - 0.15$. The dominant azimuthal mode depends on the detailed properties of the infalling flow. A large number of spirals ($\gtrsim10$) can develop when the infalling flow is axisymmetric \citep{lesur2015}. On the other hand, the $m=1$ mode can dominate when the infalling flow is asymmetric \citep{hennebelle2017}.

\subsection{Photoevaporation\index{Photoevaporation}}
\label{sec:photoevaporation}

Photoevaporation\index{Photoevaporation} is another externally driven substructure-forming hydrodynamic process. Since photoevaporation is reviewed extensively in {\it Protostars and Planets VI}, here we briefly summarize the mechanism and refer readers to \citet{alexander2014} for a more complete review.

The disk can ``evaporate'' as the gas is heated up by high-energy (UV/X-ray) stellar photons and obtains large enough kinetic energy that exceeds the gravitational binding energy \citep{bally1982,shu1993,hollenbach1994}. Whether or not photoevaporation can produce inner cavities or annular gaps depends on the mass-loss rate. Qualitatively speaking, the mass-loss rate via photoevaporation\index{Photoevaporation} has to be greater than the disk's accretion rate (at least locally). Once a hole or gap is opened, the inner edge of the outer disk becomes directly exposed to the stellar irradiation. Time-dependent models show that the disk gas is then rapidly cleared from the inside-out on timescales of $10^5 - 10^6$ years, leaving an inner cavity or gap of a few $\times$ 10~au in size \citep{morishima2012,bae2013,kunitomo2020,ercolano2021}. When the X-ray-heated inner edge of the disk becomes dynamically unstable, it is also possible that the disk gas is cleared on much shorter, dynamical time-scales, a process called thermal sweeping \citep{owen2012}. The pressure bump at the inner edge of the outer disk is expected to be strong enough to trap dust particles, which can be observable in (sub-)mm continuum observations \citep{garate2021}. Due to the expected azimuthal symmetry of stellar irradiation, time-dependent disk evolutionary calculations are commonly carried out adopting one-dimensional radial grids or two-dimensional radial -- vertical grids. Whether the pressure bump created by photoevaporation can become unstable to the RWI and generate vortices remains to be examined using multi-dimensional time-dependent calculations including the azimuthal dimension.

\section{Magnetohydrodynamic Processes}\label{sec:MHD}

Although we do not have direct evidence that protoplanetary disks are threaded by magnetic fields\footnote{Some tentative evidence was recently found from circular polarization measurements of molecular lines \citep[see][]{teague2021b}.} and there exist few constraints on what kind of geometry magnetic fields have, the existence of magnetospheric accretion and jets/outflows suggests that magnetic fields are present at least in some parts of the disk. In addition,  theoretical studies show that magnetic turbulence and/or magnetized winds\index{Magnetized winds} likely play an important role in the dynamical evolution of protoplanetary disks. 

Even when magnetic fields are present, however, the gas disk would interact with them only if the disk is sufficiently ionized and coupled to the magnetic fields. For the most regions in protoplanetary disks, besides the innermost/outermost regions and the surface layers where a sufficient ionization level can be sustained, it turns out that the gas is only weakly ionized so that the ideal MHD assumption of a perfectly conducting fluid breaks down. To provide a broad overview, in a disk around a solar-mass star, it is generally found that ideal MHD is a good approximation within $\sim$0.1~au when the disk temperature is above $\sim 1000\,$K, while non-ideal MHD effects, namely Ohmic resistivity, the Hall effect\index{Hall effect}, and ambipolar diffusion\index{Ambipolar diffusion}, dominate in the inner ($0.1\,\mathrm{au}\lesssim R\lesssim1$~au), middle ($ 1\,\mathrm{au} \lesssim R \lesssim 10 \,\mathrm{au} $), and outer ($R\gtrsim10$~au) disk regions, respectively (see {\it Lesur et al.} in this book). The exact regions of where each non-ideal effect dominates depends on various factors including the stellar luminosity, the external ionization sources and their strength, and the amount of dust and gas.

In this section, we discuss the ideal and non-ideal MHD processes that can potentially form substructures in protoplanetary disks. Specifically, we focus on the zonal flows\index{Zonal flows} that are formed out of an initially unordered flow through the magnetic self-organization process. These banded structures are long-lived and can offer a possible explanation for the observed ringed substructures in protoplanetary disks. We note that the term zonal flow\index{Zonal flows} describes the resulting banded structure after it is produced but it does not necessarily ascribe a physical mechanism to its formation.  Below, we split the physical mechanisms that can form zonal flows\index{Zonal flows} into two subsections: those formed by magnetic turbulence in ideal MHD and those formed by non-ideal MHD processes. We then introduce magnetized winds\index{Magnetized winds}, which can arise in both ideal and non-ideal MHD regimes and can produce annular substructures.

\subsection{Ideal MHD}

\citet{balbus1991} showed that the presence of weak magnetic fields in disks can destabilize the disk through the magnetorotational instability (MRI\index{Magnetorotational instability}). The MRI drives turbulence and can potentially provide the necessary angular momentum transport to account for observed accretion rates in disks. The physical picture behind the MRI is as follows: when two neighboring fluid elements are coupled by a magnetic field, a small radially inward displacement of one element leads to a magnetic tension force that opposes its rotation, drains its angular momentum, and causes it to drift further radially inward. The angular momentum is transferred to the outer element which moves outward and the process precedes to runaway. Disks are unstable to the MRI so long as the angular velocity increases radially inward and the magnetic field is not strong enough to stabilize the initial perturbation (plasma $\beta \equiv P_{\rm gas}/P_{\rm magnetic} \lesssim 1$).

\subsubsection{Zonal flows\index{Zonal flows}}
\label{sec:zonal_flows}

In ideal MHD turbulence, ring structures were noted early in the global simulations of \citet{hawley2001}. Their simulations showed that initially small radial variations of the Maxwell stress ($B_R B_\phi/4\pi$) over the largest scales of the simulation domain can eventually lead to the concentration of gas into rings. \citet{hawley2001} explained this as increasing accretion efficiency with decreasing density. Local shearing box simulations have been carried out to explore the zonal flows\index{Zonal flows} \citep{johansen2009,bai2014}. In simulations from \cite{johansen2009}, zonal flows are initially launched by Maxwell stress fluctuations that are as small as 10\% compared to the background. The resulting pressure bumps are in a geostrophic balance with sub/super-Keplerian zonal flows\index{Zonal flows}, and they can grow to fill up the largest scale of the simulation box, as large as 10 gas scale heights. These structures can live 10 -- 50 orbital periods, and the lifetime increases with the simulation box size. It is unclear if the zonal flow structure eventually converges with the box size \citep{simon2012,bai2014}. Recent global simulations  \citep{Zhu2015,Jacquemin-Ide2021} show the formation of ring structure in disks threaded by net vertical magnetic flux. But we need to caution the potential influence from the inner boundary condition in these global simulations.

\subsubsection{Spirals}
\label{sec:mri_turbulence}
Magnetohydrodynamic turbulence in disks can generate spirals. Since turbulence is generally homogeneous along the azimuthal direction (except for large scale vortices), higher mode spirals are excited. The number and strength of excited spirals depends on the properties of the turbulence. For the MRI\index{Magnetorotational instability}, spirals are generated when a wave swings from leading to trailing \citep{heinemann2009a,heinemann2009b} and the maximum coupling occurs when the perturbation's azimuthal wavelength is $2\pi H$. If we assume that a perturbation at the $2 \pi H$ azimuthal length scale will excite one spiral arm, then the disk will excite $m \sim R/H$ spirals, which is generally of order of $\sim10$. In agreement with the prediction, global MHD simulations suggest that the kinetic energy spectra and the velocity spectra peak for an azimuthal wavenumber between $m=3$ and $\sim10$ \citep{flock2011,suzuki2014}. Note that, in general, $m=2$ modes are not the dominant mode in MRI turbulent disks. 

\subsection{Non-ideal MHD}

\subsubsection{Dead-zones\index{Dead-zone}}\label{sec:deadzone}

\citet{gammie1996} found that  Ohmic resistivity in the inner ($\sim1$~au) regions of protoplanetary disks can make disks stable against the MRI\index{Magnetorotational instability}, although the surface layers may still be sufficiently ionized for the MRI to drive accretion. The midplane region where the MRI is inactive is called the ``dead-zone\index{Dead-zone}''. 

Pressure bumps can form at the inner/outer edge of dead-zones\index{Dead-zone} due to the mismatch in the accretion efficiency between the MRI active and dead zones. Interior/exterior to the pressure bumps at the inner/outer edge, gaps can open again due to the mismatch in the accretion rate (see the dead-zone\index{Dead-zone} in the MHD section in Figure \ref{fig:cartoon}). The density contrast across the transition from the MRI-active region to the dead-zone\index{Dead-zone} is typically a few, large enough to trigger the RWI\index{Rossby wave instability} which in turn can generate vortices \citep{lyra2012,lyra2015,flock2015,flock2017b}. Simulations by \citet{ruge2016} showed that dust grains with a size of up to 1~cm can be trapped at the edge of the dead-zone\index{Dead-zone}, and large grains ($\gtrsim 50~\mu$m) eventually concentrate into vortices on timescales of $\sim$20 local orbits. The location of the inner edge of the dead-zone is determined mainly by the stellar luminosity, such that it lies close to the star at about 0.1~au for T Tauri stars while it can be as far as 1~au for Herbig stars \citep{flock2016,flock2019}. Due to the radial location and associated small physical sizes, rings or vortices at the inner dead-zone\index{Dead-zone} edge are unlikely to be observable with ALMA\index{ALMA}. The location of the outer edge of the dead-zone\index{Dead-zone} is determined by and sensitive to the ionization state of the disk and the gas and dust density \citep{dzyurkevich2013, turner2014,lyra2015,flock2015}. Depending on those parameters, rings or vortices at the outer dead-zone\index{Dead-zone} edge can be observed using ALMA\index{ALMA} \citep[e.g.,][]{flock2015}

\subsubsection{Hall effect\index{Hall effect}}
\label{sec:hall}

The ``dead-zone\index{Dead-zone}'' picture has been significantly modified with the inclusion of the Hall effect and ambipolar diffusion\index{Ambipolar diffusion} in recent years.  \citet{kunz2013} found that in the non-linear saturated state of the Hall-dominated MRI turbulence, large-scale axisymmetric pressure bumps can arise through zonal flows\index{Zonal flows}. Their simulations showed that zonal flows\index{Zonal flows} will develop when the Hall length (the ratio of the Hall diffusivity to the Alfv\'en speed, $l_H=\eta_H/v_A$) is a significant fraction of the gas scale height ($l_H\gtrsim 0.2H$). In vertically unstratified global simulations, the number of zonal flows\index{Zonal flows} is found to increase as the Hall effect\index{Hall effect} becomes stronger ($l_H>H$) and zonal flows have typical widths of $\sim H$ \citep{bethune2016}. The vertically stratified simulations of \citet{bai2015} showed that zonal flows\index{Zonal flows} are launched when the Hall length increases, but they suggested that zonal field concentrations are merely an inevitability of MRI turbulence in stratified simulations with an initial net vertical magnetic flux. The zonal flows\index{Zonal flows} formed in global simulations are capable of trapping dust grains in magnetic flux concentrations with dust density enhancements of $\sim 20$ times over the initial profile \citep{krapp2018}.

\subsubsection{Ambipolar diffusion\index{Ambipolar diffusion}}

The formation of zonal flows\index{Zonal flows} in disks where only ambipolar diffusion\index{Ambipolar diffusion} operates have been seen in the shearing-box simulations of \citet{bai2014} and in the global three-dimensional simulation of \citet{bethune2017}. The exact reason for the  formation of zonal flows\index{Zonal flows} in ambipolar diffusion\index{Ambipolar diffusion}-dominated regions is still not fully understood and deserves future studies.
It has been shown that magnetic flux concentrations can form as a result of the reconnection of radial magnetic fields along a midplane current layer where $B_\phi$ changes sign, steepened by the effects of ambipolar diffusion\index{Ambipolar diffusion} \citep{suriano2018,suriano2019,hu2019}. The reconnection of magnetic fields in a well-defined current sheet at the disk midplane is aided by the presence of strong magnetic fields and large ambipolar diffusivities \citep{brandenburg1994}. 

In non-ideal MHD simulations with strong ambipolar diffusion\index{Ambipolar diffusion}, \citet{bai2014} found that magnetic flux concentrates into thin shells whose width is typically less than $\sim 0.5H$ \citep[see also][]{simon2014}. In shearing box simulations that include dust grains, grains are effectively trapped in the zonal flows\index{Zonal flows} and are concentrated into thin radial bands. Small grains ($\mathrm{St}=0.001$) have dust density contrasts of $\sim$2, while large grains ($\mathrm{St}=0.1$) can have density contrasts of order $10^3$ \citep{riols2018}. The width of the rings is about $1-2H$ and the separation between the rings is of order of $3-4H$; however, these properties depend on the plasma $\beta$. In their simulations, zonal flows\index{Zonal flows} and radial pressure profiles are found to remain almost steady within the simulation timescale ($\sim100$ orbits). Using 3D shearing box simulations, \citet{riols2019} showed the development of a linear and secular instability driven by MHD winds\index{Magnetized winds}, which gives birth to long-lived (a few hundred to thousand orbits) rings of sizes $> H$ with typical ring separations of $\sim10H$. The linear instability for the spontaneous formation of rings in disks with large-scale vertical magnetic fields is able to predict ring density contrast at given magnetic field strengths \citep{riols2019}.

The global axisymmetric simulations by \citet{riols2020} showed that 0.1~mm and 3~mm dust grains accumulate in the gaseous rings formed from MHD winds\index{Magnetized winds} with ambipolar diffusion\index{Ambipolar diffusion}, and the separation between dust rings is $2.5H$ for $\beta = 10^4$  and $4H$ for $\beta = 10^3$ near $R = 20$~au. They also find the half-width of the 3-mm dust rings is $0.3H$ at $R = 20$~au for $\beta = 10^3$ and $0.15H$ for $\beta = 10^4$, whereas the gas rings have a half-width of $1.2H$. Gaseous rings are formed at smaller contrast when $\beta=10^5$, but dust grains do not accumulate in them. \citet{cui2021} also observe the spontaneous concentration of magnetic flux into axisymmetric bands in global 3D simulations of the outer regions of protoplanetary disks with ambipolar diffusion\index{Ambipolar diffusion}. The simulations are capable of resolving the MRI, and the disks both launch winds and have active MRI\index{Magnetorotational instability} turbulence. Zonal flows\index{Zonal flows} form stochastically and are not as long-lived as previously seen in less resolved simulations that explore a similar parameter space. In their fiducial model they find ring/gap widths of $1.5-2.5H$ and surface density contrasts between rings and gaps of 15--50\%, while ring/gap widths can vary from 1 to $5H$ over a range of ambipolar Els\"asser numbers and for $\beta=10^4$ and $10^3$. Above a magnetic field strength corresponding to $\beta>10^5$, the zonal flows\index{Zonal flows} vanish. Similar zonal flows\index{Zonal flows} and particle concentration have also been seen in \cite{Hu2022}. In addition, \cite{Hu2022} found significant meridional circulation that resemble to the observed meridional flows in the HD~163296 disk \citep{teague2019}.

\subsection{Magnetized Winds\index{Magnetized winds}}

For both ideal and non-ideal MHD simulations, a magnetocentrifugal wind can be launched if the disk is threaded by net vertical magnetic fields \citep{blandford1982}. Magnetocentrifual winds can extract angular momentum by exerting a torque in opposition to the disk rotation, called the magnetic breaking torque. Although this breaking torque may not be as important as the MRI turbulent stress that transports angular momentum in ideal MHD disks \citep{ZhuStone2018}, it may be essential for accretion in disks dominated by non-ideal MHD effects \citep{bai2013}. 

Accretion due to wind launching can potentially lead to structure formation \citep{moll2012}. For disks in both ideal and non-ideal MHD limits, a linear magnetic wind instability has been found to explain the formation of rings and gaps in stratified shearing box simulations \citep{riols2019}. Qualitatively, the instability can be described as follows. If magnetic diffusivity is not too large, magnetic flux will be transported with the viscous flow radially away from a local overdensity in the disk, and therefore concentrate towards disk regions with lower density. Thus, the increased magnetic flux at less diffuse gaps will also lead to an increased angular momentum removal rate in the wind, and the instability grows if more mass is removed in the wind than can be replenished into the gap.

An MHD disk wind\index{Magnetized winds} can lead to a localized gap if the disk wind depletion timescale is shorter than the viscous accretion timescale, i.e., gas is lost in the wind before it can be replenished from larger radii \citep{suzuki2016}. Then, dust can concentrate in the resulting pressure maximum outside of the gas gap only when the radial dust drift timescale is comparable or shorter than the wind depletion time. \citet{takahashi2018}, for example, found that this can be satisfied for grains of size $\mathrm{St}=0.1$ which form a $\sim$10~au wide gap near 20~au, given a disk viscosity of $\alpha\sim10^{-4}$.

\section{Tidal Interaction with Perturbers}
\label{sec:companion}

The ubiquity of planetary systems around main sequence stars suggests that planet formation must commonly happen. As a planet grows, the tidal interaction between the planet and its natal disk perturbs the gas and solids in the disk. This perturbation is observable with modern observing facilities. Tidal interaction with stellar-mass bodies -- both gravitationally bound companions and gravitationally unbound, flybying perturbers\index{Stellar flyby} -- can also produce observable signatures. In this section, we summarize the mechanism and outcome of tidal interactions with perturbers.

\subsection{Gravitationally Bound Planetary and Stellar Companions}
\label{sec:planetary_companion}

\subsubsection{Excitation of spiral waves}
\label{sec:companion_spiral}
The interaction between a companion and its host disk starts with the excitation of spiral waves, which can be described well with linear theory \citep{goldreich1978,goldreich1979,ogilvie2002,bae2018a,miranda2019}. A companion launches wave modes at the Lindblad resonances through the resonance between the rotation of the companion's potential and the epicyclic motion of the disk gas \citep{goldreich1978,goldreich1979}. In Keplerian disks, the propagation of wave modes depends on the azimuthal wavenumber in such a way that different azimuthal wave modes can constructively interfere with each other and create a coherent structure -- a spiral arm \citep{ogilvie2002}. It is also possible that a companion can excite multiple spiral arms because such constructive interference can occur for multiple sets of wave modes \citep{bae2018a,bae2018b}. 

Using two-dimensional hydrodynamic simulations, \citet{bae2018b} investigated how the number and pitch angle of companion-driven spirals vary as a function of the disk temperature and companion mass. In general, a smaller number of spirals form with larger companion mass and higher disk temperature \citep{bae2018b}. For companions with at least a Saturn-mass (around a solar-mass star), it is expected that $2-3$ spirals are excited inside the companion's orbit and $1-2$ spirals are excited outside the companion's orbit, with the exact number depending on the disk temperature. Important characteristics of Lindblad spirals include that the pitch angle peaks at the location of the planet and quickly decreases away from the planet \citep{rafikov2002,muto2012}, and the spirals are more opened up with a higher mass planet due to the non-linear wave propagation \citep[][see also Figure \ref{fig:cartoon} of this Chapter]{zhu2015b,bae2018a,bae2018b}. By measuring the pitch angle of a spiral over a broad range of disk radii, these characteristics can thus help to determine the location and mass of the planet.

While many seminal works on planet-disk interaction\index{Planet-disk interaction} adopted an isothermal assumption (see PPVI review by \citealt{Baruteau2014}), recent studies have highlighted the importance of using more realistic disk thermal properties (see {\it Paardekooper et al.} in this book). Using linear theory and two-dimensional hydrodynamic simulations, \citet{miranda2020} and \citet{zhang2020} showed that disk's cooling timescale can play an important role in the dissipation of companion-driven spirals. They showed that when the cooling timescale is of order of the dynamical timescale ($t_{\rm cool} \sim 1/\Omega$), the amplitude of companion-driven spirals greatly decreases as the density waves are subject to a strong radiative dissipation. The observational implications of spirals subject to different cooling timescales are presented in \cite{Speedie2022}. Although the intrinsic number of spirals a companion excites does not depend on the disk's cooling timescale, these studies suggest that the number of {\it observable} spirals can be smaller depending on the cooling timescale.

Another important development in recent years is the use of three-dimensional simulations. As we mentioned in Section \ref{sec:spirals_obs}, spirals are more frequently observed in NIR than mm continuum observations. It is thus important to understand that the pitch angle of companion-driven spirals in the surface layers (probed with NIR observations or optically thick molecular lines in mm wavelengths) can be different from the pitch angle in the midplane (probed with mm observations). Using three-dimensional numerical simulations adopting vertically stratified temperature structure, \citet{juhasz2018} showed that the pitch angle of the spirals in thermally stratified disks is the smallest in the disk midplane and increases towards the disk surface, by approximately the ratio of the sound speed between the layers: $\psi(z)/\psi(z=0) \simeq c_s(z)/c_s(z=0)$. 
When the disk is thermally stratified and/or cooling is slow, the disk can have non-zero vertical buoyancy frequency. In such a case, companions can excite spirals via buoyancy resonances \citep{zhu2012,lubow2014,mcnally2020,bae2021,yun2022}, in addition to those driven by Lindblad resonances. Buoyancy resonances occur when the companion's orbital frequency and the buoyancy frequency match with each other. Buoyancy spirals have several characteristics which distinguish themselves from Lindblad spirals. These include that (1) buoyancy spirals are much more tightly wound compared with Lindblad spirals, with a pitch angle $\lesssim 10^\circ$ in the vicinity of the companion; (2) their pitch angle monotonically decreases over the disk radius; and (3) they predominantly produce vertical velocity perturbations. 

Regardless of the companion mass and disk properties, companion-driven spirals (via both Lindblad and buoyancy resonances) orbit at the companion's orbital frequency. As such, only small particles with a sufficiently short stopping time may be trapped by companion-driven spirals (see Section \ref{sec:drift}). This might support the fact that spirals are more commonly seen in NIR scattered light observations than mm continuum observations. While larger grains generally do not couple to spiral, they (as well as small grains) can experience vertical stirring arising from the turbulence and vertical flows induced by companion-driven spirals. \citet{bae2016b,bae2016c} showed that companion-driven spirals can undergo the spiral wave instability (SWI), a parametric instability arising from the resonant coupling between spiral waves and pairs of inertial waves. When the SWI is saturated corrugated vertical flows develop, which can loft dust particles from the midplane \citep[][see also the vertical slice in the tidal processes section in Figure \ref{fig:cartoon}]{bae2016b,bae2016c}. \citet{bae2016c} showed that a Jupiter-mass planet in a MMSN-like disk can produce significant vertical diffusion of dust particle, corresponding to a diffusion coefficient $\alpha_Z \simeq 10^{-3} - 10^{-2}$. More recent three-dimensional multi-fluid simulations also showed that giant planets can induce significant vertical stirring of particles \citep{bi2021,binkert2021}, although whether the SWI is the origin of the vertical stirring seen in these simulations is unclear at this point.

Finally, spirals excited by an eccentric perturber have been studied recently \citep{zhuzhang2022,FairbairnRafikov2022}. These spirals are highly complicated: they can detach from the planet, bifurcate, break, or even cross each other. The pitch angle and pattern speed are different between different spirals and can vary significantly across one spiral. These properties are very different from those of spirals driven by a planet on a circular orbit, and can help us to find  perturbers on eccentric orbits in future observations. 

\subsubsection{Gap opening}
\label{sec:companion_gap}

While the initial excitation of companion-driven spirals is well described with linear theory, companion-driven spirals non-linearly steepen to shocks as they propagate \citep{goodman2001,cimerman2021}. As spiral waves generate shocks, angular momentum is transferred to the disk gas, opening a gap \citep{rafikov2002}. For low-mass companions whose masses are smaller than the so-called thermal mass, defined as $M_{\rm th} \equiv (H/R)_p^3 M_*$ where $(H/R)_p$ is the disk aspect ratio at the location of the companion, spiral waves have to propagate before they steepen into shocks \citep{goodman2001}. This means that gaps open at finite distance from the companion, while the planet is still embedded in its co-orbiting region \citep{dong2017b}. When a companion excites multiple spirals, it is possible that the companion carves multiple gaps in the disk as each of its spiral opens a gap \citep{bae2017,dong2017b,dong2018c}. When a companion reaches a thermal mass, the spiral waves launched by the companion are already non-linear at excitation. Such massive companions can thus carve a gap around their orbit. As a reference, for a solar-mass host star, a thermal mass is about 0.5 times Saturn's mass when $(H/R)_p=0.05$, about a Jupiter mass when $(H/R)_p=0.1$, and about 3 Jupiter masses when $(H/R)_p=0.15$. 

The depth and width of the gap a companion opens depend on the mass of the companion, in addition to the disk temperature and viscosity. Thus, when gap properties are constrained from observations, one can use the properties to infer the mass of the companion (see Section \ref{sec:implications}). The scaling relation between the companion mass, disk temperature, and disk viscosity, and the gap width/depth can be derived analytically by considering the angular momentum exchange between the disk and the companion \citep{fung2014,kanagawa2015a,kanagawa2015b,ginzburg2018} or empirically by carrying out numerical simulations \citep[e.g.,][]{duffell2013,fung2014,kanagawa2016,dong2017,kanagawa2017,zhang2018,yun2019}. Under the assumption that the disk accretion can be described with the viscous theory adopting $\alpha$ viscosity \citep{shakura1973}, \citet{kanagawa2015a} analytically derived the gap depth as
\begin{equation}
\label{eqn:gap_depth}
    \Sigma_p/\Sigma_0 = {1 \over 1+0.04K_d},
\end{equation}
where $\Sigma_p$ and $\Sigma_0$ denote the perturbed and unperturbed surface density, respectively, and $K_d \equiv \alpha^{-1} (H/R)_p^{-5} (M_p/M_*)^2$. While the above relation is derived assuming that the angular momentum deposition occurs locally, within about a scale height from the companion's radial location, it is also possible to consider additional non-local deposition of angular momentum in the gap width and depth derivation \citep{ginzburg2018}. 
Strictly speaking, the scaling relations derived analytically can be applicable to low-mass planets only. For high-mass planets, non-linear effects can become important, and numerical simulations can complement the scaling relations. In general though, empirical scaling relations obtained from numerical simulations show a reasonable match to the above analytic formula, within the planetary mass regime (see e.g., \citealt{fung2014,kanagawa2015b,zhang2018,yun2019}). When the planet is misalgined with the disk, the gap will be shallower and a similar gap depth formulae can be derived and tested against simulations \citep{zhu2019}.

Beyond the gap within which a companion is embedded, a local pressure maximum forms, trapping grains. Numerical simulations show that the width of the gas ring, measured as the full width at half maximum in the perturbed surface density profile $\Sigma/\Sigma_0$, is $\sim2 - 4~H$. In the PDS~70 disk, where two giant planets are detected within its inner cavity \citep{keppler2018,haffert2019}, the continuum ring beyond the orbit of PDS~70c has a width of about 35~au measured at the half-maximum of the peak flux density, corresponding to about 3 gas scale heights \citep{keppler2019}. This agrees well with the gas ring width seen in numerical simulations.

Companion-induced rings and gaps are generally stable but the number and location of rings and gaps can vary when planets undergo orbital migration, with the details depending on the migration speed \citep{meru2019,nazari2019,weber2019,kanagawa2020,kanagawa2021}. For a planet migrating slower than the dust, a dust ring forms outside of the planet’s orbit, whereas if the planet migrates faster than the dust the ring appears inside of the planet’s orbit \citep{meru2019,nazari2019}.  If the radial drift velocity of the dust can be estimated, for example via the spectral index of the millimeter emission, the continuum morphology can thus be used to constrain the planet migration speed. When planets undergo intermittent orbital migration, it can also leave multiple rings and gaps behind them \citep[][see also {\it Paardekooper et al.} in this book]{wafflard-fernandez2020}.

While it is generally believed that stellar companions are too massive to create narrow, axisymmetric gaps in protoplanetary disks, stellar companions can tidally truncate the inner disk, open  a large inner cavity, and create a ring beyond the cavity \citep{artymowicz1994,artymowicz1996}. The size of the cavity depends upon the eccentricity of the binary and the inclination of the disk to the binary orbital plane \citep{miranda2015,lubow2015,lubow2018,franchini2019b}. For a coplanar disk, the inner cavity size is in the approximate range $1.8$--$3.3\,a_{\rm b}$ for eccentricity in the range $0$--$1$, where $a_{\rm b}$ is the semi-major axis of the binary orbit.

\subsubsection{Disk breaking and misalignment\index{Disk breaking and misalignment}}
While companions in highly viscous disks can cause the disk to warp \citep[e.g.][]{Xiang-Gruess2013,Arzamasskiy2018,nealon2019}, 
massive planetary companions and stellar companions in less viscous disks can induce deep gaps to break the gas disk and produce misalignments\index{Disk breaking and misalignment} between different parts of the disk \citep[e.g.][]{zhu2019,bi2020}. 
With a deep gap satisfying the disk breaking\index{Disk breaking and misalignment} condition in \cite{zhu2019}, the planet effectively separates the disk into disjoint regions that are not in communication with each others. If the disk is broken into disjoint rings, the misaligned\index{Disk breaking and misalignment} rings can then nodally precess independently. The timescale for the precession depends upon the radius of the disk ring and so the precession can lead to a large misalignment\index{Disk breaking and misalignment} between the different parts of the disk \citep[e.g.][]{lubow2018}.

While we expect most planets to form coplanar to their parental disk, their dynamics can deviate from the disk. This is particularly efficient when the planet is sufficiently massive to open a gap in the circumstellar disk in binary systems \citep{picogna2015,lubow2016}.  Around one component of a binary, planet-disk interactions lead the inclination of a planet to increase on average. Even a small misalignment can lead to a highly inclined planet that may be able to undergo Kozai-Lidov oscillations \citep{kozai1962,lidov1962,martin2016,franchini2020}. In a circumbinary disk, planet-disk interactions have similar effects, although an outer disk generally leads to the planet tilt decreasing on average \citep{pierens2018}.

The disk inclination relative to an inner binary or an outer binary companion can evolve over the lifetime of the disk. Circumstellar disks and low inclination circumbinary disks nodally precess about the binary angular momentum vector \citep{papaloizou1995,lodato2013}. Circumstellar disks generally precess as a solid body since the sound crossing timescale is short compared to the global disk precession timescale \citep{larwood1996}. Circumbinary disks that are sufficiently extended in radius may be subject to disk warping or even breaking\index{Disk breaking and misalignment} as a result of the precession \citep{nixon2013,facchini2013,juhasz2017,facchini2018,price2018}. For highly inclined circumbinary disks around an eccentric binary the precession is about the binary eccentricity vector (or semi-major axis) \citep{verrier2009,farago2010,doolin2011,aly2015}.   
 
A circumstellar disk with a sufficiently high inclination can undergo Kozai-Lidov \citep{kozai1962,lidov1962} oscillations where the disk inclination and eccentricity are exchanged \citep{martin2014,fu2015}. This can lead to highly eccentric disks and mass transfer between binary components \citep{franchini2019}. Since a highly misaligned circumbinary disc provides a source of high inclination material to the circumstellar discs, these oscillations may be long lived \citep{smallwood2021}.
 
Dissipation within a disk during nodal precession causes a circumstellar disk to move towards coplanar alignment \citep{lubow2000,nixon2011,foucart2013}.  A circumbinary disk moves either towards coplanar alignment or towards polar alignment depending on its initial inclination \citep{martin2017,zanazzi2018}. A polar aligned low-mass disk has its angular momentum vector aligned to the binary eccentricity vector. The disk is perpendicular to the binary orbital plane. A massive circumbinary disk reaches a polar aligned stationary state at lower levels of misalignment\index{Disk breaking and misalignment} \citep{zanazzi2018,martin2019,chen2019}. There have been two observed polar disks, a polar gas disk \citep{kennedy2019} and a polar debris disk \citep{kennedy2012,smallwood2020}.  

The alignment timescale for a disk depends upon a few parameters. First, it depends on the nodal precession timescale. The shorter the precession timescale, the shorter the alignment timescale.   For an external binary companion, the closer the companion, the faster the precession and the shorter the alignment timescale. This is why disks in closer binaries tend to be aligned while those in wider binaries may be misaligned\index{Disk breaking and misalignment} \citep{czekala2019}. The more radially extended a circumbinary disk is the longer the alignment timescale. Further, the alignment timescale also increases with the disk aspect ratio and decreases with the Shakura \& Sunyaev $\alpha$ parameter \citep{lubow2000,bate2000}. Thus, depending on the properties of the disk and binary, the lifetime of the disk may be shorter than the alignment timescale meaning that planet formation may occur in misaligned disks\index{Disk breaking and misalignment} \citep{martin2016}.

\subsubsection{Vortex formation}

As a planet opens a gap, the rotational velocity of the protoplanetary disk evolves to maintain the radial force balance (Equation (\ref{eqn:radial_balance})). The resulting sharp velocity shear can trigger the Rossby wave instability\index{Rossby wave instability} and form vortices. The properties of vortices formed through this process are expected to be similar to the vortices formed in more generalized simulations of RWI\index{Rossby wave instability} adopting a Gaussian density perturbation which we discussed in Section \ref{sec:RWI}. 

Vortices can also be generated at the inner edge of circumbinary disks \citep[e.g.,][]{price2018b}. Although vortices in circumbinary disks share some similarities with the ones created by planets at the gap edge, they seem to be more closely linked to the spirals launched by the binary stars and the eccentric gap edge \citep{shi2012,Ragusa2017}. When the vortices move close to the binary, material streams to the binary and then flung out back to the disk edge, enhancing the vortices. Even in disks with high viscosity or MHD turbulence and even when the binary orbit is circular, asymmetric over-densities can appear when the gap edge starts to become eccentric \citep{munoz2020,ragusa2020}. The discovery of the crescents\index{Protoplanetary disk substructure!crescents} in circumbinary disks (Figure \ref{fig:crescent}a) seems to support that at least some crescents could be induced by the binary.

\subsection{Stellar Flybys\index{Stellar flyby}}
\label{sec:flyby}

The great majority of local star formation happened in groups and clusters of tens to hundreds members, embedded within their natal (giant) molecular clouds \citep[e.g.,][]{carpenter2000,lada2003,evans2009}. It is also  believed that our very own star, the Sun, was born in a cluster (see e.g., reviews by \citealt{adams2010}). The presence of nearby stars can greatly affect the dynamics of the stars themselves, but also the formation and evolution of their disks  because the disks can be tidally perturbed or even truncated during close encounters.

The most common observable from flyby events is the formation of spirals. The spiral formation mechanism via stellar flybys\index{Stellar flyby} can be understood in a similar way that a gravitationally bound companion excites spirals, except that the location of the external potential relative to the disk can rapidly change. In general, numerical simulations show that a stellar flyby produces grand-design, two-armed spirals, consisting of one that is connected to the perturbing star and the other that forms on the opposite side. The fact that $m=2$ mode dominates is sensible given the large perturber mass \citep{bae2018b}. Flyby-induced\index{Stellar flyby} spirals generally have large pitch angles of $20-30^\circ$ during the encounter, which subsequently moderates to $\sim10-20^\circ$ \citep{cuello2019}. Spirals driven by a stellar flyby\index{Stellar flyby} are expected to disappear once the close encounter is over, typically after a few thousand years \citep{cuello2019}.

\section{The Effect of Self-gravity}
\label{sec:self-gravity}

So far, we have ignored the effects of the disk's self-gravity. However, when the disk is sufficiently massive, self-gravity of the disk can trigger gravitational instability\index{Gravitational instability}, change the disk's underlying structure (e.g., make the disk thinner), or affect the outcome of other substructure-forming processes. In this section, we first describe gravitational instability in Section \ref{sec:GI}, and then we discuss the influence that self-gravity can have on other substructure-forming processes in Section \ref{sec:influence_of_self-gravity}.

\subsection{Gravitational Instability\index{Gravitational instability}}
\label{sec:GI}

In a uniform gaseous medium, perturbations on large scales can grow and collapse when the self-gravity of the gas overcomes the gas pressure -- the so-called Jeans instability. In rotating systems like protoplanetary disks, gas pressure can stabilize perturbations on small scales, similar to the Jeans instability, while rotation can stabilize perturbations on large scales. 
When the disk is sufficiently massive such that the so-called Toomre $Q\equiv c_s\Omega/(\pi G \Sigma)$ parameter \citep{toomre1964} is $\lesssim$1, the small and large scales separate from each other and perturbations at intermediate scales are subject to GI (see review by \citealt{kratterlodato2016}). 

In order to have a small $Q$ close to unity, the disk needs to be massive (large $\Sigma$) and/or cold (small $c_s$). Since  protoplanetary disks experience significant mass infall during early protostellar stages and lose mass over their lifetime, it is likely that protoplanetary disks are more prone to GI while they are young. If we assume that the disk temperature is dominated by stellar irradiation, then the temperature drops as $T\propto R^{-1/2}$. Adopting a radially decreasing surface density profile $\Sigma\propto R^{-1}$, $Q$ therefore varies as $Q\propto R^{-3/4}$, indicating that the outer disk is generally more prone to GI. Indeed, numerical simulations of protoplanetary disk subject to infall suggest that young protoplanetary disks ($<$1 Myr) can be subject to GI at large radii of $R \gtrsim$ 10~au  \citep[e.g.,][]{vorobyov2005,bae2014,XuKunz2021a,XuKunz2021b}.

For a disk with $Q \lesssim 1$, GI quickly leads to the formation of non-axisymmetric spirals which can efficiently transport angular momentum through gravitational torques. GI-driven spirals can also produce turbulence which can transport angular momentum. Angular momentum transport is accompanied by energy dissipation in the disk so that the disk heats up, increasing $Q$ and stabilizing the instability. When $Q$ becomes much larger than 1, GI does not operate in the disk, the disk cools down, and $Q$ decreases. Eventually, the heating and cooling are balanced such that $Q$ is maintained around 1. In this regard, GI can be seen as a thermally self-regulated instability with the Toomre $Q$ staying around 1.  
However, GI does not always lead to stable spirals in disks. Depending on the disk's cooling rate, the spirals can collapse to form bound objects (e.g. giant planets; \citealt{gammie2001,rice2003,rice2005,cossins2009}).  Traditionally, the orbital cooling parameter ($\beta$ cooling) has been adopted to quantify the cooling efficiency \citep{gammie2001}.  When the cooling timescale is shorter than the orbital timescale, a condition that is generally easier to be met at large distances $\gtrsim 50$~au, the GI-driven spirals can collapse to form bound clumps \citep{Rafikov2005,Cossins2010a, zhu2012c}. This is the competing mechanism for giant planet formation far away from the central star \citep{boss1997}, besides the core accretion mechanism. The exact cooling criterion for clump formation is still being actively studied, including its numerical convergence \citep{meru2011,lodato2011, paardekooper2012, booth2019}, different viscosity and cooling treatments \citep{rice2012,michael2012,rice2014,baehr2015,hirose2019,vorobyov2020}, and relationship to the disk's local condition \citep{,rogers2012,takahashi2016}. 

The main observable outcome of GI is spirals. Similar to spirals driven by other processes, one can characterize GI-driven spirals using the pitch angle and multiplicity. \citet{cossins2009} carried out smoothed particle hydrodynamics (SPH) simulations to study GI-driven spirals in disks with different masses and cooling timescales. The dominant azimuthal mode and the pitch angle are found to be more sensitive to the disk mass than the cooling time. When  $M_d/M_* = 0.05$, the  $m=14$ mode dominates with a pitch angle of about $10^\circ$. When $M_d/M_* = 0.125$, the disk has $m=9$ with a pitch angle of $14^\circ$. When the disk becomes more massive with $0.25<M_d/M_*<1.5$, the $m=2$ mode dominates \citep{forgan2011,hall2019}. Thus, $M_d/M_* \sim 0.2$ is roughly the boundary above which $m=2$ mode dominates. 

GI-driven spirals are generally transient. Spirals can constantly form and disappear, typically within the orbital timescale. Even within one orbital timescale, the spirals can self-adjust so that the normal flow speed into spiral the arms is sonic \citep{cossins2009,bethune2021}. This suggests that the pattern speed of the spirals $\Omega_p$ is generally close to the local angular velocity of the flow $\Omega \simeq \Omega_K$. Indeed, numerical simulations showed that  $|\Omega_p - \Omega|/\Omega < 15~\%$ for disks with $M_d/M_* \leq 0.125$, although $|\Omega_p - \Omega|/\Omega \gtrsim 50~\%$ when the disk-to-star mass ratio exceeds 0.5  \citep{forgan2011}. However, we note that most GI simulations start with a globally gravitationally unstable configuration as an initial condition. If a disk is gravitationally unstable only locally within a narrow range of radius, due for instance to the concentrated mass addition from the envelop near the centrifugal radius \citep[e.g.,][]{bae2014}, the spirals are launched only at the gravitationally unstable region of the disk while they can propagate radially to the gravitationally stable region. In such a case, the pattern speed of the GI-driven spirals is the angular velocity at the excitation radius of the spirals, a picture similar to planet-driven spirals as explained in Section \ref{sec:planetary_companion}. 

Spirals generated by gravitational instability\index{Gravitational instability} can trap dust particles \citep{rice2006b,gibbons2014}.
Shearing-box simulations including dust particles \citep{baehr2021a} suggest that
dusty spirals have universal opening angles ($\sim$10$^\circ$), independent of the computational domain, the cooling time, or the particle size. Particles with ${\rm St} \sim1$ drift fastest and thus have the highest concentration in spirals. Particles with ${\rm St} > 1$ can also be trapped in spirals, but in this case it is the gravity from the gaseous spiral arms that enables trapping, instead of aerodynamic drag\index{Aerodynamic drag}. It is thus possible that GI-driven spirals are observed with mm continuum observations \citep{cossins2010,evans2017,forgan2018,hall2018}

In addition to continuum observations, molecular line observations can also probe the spirals induced by GI \citep{hall2020}. \cite{forgan2012} suggest that the in-plane velocity broadening can be one order of magnitude higher than the perpendicular velocity broadening. \cite{shi2014} carried out 3-D shearing box simulations and found that turbulent velocities are nearly uniform vertically, increasing by just a factor of 2 from the midplane to the surface. They found a peak at around 0.2--0.3 $c_s$,  higher than \cite{forgan2012}. These results can be tested by future disk kinematics observations. 

To summarize, gravitational instabilities\index{Gravitational instability} can produce 2 to $>10$ spirals, although high-order modes are likely more challenging to observe due to the smaller density perturbations that they produce. The pitch angle of GI-driven spirals is typically $5 - 15^\circ$. The spirals can be transient and the pattern speed is expected to be close to the local Keplerian speed.

\subsection{Influence of Self-gravity on Other Substructure-forming Processes}
\label{sec:influence_of_self-gravity}

When the disk is gravitationally stable ($Q>1$), disk self-gravity is ignored by most theoretical works for several reasons. First, gravitational instability\index{Gravitational instability} has a sharp turn-on/off at $Q\sim1$ -- gravitational instability does not operate even in a $Q=2$ disk. Second, Class II protoplanetary disks are thought to be far less massive than a $Q=1$ disk. Third, it is computationally expensive to include disk self-gravity in simulations. However, even in gravitationally stable disks ($Q>1$), disk self-gravity can be important for many physical processes, including dust settling, vortex formation, and planet-disk interactions\index{Planet-disk interaction}. 

In a self-gravitating disk, both the gravity from the gas to the dust and the anisotropic turbulence lead to a more settled dust disk  \citep{baehr2021b}. This can potentially explain the thin dusty layer in the moderately accreting HL Tau \citep{pinte2016}. When the interaction between the dust and the gas is taken into account, the dusty disk can also be subject to the secular gravitational instability \citep{youdin2011,takahashi2014,takahashiinutsuka2016,tominaga2018,tominaga2019}, which can generate ring substructures.  

Self-gravity can significantly weaken vortices in protoplanetary disks \citep{LinPapaloizou2011}. \cite{lovelace2013} suggested that disk self-gravity can suppress the development of RWI modes with mode number $m<(\pi/2) (R/H) Q^{-1}$. Thus, for all mode numbers $m \geq 1$ the vortex formation by the RWI\index{Rossby wave instability} will be significantly affected when $Q < (\pi/2) (H/R)^{-1}$. For a disk with $H/R=0.1$ as an example, this suggests that even for a disk with $Q \simeq 10$ self-gravity can play a significant role in vortex formation \citep{zhu2016}. 

Finally, the outcome of planet-disk interactions\index{Planet-disk interaction}, such as the number and strength of planet-driven spirals and the width and depth of gaps around the planet, can also be affected in $Q>1$ disks. \cite{zhang2020} found that in disks with $Q \simeq 2$, planets generate stronger and more tightly wound spirals and the gaps are also deeper (see also \citealt{Pohl2015}).

\section{Processes Induced by Dust Particles}\label{sec:dust}

Until now, we have discussed processes where the grains' spatial distribution passively evolves depending on the dynamics of the gas. However, there are situations where dust plays a dominant role in shaping the continuum emission morphology. We separate such processes into two broad categories:  icelines and dust-induced instabilities.

\subsection{Icelines\index{Icelines}}\label{sec:icelines}

Icelines\index{Icelines} are the radial locations where volatiles sublimate and condense. Because the sublimation temperature is material-dependent, different volatile species have icelines at different radial locations. For instance, N$_2$, CO, CO$_2$, and H$_2$O have sublimation temperatures of $\sim$ 10--20, 20--30, 60--70, and 130--160 K at typical disk densities, respectively \citep{zhang2015}. Because the disk temperature usually decreases with distance from the central star, species that are more volatile have icelines at larger radial distances. 
Because dust particles drift toward the central star, they inevitably pass through icelines and lose ice. 

Around the icelines\index{Icelines} of major volatiles, the inward flow of icy aggregates can produce dust substructures in three different ways. First, aggregates may break up into fragments when they lose ice. Because small fragments drift slowly, they pile up inside the icelines \citep{saito2011,ida2016,hyodo2019}. 
Second, the vapor released from the aggregates can be transported outward by turbulence and  condense onto other aggregates outside the icelines, which can result in a local enhancement in the solid surface density \citep{ros2013,schoonenberg2017,drazkowska2017,stammler2017,hyodo2019}. 
Third, the aggregates' stickiness can change as their composition changes at the icelines. This can cause a steep variation of the grain size and dust surface density across the icelines  \citep{birnstiel2010,banzatti2015,pinilla2017_icelines,okuzumi2019}. However, we note that there are currently large uncertainties in the stickiness of silicates and ices. It was previously believed that H$_2$O ice is sticker than silicates \citep{dominik1997,blum2000,wada2009,gundlach2015}, but some recent studies call this into question  \citep{kimura2015,gundlach2018,musiolik2019,steinpilz2019}. Other recent experiments show that CO$_2$ ice is less sticky than H$_2$O ice \citep{musiolik2016a,musiolik2016b}.

\citet{zhang2015} proposed that annular gaps seen in the HL Tau disk could be related to the icelines\index{Icelines} of some volatiles. They assumed that icy particles outside the icelines have grown beyond millimeter sizes through ice recondensation. Such particles have smaller millimeter opacities and therefore produce less thermal emission per mass than those smaller than millimeters. \citet{zhang2015} explained the dips seen in the dust continuum images of the HL Tau disk as a result of the millimeter optical depth reduction due to ice condensation growth. However, ice recondensation can cause an increase in the solid surface density, and therefore whether or not ice condensation growth results in a reduction of millimeter dust emission has to be further tested.

Outside icelines\index{Icelines}, icy aggregates may also experience another important mechanical process: sintering. This is a phenomenon where the grains inside an aggregate fuse together, at temperatures slightly below the sublimation and melting temperatures of constituting materials. An important consequence of sintering is that it makes aggregates harder but less sticky \citep{sirono2011,sirono2017}. Combined with the size-dependent radial drift, sintering can also result in an enhancement of the dust surface density outside of the icelines \citep{okuzumi2016}. Unlike the three iceline-related substructure formation processes described above, sintering-induced dust ring formation may occur at the icelines of moderately abundant volatiles such as NH$_3$ and CH$_4$, in addition to the icelines of more abundant H$_2$O, CO$_2$, and CO \citep{sirono2017}.   

\citet{okuzumi2016} proposed that the prominent dust rings in the HL Tau disk can be explained by aggregate sintering outside of major icelines. According to their scenario, the icelines define the rings' inner edges, whereas the sintering timescale determines the rings' radial extent. The ring widths can be either larger or smaller than the gas scale height, depending on the size of grains that constitute the aggregates. \citet{okuzumi2016} list several volatiles (including H$_2$O, NH$_3$, CO$_2$, CH$_4$, and CO$_2$) whose abundance may be high enough to induce aggregate sintering. The number of dust rings predicted from this scenario can be smaller than the number of the major volatiles because some rings can overlap.  

The iceline\index{Icelines} scenarios do not require gas ring/gap substructures to produce dust ring/gap substructures. However, this does not mean that the gas surface density must be radially smooth across icelines. For example, \citet{hu2019} show that sintering-induced dust rings at icelines, when combined with non-ideal MHD effects, can produce gas rings at the same locations. Specifically, the most prominent ring at the CO$_2$ iceline shows a gas surface density contrast of 17\% compared to the expected smooth disk profile. This occurs because a large amount of dust in the dust rings leads to a lower ionization level \citep{sano2000,wardle2007,okuzumi2009,hu2021} that results in slower disk gas accretion.

The most important prediction from the iceline\index{Icelines} scenarios is that the locations of the iceline-induced substructures should be determined by the disk radial temperature profile. This prediction has been widely used to observationally test the scenarios \citep{long2018,Huang2018a,vandermarel2019}. 
The current observational data show no clear correlation between the locations of the observed rings and the expected locations of the icelines of major volatile species.
This can be seen in  Figure~\ref{fig:rings}(e), where we compare the radial locations of observed rings with the expected locations of some of the major icelines (N$_2$, CO, CO$_2$, and H$_2$O). Moreover, the radius ratios of substructure pairs do not show clustering at particular values (not shown in Figure \ref{fig:rings} but see \citealt{Huang2018a,vandermarel2019}), which would be expected if the substructures are of iceline origin and if the disk radial temperature profile obeys a power law. 
Existing continuum observations thus disfavor iceline\index{Icelines} scenarios as the universal explanation for the observed ring/gap substructures. Yet, the latest high-resolution observations of dust continuum and CO emission lines from the Molecules with ALMA\index{ALMA} on Planet-forming Scales (MAPS) ALMA\index{ALMA} Large Program show that two of five observed disks (HD 163296 and MWC 480) have bright dust rings at the midplane CO iceline \citep{zhang2021}. 

Caution is needed in testing iceline\index{Icelines} scenarios using a simple disk temperature model.   \citet{owen2020} shows that icelines can be thermally unstable and suggests that a simple (steady and power-law) temperature profile may not be applicable to inferring iceline locations. 
Even without dust rings, centrally irradiated disks can be thermally unstable \citep{dullemond2000,watanabe2008,wu2021} and icelines may migrate back and forth \citep{ueda2021,okuzumi2022}. Accretion outbursts may also affect iceline positions  \citep{martin2012}.

\subsection{Dust-induced Instabilities}
\label{sec:dust_instability}

The streaming instability\index{Streaming instability} can produce dust substructures in disks \citep{youdin2005,johansen2007}. As explained in Section~\ref{sec:drift}, dust grains in a gas disk with a negative radial pressure gradient experience headwinds. When dust overdensities are present, the back-reaction from grains on the gas causes the gas to orbit faster than otherwise, weakening the headwind on the grains. The weaker headwind reduces the radial drift of grains, further enhancing the concentration of grains. Numerical simulations show that when the streaming instability is saturated, the resulting grain rings have a width that ranges from a fraction up to 10 percent of the gas scale height and the spacing between the rings is 10 to 20 percent of the gas scale height \citep{yang2014,yang2017,li2018,carrera2021}. 

When the concentration of grains becomes sufficiently large (dust-to-gas mass ratio $\sim$ 1), the gas dynamics can be dominated by the grains. Under such circumstances, the drag from the grains on the gas can overcome the viscous diffusion of the gas, creating a local gas pressure maxima \citep{gonzalez2017}. When the gas accretion is governed by magnetic processes, inhomogeneous grain distributions can result in spatial variations in the degree of coupling between the gas and the magnetic field, by changing the disk ionization structure \citep{hu2019} or the conductivity of the gas \citep{dullemond2018} across the over-/under-dense regions. This, in turn, leads to a non-uniform accretion of the gas, creating local gas pressure maxima within which grains can be efficiently trapped. 

Because the rate of collisional grain growth scales with the grain surface density, grains grow faster (slower) in a positive (negative) grain surface density perturbation. The resultant radial variation of the grain size yields a radial variation of the grains' radial drift velocity, which in turn alters the radial surface density variation. \citet{tominaga2021} showed that this feedback can be positive and produce dust rings and gaps. 

The dust concentration within a vortex can feedback to the gas and weaken the vortex. Both analytical and numerical works \citep{fu2014b,railton2014} suggest that a dust-loaded vortex can be subject to instabilities, especially at the vortex center. However, these instabilities may not be able to destroy the vortex. The dust feedback can be even more complicated if multiple dust species can feedback to the vortex \citep{crnkovic-Rubsamen2015}, potentially leading to anti-correlations between small and large dust grains within vortices. Such instability studies involving multiple dust species just start to emerge \citep{krapp2018, Paardekooper2020, yang2021}. They can potentially lead to interesting observational effects, which deserve future studies. 

\section{\textbf{Connecting Observed Substructures with Physical Processes}}
\label{sec:obs_theory_connection}

Now that we have summarized observed disk substructure properties and introduced various potential substructure-forming processes, in this Section, we aim to connect the observations and the theory. We highlight the successes and challenges of the physical processes in explaining rings and gaps (Section \ref{sec:rings}), spirals\index{Protoplanetary disk substructure!spirals} (Section \ref{sec:spirals}), and crescents\index{Protoplanetary disk substructure!crescents} (Section \ref{sec:crescents}). In each subsection, we start by discussing companions as the potential substructure-forming mechanism because it is the only mechanism for which we have clear observational evidence (e.g., cavity in PDS~70, spirals in AS~205; see Figure \ref{fig:substructures} and subsections below). We then discuss what advantages and disadvantages other substructure-forming processes have in comparison to the companion scenario. As we discuss each mechanism, we also highlight future research directions (both theoretical and observational) that can help to distinguish between the different origins. We summarize suggested future observations in Table \ref{tab:future_prospects}.

\begin{figure*}[ht!]
\centering
\includegraphics[width=1\textwidth]{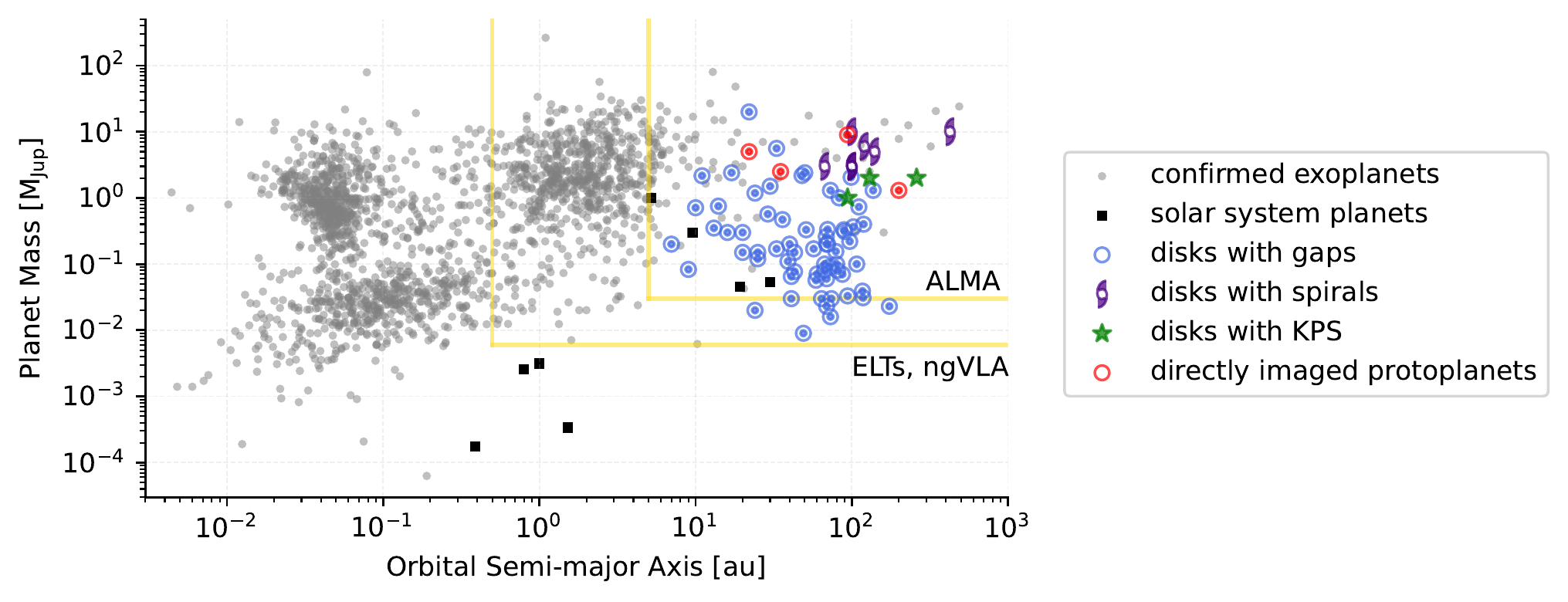}
\caption{
The distribution of the mass and orbital semi-major axis of (grey circles) confirmed exoplanets, (black squares) solar system planets, and (color symbols) protoplanets proposed to reproduce the observed disk substructures (blue $\odot$ for annular substructures, purple spiral symbols for spirals, and green stars for kinematic planetary signatures). Directly images protoplanets (PDS 70b/c, AB Aur b, AS 209b) are shown with red $\odot$ symbols.  Uncertainties in the estimated planet mass are not presented in this figure, though it is typically a factor of a few to ten, which are mainly arising from the ill-constrained disk properties including the disk viscosity. The yellow lines show illustrative regions where current and future observing facilities are expected to find protoplanets directly or through disk substructures they create. Data used in this plot and references are available at \url{http://ppvii.org/chapter/12/}.}
\label{fig:protoplanets}
\end{figure*}

\subsection{Rings and Gaps\index{Protoplanetary disk substructure!rings and gaps}}
\label{sec:rings}

The most immediate observables of annular substructures are the number of rings/gaps, their radial location, and their radial width. These properties of observed rings and gaps are summarized in Section \ref{sec:disk_substructures:rings}, Figure \ref{fig:rings}, and Table \ref{tab:disks_with_rings}. From the observations carried out so far, we do not find a concentration of rings around certain radial locations or correlations between ring properties and stellar/disk properties (Figure \ref{fig:rings}). Before we further discuss what the absence of patterns or correlations may indicate regarding the origin of annular substructures, it is worth pointing out that ring properties are inferred mostly from (sub-)millimeter continuum observations and thus caution is needed when interpreting the observations, in particular for the radial width. Large grains probed by continuum observations are likely subject to strong aerodynamic drag\index{Aerodynamic drag}, the degree of which depends on the Stokes number of the grains (Section \ref{sec:coupling}). For a given grain size, the Stokes number can vary significantly depending on the column density of the protoplanetary disk. This means that one shall not directly compare the width of dust rings located at different radii within the same disk or the width of dust rings found in different disks. Instead, for a proper comparison, we  have to convert the observed width of dust rings to the width in the underlying gas distribution in order to remove the dependency on differing Stokes numbers. Of course, this is not a straightforward task. As described by Equation (\ref{eqn:stokes_number}), the Stokes number is a function of grain's internal density and the gas column density of the disk both of which are, needless to say, difficult to constrain. The fact that most ring properties are constrained by continuum observations is a great challenge when it comes to comparing observations with simulations, too, because most (but not all) theoretical predictions come from gas-only simulations. Given these difficulties, it may thus be too early to conclude that correlations between ring properties and stellar/disk properties are lacking. In the future, directly constraining the radial width of gas rings from the radial variations in the rotational velocity profile \citep[e.g.,][]{teague2018} and from the observations of optically thin gas tracers \cite[e.g.,][]{facchini2020} for a large disk sample would help to perform a more proper statistical analysis. 

Among the physical processes capable of creating rings and gaps, the companion scenario is most flexible in explaining the broad range of ring/gap properties. This flexibility is due in part to ill-constrained disk viscosity and temperature (see e.g., Equation \ref{eqn:gap_depth}), but nevertheless one can almost always adjust the companion mass and orbital radius to reproduce the observations. In fact, there has been a lot of effort from the community to explain observed rings and gaps using planets. Figure \ref{fig:protoplanets} shows the population of such hypothesized planets that are proposed to reproduce the observed disk substructures using either hydrodynamic planet-disk interaction simulations or scaling relations between the planet mass, disk properties, and gap width/depth. This figure suggests that the observed annular substructures\index{Protoplanetary disk substructure!rings and gaps} could have been created by a population of planets with masses between about 0.1--10 Jupiter masses and orbital radii of about 10-500 au. 

However, challenges do exist for the planet scenario. Part of the complication arises from the possibility that a single planet can open multiple gaps, some (or all) of which do not have a planet within them \citep[][see also Section \ref{sec:companion_gap}]{bae2017,dong2017b,wafflard-fernandez2020}. This suggests that, even in the case that all of the observed rings and gaps are created by planets, some of the hypothesized planet population presented in Figure \ref{fig:protoplanets} can be false positives. After all, we need more direct evidence of planets in order to unambiguously connect companions to annular substructures. The best ways include searching for thermal emission in the near and mid-infrared where planets are the brightest, line emission from accreting gas (e.g. H$\alpha$), and emission from circumplanetary disks (CPDs). Encouragingly, there are a small, yet increasing number of young planets directly detected. PDS~70b and c are detected in near-infrared imaging and H$\alpha$ observations \citep[][see also {\it Benisty et al.} in this book]{keppler2018,wagner2018,haffert2019,mesa2019}, and a CPD is detected in sub-mm around PDS 70c \citep{isella2019,benisty2021}. More recently, AB Aur b is detected in near-infrared imaging \citep{currie2022} and AS 209b is detected through its CPD emission in $^{13}$CO \citep{bae2022}. The detection of these planets has proven that direct detection of young, forming planets is possible with existing facilities. Furthermore, these observations provide direct evidence that companions can indeed create  substructures: a cavity created by PDS~70b and c, spirals excited by AB~Aur~b, and a gap and velocity kinks (see Section \ref{sec:spirals}) produced by AS~209b. In the very near future, the James Webb Space Telescope (JWST) should provide the capability to detect young, forming planets  predicted via ALMA\index{ALMA} observations. Indeed, at the time of writing this Chapter, there are accepted JWST programs which specifically aim to search for planets responsible for the substructures previously observed with ALMA\index{ALMA} and ground-based telescopes. In the further future, 30-m class ground based telescopes and the next generation Very Large Array (ngVLA) will be key in the search for planets around young stars. 

Icelines\index{Icelines} are another popular scenario proposed to explain rings and gaps. Icelines must exist since temperature changes across the disk are unavoidable and the phase transition of volatiles should occur at appropriate locations. Until now, a common practice to test the iceline scenario is to infer the disk temperature, from either simplified approximations of the midplane temperature using the stellar luminosity (Figure \ref{fig:rings}e; see also e.g., \citealt{Huang2018a}) or disk specific radiative transfer modeling \citep{vandermarel2019}, and compare the predicted condensation front locations with observed ring/gap locations. Most previous attempts did not find a strong correlation between the position of the observed rings/gaps and the expected location of ice-lines, although it is worth noting that some of the observed continuum rings in the disks around HL Tau \citep{zhang2015,okuzumi2016}, MWC 480, and HD 163296 \citep{zhang2021} lie close to the expected positions of midplane icelines.
Another indirect way to test the iceline scenario is to map the gas density and pressure profiles across rings and gaps. Unlike most other annular substructure-forming mechanisms which generate pressure bumps that trap grains, pressure bumps are not a prerequisite for rings formed by icelines. Therefore, if we find a case where the grain size changes across a continuum ring/gap but there is no associated gas pressure maximum/minimum at the ring/gap location, icelines would offer a compelling explanation. So far, the capability of measuring the gas density profile has been limited by the sensitivity and angular resolution of optically thin gas tracers, but some progress has been made. As an example, five disks with continuum rings/gaps have been recently targeted by the ALMA\index{ALMA} large program MAPS \citep{oberg2021}. \citet{zhang2021} found that dust gaps are coincident with gas gaps in some of the disks, as expected in the case of radial migration of grains due to gas pressure gradients, whereas depletion of the gas is minimal or lacking within some of the dust gaps in IM~Lup, GM~Aur, and MWC~480. A more direct way to test the iceline\index{Icelines} scenario would be to measure the disk midplane temperature at the location of the rings and gaps directly from the observations and compare it with the condensation temperature of volatiles. So far, such a direct measurement has been applied only to the outer disk ($>50$ au) of HD~163296 \citep{Dullemond2020}, where it is found that the temperature is nearly constant around 20K across a region characterized by the three dust rings in the disk. This temperature is roughly consistent with the condensation temperature of CO. Clearly, ALMA\index{ALMA} observations of a larger disk sample at both higher angular resolution and sensitivity are required to understand the role that icelines play in the formation of dust rings and gaps.

While the companion and iceline scenario offers  predictions that can be tested with observations, most of the discussed (magneto-)hydrodynamic processes either have somewhat similar predictions in terms of the gas distribution and kinematics, or are difficult to observationally test. For example, numerical simulations showed that both VSI\index{Vertical shear instability} and MHD zonal flows\index{Zonal flows} tend to produce multiple sets of rings and gaps with typical radial length scales of order of a gas scale height. For VSI\index{Vertical shear instability} and magnetically-driven zonal flows\index{Zonal flows}, it is expected that they induce corrugated vertical flows with velocities as high as 10-100 m s$^{-1}$ on spatial scales comparable to the disk pressure scale height \citep{barraza2021,Hu2022}. If such flows can be identified and their radial locations coincide with continuum rings and gaps, this can support VSI or zonal flows as the origin of the annular substructures. For magnetically-driven zonal flows, the detection of magnetic fields and the variation of the magnetic field strengths across observed annular substructures at sufficiently high spatial resolution might support the link between zonal flows\index{Zonal flows} and annular substructures. However, measurements of the magnetic field morphology and strength in disk are extremely difficult. In particular, constraining the field morphology via mapping of the dust continuum polarization is hampered by dust scattering \citep[see, e.g.,][]{kataoka2015} and the direct measurement of the field strength via observations of Zeeman splitting of molecular lines require a sensitivity not yet achieved by ALMA\index{ALMA} observations \citep{harrison2021}. Despite these difficulties, a detailed study of the gas kinematics would undoubtedly provide important information on the origin of rings and gaps, and, in general, on the processes that control the evolution of circumstellar disks. So far, ALMA\index{ALMA} observations have either focused on mapping the dust continuum emission at very high angular resolution or mapping the gas kinematics at high velocity resolution (50-100 m s$^{-1}$) but on insufficient angular scales ($>0.1"$). Mapping line emission at both high angular and high velocity resolution with ALMA\index{ALMA} is thus highly desired.

We would like to conclude our discussion of rings and gaps by mentioning two other physical processes, photoevaporative/MHD winds\index{Photoevaporation}\index{Magnetized winds} and streaming instability\index{Streaming instability}. Proving winds as a viable mechanism for gap (or cavity) formation requires spatial information as to where the winds originate from, which is challenging. Nevertheless, recent high resolution spectroscopic observations of molecular vibrational lines in infrared wavelengths showed some promising results. \citet{banzatti2022} reported that spectra of CO vibrational lines in a fraction of their target disks have a combination of double-peak and triangular line shapes that can be interpreted as emission from a Keplerian disk plus disk winds. Based on their analysis, the CO emission in some disks, including transition disks SR~21 and IRS~48, appears to arise from a wind originating within the dust cavity observed in the mm. Although this does not prove that dust cavities are created by winds, it at least demonstrates that winds might be effective in removing gas within 1-10 au from the central star. 

Although streaming instability\index{Streaming instability} may operate at very small scales, the structures it produces may merge over time and potentially be observable. Considering the large amount of numerical resource it needed to capture both the instability and the large scale structure over a long period of time, the largest 3-D simulations so far only cover the disk region within 1-2 disk scale heights \citep{yang2014,LiYoudinSimon2018}. However, these simulations already reveal some axisymmetric ring formation at the radial scale of a fraction of the disk scale height. With bigger simulation boxes and longer simulation duration, these structures may be more prominent, which deserves future studies. Since dust-induced instabilities, including the streaming instability, generally require high dust-to-gas mass ratio close to unity, measuring spatially-resolved gas and dust column density would be helpful to assess whether the disk is in a condition that can trigger dust-induced instabilities. Searching for clumps along continuum rings on scales comparable to or smaller than the gas scale height will be also helpful to infer whether streaming instability is in the act.

\subsection{Spirals\index{Protoplanetary disk substructure!spirals}}
\label{sec:spirals}

The most immediate observables of spirals are the number of spirals, their radial location/extent, and their pitch angle. These properties of observed spirals are summarized in Section \ref{sec:spirals_obs}, Figure \ref{fig:spirals}, and Table \ref{tab:disks_with_spirals}. In addition to these properties, the pattern speed of spirals can be inferred via long-term monitoring observations, which can put meaningful constraints on the origin of the spirals as we will discuss below. As we pointed out in Section \ref{sec:spirals_obs}, the occurrence rate of spirals is significantly lower than that of annular substructures. In particular, a relatively large fraction of disks observed at high effective angular resolution of $\theta_D/\theta_{res} > 3$  does not have spirals (Figure \ref{fig:spirals} (b)), suggesting that the low spiral occurrence is unlikely due to insufficient angular resolution. Instead, it may well be that spirals are intrinsically rarer than annular substructures in protoplanetary disks. Alternatively, the low spiral occurrence could be associated with poor coupling of grains within spirals. When the pattern speed of a spiral $\Omega_p$ and the background Keplerian speed $\Omega_K$ differ, only small particles with a stopping time shorter than the spiral crossing time can be trapped by the spiral (Section \ref{sec:drift}). Indeed, recent multi-fluid (gas + dust) planet-disk interaction simulations showed that grains with Stokes number of $\gtrsim 0.1$ are poorly coupled to planet-driven spirals \citep{Sturm2020,Speedie2022}. Along this line, it is interesting to note that spirals are preferentially detected in NIR observations (16 out of 22 disks with observed spirals) than in mm continuum observations (7 out of 22 disks with observed spirals; see Table \ref{tab:disks_with_spirals}). A direct test could be done if we measure the pattern speed of spirals through long-term monitoring observations, which can tell us whether or not the spirals orbit at the local Keplerian frequency. Spirals driven by companions, RWI-driven\index{Rossby wave instability} vortices, or eccentric modes\index{Eccentric modes} are expected to orbit at faster or slower speed than the local Keplerian speed, whereas spirals driven by GI\index{Gravitational instability} or MRI\index{Magnetorotational instability} are expected to orbit at the local Keplerian frequency. 

As in the case of annular substructures, tidal interaction with companions is the only mechanism so far that has firm evidence for producing spirals. The discovery of spirals in wide-separation binary systems AS~205N/S, HT~LupA/B \citep{Kurtovic2018}, HD~100453A/B \citep{wagner2015}, and UX~TauA/C \citep{Menard2020,Zapata2020} offer strong evidence that companions can indeed excite observable spirals. In all three systems two-armed spirals are observed in the disk around the primary star. This is consistent with the prediction that the number of spirals driven by a companion would decrease with growing companion mass, eventually converging to two inward of the companion's orbit when the mass ratio is greater than a few $\times10^{-3}$ \citep{bae2018b}, equivalent to a few Jupiter masses around a solar-mass star. It is worth pointing out that for AS~205N and HT~LupA, spirals are detected in continuum observations with very low arm-to-trough contrast (no spirals are detected in continuum for UX~TauA/C). The low contrast supports the idea that mm grains are poorly coupled to the companion-induced spirals due to the differential rotation between the spirals and the background disk gas. In summary, all of these pieces of information are consistent with the theoretical predictions for companion-driven spirals. However, whether or not these binary systems are gravitationally bound is not entirely clear at this point \citep{salyk2014,Kurtovic2018,Zapata2020}. Future long-term monitoring observations that can characterize the orbital motion of the binary systems would enable us to determine whether these systems are gravitationally bound or they are currently experiencing flyby events\index{Stellar flyby}. 

For given underlying disk properties, spirals become weaker and more tightly wound with decreasing companion mass. This means that detecting spirals driven by planetary companions is generally more challenging than detecting spirals driven by stellar companions. As we mentioned earlier, the number of spirals a companion creates increases with decreasing companion mass and two-armed spirals are expected for multi-Jovian mass planets \citep{bae2018b}. As a result, hydrodynamic simulations often require a few to 10 Jupiter-mass planets to reproduce observed two-armed spirals (Figure \ref{fig:protoplanets}). One may ask why such large-mass planets at large orbital separation have not yet been detected since current high-contrast direct imaging observations offer sufficient sensitivity in most optimistic cases \citep[e.g.,][see also {\it Benisty et al.} in this Book]{zurlo2020,asensio-torres2021}. There are a few possible explanations to this, which are not exclusive to each other. {\it First}, the predicted planet masses from hydroynamic simulations could be larger than the actual planet masses. Most previous numerical simulations of protoplanetary disks (not just for planet-disk interaction simulations, but in general) adopted a vertically isothermal temperature structure whereas in reality, disks have stratified vertical temperature structures such that the disk surface is hotter than the disk midplane in the regions of our interest (i.e., $\gtrsim 10$~au; \citealt{law2021a}). Because planet-driven spirals form via superposition of {\it acoustic/shock} waves (Section \ref{sec:companion_spiral}), their propagation (and thus their morphology) is dependent on the underlying temperature structure. \citet{juhasz2018} showed that, for a given planet mass, the spirals launched by the planet are more opened in the surface layers when vertical temperature stratification is considered. It is thus possible that the current predictions based on the spiral morphology could overestimate the required planet masses due to the use of a vertically isothermal temperature structure. {\it Second}, the predicted flux of the planets could be overestimated. The two most widely-used direct imaging approaches are searching for infrared thermal emission and accretion tracers (e.g., H$\alpha$ line emission) from young planets. For high-contrast optical/infrared imaging, the most crucial factor that determines the observability of the planet is the luminosity of the planet. The present-day luminosity of the planet is determined by the combination of the initial luminosity and the cooling rate. Depending on the formation pathway and detailed post-formation evolution, the present-day luminosity can vary by orders of magnitude \citep[e.g.,][]{marley2007,spiegel2012,mordasini2017}. Current observing facilities can detect {\it hot-start} planets, but they are unable to detect {\it cold-start} planets \citep[e.g.,][]{asensio-torres2021}. For accretion tracers, the instantaneous accretion rate at the time of observations is crucial in determining the observability \citep{Zhu2015,Szulagyi2019,Szuagyi2020}, so whether young planets have completed gas accretion and whether planets' accretion is steady or episodic would be the question. {\it Lastly}, even if model predictions are accurate, we could still suffer from the extinction by circumplanetary and/or circumstellar disk materials along the line-of-sight to the planets \citep[e.g.,][]{sanchis2020}. 

As mentioned in Section \ref{sec:rings}, JWST observations in the mid-infrared wavelengths can offer a promising way to find young planets associated with the observed spirals. Alternatively, it has been demonstrated that velocity perturbations created by planet-driven spirals, so-called velocity kinks, can be observed with molecular line observations using ALMA\index{ALMA} (\citealt{pinte2018,teague2018,casassus2019}; see also {\it Pinte et al.} in this book). One advantage of using kinematic substructure is that the strongest velocity perturbations arising in the immediate vicinity of the planet. It can thus help not only to infer the presence of embedded planets, but also to localize them. This property can be very powerful when kinematic substructures are found along with other morphological substructures. In the case of HD~97048, velocity perturbations in CO emission are found within the continuum gap which offer promising evidence of the presence of a planet in the disk \citep{pinte2019}. In the case of AS~209, velocity perturbations are seen with an optically thick $^{12}$CO line, which coincide with a CPD seen with an optically thin $^{13}$CO line \citep{bae2022}.

Besides the companion scenario, gravitational instability\index{Gravitational instability} is another popular scenario that is often proposed to explain observed spirals. Numerical simulations predicted that GI operating in more massive disks generate fewer, more opened spirals. However, we find no clear trend that the number of spiral arms decreases with the disk mass or the pitch angle of spiral arms increases with the disk mass. Figure \ref{fig:spirals} (e) actually suggests smaller pitch angles in more massive disks. This was first noted by \cite{yu2019}. \cite{chen2021} explained this trend with the linear normal mode analysis showing that the spiral's pitch angle should be proportional to $c_s^2/M_d$. It will be important to confirm the linear normal mode analysis using direct numerical simulations in future. For a small number of objects, detailed comparisons between GI simulations and observations have been carried out. The most promising case for the GI induced spirals is Elias 2-27 \citep{hall2018,forgan2018,paneque2021}. For MWC 758 and SAO 206462, \cite{dong2015c} successfully reproduce the contrast of the observed spirals at near-IR (a factor of $\sim$3) and the pitch angle (10$^\circ$-15$^\circ$) using self-gravitating disks. These numerical simulations commonly suggest that the disk needs to be very massive with $M_d/M_* \gtrsim 0.25$ in order to explain the observed spiral morphology with GI. A direct observational test would thus be to measure the disk mass. When spatially-resolved gas surface density and gas temperature can be measured, we can also compute the Toomre $Q$ parameter to check whether and at which radii the disk is gravitationally unstable.

From the data we collected, we found that there is a potentially strong link between the presence spirals\index{Protoplanetary disk substructure!spirals} and crescents\index{Protoplanetary disk substructure!crescents} in disks. Out of 12 disks with mm crescents, 9 disks turned out to have spirals (see Tables \ref{tab:disks_with_spirals} and \ref{tab:disks_with_crescents}). One possible explanation to this potential link is that vortices have excited the spirals. In this scenario, spirals should be connected to crescents, and the two orbit the central star at the same frequency. It will be worth studying the connection between vortices and spirals for the 9 disks. Another possibility is that one physical process has created both crescents and spirals. The companion scenario is one such process, generating crescents via the RWI\index{Rossby wave instability} while exciting spirals via Lindblad resonances. In this case, vortices and spirals are not necessarily connected to each other, and they can orbit the central star at different frequencies.

Some recent NIR and (sub-)mm molecular line observations discovered asymmetric and extended streamer around some young disks  \citep{yen2019,alves2020,pineda2020,Ginski2021,garufi2021}. When the specific angular momentum of infalling materials is sufficiently different from that of the disk, infalling materials can create spirals \citep{lesur2015,kuffmeier2017,kuffmeier2018,kuznetsova2022}. Detecting spirals in the disks having large-scale streamers (and vice versa) can support the idea of infall creating spirals. In addition, because spirals are excited at the location where infalling mass\index{Infall} lands on the disk, it will be helpful to identify the location of mass landing using shock tracers, such as SO and SO$_2$. 
 
Because sound waves in differentially rotating disks propagate in the form of spirals, any perturbation in protoplanetary disks can naturally produce spirals. This also applies to turbulence driven by the hydrodynamic/MHD instabilities. In case of the MRI\index{Magnetorotational instability}, direct detection of magnetic fields in the disk via continuum and/or line polarization observations would support the MRI being the origin of observed spirals.

Spirals driven by eccentric modes\index{Eccentric modes} in marginally gravitationally unstable disks have perhaps the most distinct properties, since they are not propagating but trapped waves. Eccentric modes produce a single spiral that has a very small pitch angle of $\lesssim 5^\circ$. Also, the spiral orbits around the star at a much slower orbital frequency than the Keplerian frequency. Given that the majority of disks in our sample has more than one spiral, eccentric modes are less likely to be the dominating spiral-forming mechanism for the sample.

\subsection{Crescents\index{Protoplanetary disk substructure!crescents}}
\label{sec:crescents}

The most immediate observables of crescents are the number of crescents, their radial location, radial/azimuthal extent, and azimuthal intensity contrast. These properties of observed crescents are summarized in Section \ref{sec:crescents_obs}, Figure \ref{fig:crescent}, and Table \ref{tab:disks_with_crescents}. 

While rings and spirals have many possible origins, there are only two leading mechanisms for crescents: vortices and lumps. In general, hydrodynamic instabilities (RWI\index{Rossby wave instability}, VSI\index{Vertical shear instability}, COS\index{Convective overstability}, and ZVI\index{Zombie vortex instability}) create vortices which can trap particles therein. Lumps can form in eccentric disks as particles spend longer time near the apocenter than the pericenter. L4 and L5 Lagrangian points within the horseshoe region of a companion is another example of lumps. When vortices and lumps are compared, lumps have significantly less rotating motion around the lump center than vortices \citep{Ragusa2017}. Thus, if strong rotating gas motions around the core of a crescent can be identified via molecular line observations, it will be strong evidence that vortices driven by hydrodynamic instabilities could be the origin of the crescent. \citet{boehler2021} presented a tentative evidence that the kinematics of the disk gas around the crescent in the HD~142527 disk is consistent with the presence of a large vortex around the dust crescent, although they cautioned that the beam smearing effect may instead create a similar kinematic pattern. Future high angular/velocity resolution observations will help distinguish vortices and lumps. 

Among all 13 disks that we compiled, there is no case where multiple crescents are observed at the same radial location within a single annular structure. Because hydrodynamics instabilities generally form multiple vortices within a single ring, which subsequently merge with each other (see Section \ref{sec:hydro}), the fact that we do not see multiple crescents at the same radial location in a single disk may imply that (albeit a small number statistics) the vortex merging timescale could be relatively short. 

Unlike annular substructures or spirals, crescents do not have a case where a clear connection with a companion is confirmed.  The continuum ring beyond the orbit of PDS~70b and c has an asymmetry in the intensity, but the azimuthal intensity contrast is only moderate and it is unclear at this point whether the asymmetry is associated with a vortex, potentially driven by the RWI\index{Rossby wave instability}. In addition to crescents that can form in the gap edge via the RWI, companions can create crescents within the horseshoe region, in L4 and L5 Lagrangian points. L4 and L5 are $60^\circ$ ahead of and behind the companion, so dust concentrations that are $\simeq 120^\circ$ apart at the same radial distance to the central star may be an indication of dust trapping within Lagrangian points.  \citet{long2022} recently presented a promising evidence that dust could be trapped around Lagrangian points of yet-unseen planet embedded in the LkCa~15 disk. 

While COS\index{Convective overstability} and ZVI\index{Zombie vortex instability} are known to be able to create vortices, they are unlikely to be the cause of the crescents observed so far because they are expected to operate in the very inner part of protoplanetary disks (Section \ref{sec:cos} and \ref{sec:zvi}). In the future, ngVLA will offer the capability to probe the inner few AU regions where COS\index{Convective overstability} and ZVI\index{Zombie vortex instability} are expected to operate.  

We conclude by discussing some tentative connections between the crescent\index{Protoplanetary disk substructure!crescents} observations (Figure \ref{fig:crescent}) and theory. First, most observed vortices have aspect ratios between 2 and 5. Theoretical works suggest that only the vortices with the aspect ratio between 4 and 6 can survive the elliptical instability in unstratified disks \citep{lesur2010}. In this regard, the small aspect ratio of observed crescents (Figure \ref{fig:crescent} (a)) seem to be in tension with vortices being the origin, although we note that many of the crescents are poorly resolved in the radial direction, and future high angular resolution observations are required to obtain more accurate aspect ratio. It is also worth pointing out that crescents are observed in continuum observations and the aspect ratio in the gas could be different from what is measured with continuum. Second,
although the origin of these crescents is still unknown, some crescents are discovered in the circumbinary disks. Considering that the theory predicts that lumps in circumbinary disks are different from vortices in several ways, we can search for these differences in observations to understand the nature of these crescents. Our current statistics suggests that crescents in circumbinary disks are normally further away from the star (Figure \ref{fig:crescent} (a)), and they are also wider (with respect to the local scale height, Figure \ref{fig:crescent} (c)). 
Third, the submm intensity contrast between the vortex center and the background disk can be much larger than 10 (Figure \ref{fig:crescent} (b)), suggesting that particle trapping is likely occurring in these crescents. Finally, the crescents seem to have larger aspect ratios in more massive disks (Figure \ref{fig:crescent} (e)). Considering that a larger aspect ratio corresponds to a weaker vortex, this seems to agree with the theory that disk self-gravity weakens the vortex.

\section{\textbf{Implications to Protoplanetary Disk and Planet Formation Theory}}
\label{sec:implications}

One of the reasons disk substructures have drawn much attention from the community is because of the possibility that planets could have created them. As discussed in Section~\ref{sec:companion}, the gravitational interactions with planetary mass companions can create rings, gaps, spirals, and crescents. If the observed substructures are caused by forming planets, then there must be a large number of planets with $\sim0.1-10$ Jupiter masses at orbital distances of $10-500\,\rm au$ from their host star (Figure \ref{fig:protoplanets}). While there is little evidence for  a counterpart planet population around older, main-sequence stars, note that most of the current exoplanet detection methods are unable to find this planet population due to the short time baseline and/or insufficient sensitivity. If the lack of mature planets at large orbital distance is real, orbital migration through planet-disk interactions\index{Planet-disk interaction} may be one possibility to reconcile the potential difference in protoplanet and mature planet populations \citep{lodato2019}. However, even in that case, the question remains as to how to form giant planets at large orbital radii. In the core accretion theory, the timescale to build a planet scales roughly as $a^2$ \citep{pollack1996}, so building Jovian-mass planets via core accretion is challenging. Gravitational instability\index{Gravitational instability} offers an alternative \citep{boss1997}, but planets formed via GI tends to either grow far beyond Jupiter mass \citep[but see][]{deng2021} or subject to rapid inward migration, so the formation of planets at large orbital radii is still an open question.

If most of the observed substructures turned out to be created by other mechanisms than planet-disk interaction\index{Planet-disk interaction}, this implies that the building blocks of planets are confined in narrow regions and that the planet-forming environment could be very different from what have been assumed in planet formation models. Indeed, planetesimal and planet formation models have started to take non-smooth density distribution into account (see e.g., \citealt{carrera2021} for planetesimal formation by the streaming instability\index{Streaming instability} in pressure bumps and \citealt{chambers2021} for rapid formation of Jupiter and wide-orbit exoplanets in disks with pressure bumps). Although it is still unclear how early disk substructures form, there has been an emerging picture that they could form well before the Class II phase \citep[e.g.,][]{segura-cox2020}, corresponding to stellar ages younger than about 1 Myr. If substructures form so early during protostellar phases, it is possible that they might be related through infall\index{Infall}  and/or (magneto-)hydrodynamic/gravitational instabilities. It is then reasonable to envision a scenario in which a first generation of planets might form with the help of substructures formed via non-planetary origins, while a second generation of planets may form from the pressure bumps created by the first generation of planets.

Regardless of the formation mechanism, the ubiquity of dust rings confined within pressure maxima has demonstrated that the long-standing radial drift problem can be solved, or at least mitigated. Based on the analysis of DSHARP data, \citet{dullemondbirnstiel2018} reported that the brightest rings in AS~209, Elias~2-24, HD~163296, GW~Lup and HD~143006 contain a substantial amount of dust, between a few to about 200 Earth masses of solid material. Because ALMA\index{ALMA} observations are insensitive to solids larger than a few centimeters and protoplanetary disks can be optically thick even in mm wavelengths \citep{sierra2019,zhu2019b}, these estimates are probably lower limits for the total amount of solids in these regions. Because large dust grains and pebbles are thought to play a key role in the assembly of planets via the pebble accretion process (see {\it Drazkowska et al.} in this book), disk regions where these dust particles are enhanced relative to the gas are preferred places for the formation of rocky and perhaps giant planets. Furthermore, dust rings have been observed on a very wide range of distances from the central star, perhaps suggesting that formation of planets via core accretion might happen much further away form the central star than previously thought. Of course, as we pointed out in Section \ref{sec:substructure_occurrence}, whether substructures are similarly common in fainter disks around lower-mass stars is yet to be confirmed.

For the streaming instability\index{Streaming instability} and pebble accretion to form planets or cores of giant planets within the typical protoplanetary disk lifetime, a sufficient amount of materials should exist but also a low level of turbulence is required so that the dust density near the midplane remains sufficiently high. Edge-on disks and inclined protoplanetary disks with annular substructures offer a great opportunity to infer the level of turbulence, as the vertical thickness of the dust disk can be measured \citep[see, e.g.,][]{isella2016, pinte2016, villenave2020}. These works indicate that millimeter-size grains are confined within a vertical region whose extent is less than about 10\% of the vertical extent of small grains and gas. As discussed in Section \ref{sec:coupling}, these results suggest that disks may generally have a relatively low turbulence level, characterized by $\alpha \lesssim 10^{-3}$. In agreement with the level of turbulence inferred from the millimeter continuum observations, it is worth noting that a low level of turbulence in a small sample of protoplanetary disks was derived from measurements of non-thermal broadening of molecular lines \citep{flaherty2017,flaherty2018,flaherty2020,teague2018c}.  In conclusion, the low level of turbulence and the substantial mass confined within observed ringed substructures suggest that, if the disks with annular substructure have not yet formed any planets, these disks might be places of ongoing/future planet formation.

\section{\textbf{Conclusion}}
\label{sec:conclusion}

The prevalence of substructures in protoplanetary disks indicates that their formation likely does not require fine-tuned conditions, at least in the disk population observed so far. On the other hand, the diversity of substructure properties, for example the number and radial location of rings and gaps, suggests that specific properties of disk substructures might be sensitive to disk properties such as the gas surface density, temperature, gas-to-dust ratio, and perhaps the chemical inventory. 

While there are some suggestive findings, we find that connecting observed substructure properties with theoretical prediction is not always straightforward even with the data taken with state-of-the-art observing facilities. This could be because the quality of the data is still not sufficient or too inhomogeneous to differentiate between various mechanisms. It must be also due in part to the fact that theoretical/numerical studies often consider the gas only, while a large fraction of substructures have been observed in continuum observations, making one-to-one comparison between theoretical predictions and observations less straightforward. In addition, theoretical studies tend to prefer simplified models to highlight the fundamental physical processes. However, in order for more proper comparison with observations, theoretical models have to include all relevant physical processes. The challenges in connecting substructure properties to theoretical predictions can also be due to that different mechanisms operate in different disks so that we do not see clear correlation among the substructure properties. In the end, this all may suggest that we need to find the cause of substructures more directly, rather than relying on substructure properties to infer their cause. 

Observations of protoplanetary disks in the last few years have focused on the discovery of substructures. The new discoveries certainly have infused new energy in the study of disk evolution and planet formation. As we discussed, however, it is very challenging to infer the origin based on the morphology only. In the coming years, better morphological characterization of and kinematics associated with individual substructures using high angular/spectral resolution and sensitivity, across a larger sample including small and faint disks, will help to determine the cause of disk substructures.

\bigskip
\textbf{Acknowledgments.} We thank reviewers, Jonathan Williams, Xue-Ning Bai, and an anonymous reviewer, for carefully reading the Chapter and providing us suggestions that greatly helped to improve the Chapter.

%\centerwidetable
\begin{deluxetable}{cccccccccccccc}
%\begin{deluxetable}{Sc Sc Sc Sc Sc Sc Sc Sc Sc Sc Sc Sc Sc Sc}
%\begin{deluxetable}{Sl Sl Sl Sl Sl Sl Sl Sl Sl Sl Sl Sl Sl Sl}
\rotate
\tabletypesize{\scriptsize}
\tablecaption{Systems with rings\index{Protoplanetary disk substructure!rings and gaps}}
\renewcommand{\arraystretch}{1.2}
%\tablehead{
%\colhead{2MASS Name} &
%\colhead{Alt. Name} &
%\colhead{d} &
%\colhead{$M_\star$} &
%\colhead{$L_\star$} &
%\colhead{Class} &
%\colhead{$M_d$} &
%\colhead{$\lambda$} &
%\colhead{n} &
%\colhead{Ring location} &
%\colhead{Ring FWHM} &
%\colhead{FWHM} &
%\colhead{sep} & 
%\colhead{ref} \\
% & & \coldhead{(pc)} & \coldhead{($M_\odot$)} & \coldhead{($L_\odot$)} & & \coldhead{($0.01M_\odot$)} &  &  & \coldhead{(au)} & \coldhead{(au)} & \coldhead{(au)} & \coldhead{($\arcsec$)} &
% }
\tablehead{2MASS Name & Alt. Name & d & $M_\star$ & $L_\star$ & Class & $M_d$ & $\lambda$ & n & Ring location & Ring FWHM & FWHM & binary sep. & ref \\
& & (pc) & ($M_\odot$) & ($L_\odot$) & & ($0.01M_\odot$) & & & (au) & (au) & (au) & ($\arcsec$) & } 
\startdata
 J04345542+2428531 & AATau               & 137 &         0.5 &         0.7 & II      &     1   & mm          &   3 & 48.7,94.9,142.6                      & 2.6,28.7,26.4                     &     21 & -     & 1     \\
 J04335200+2250301 & CITau               & 159 &         0.9 &         1.6 & II      &     2.3 & mm          &   4 & 27.7,62.0,99.2,152.8                 & 19.3,29.4,8.7,59.7                &     19 & -     & 2,3,4     \\
 J04141760+2806096 & CIDA1               & 138 &         0.2 &         0.2 & II      &     0.2 & mm          &   2 & 16.2,21                              & 3.5,9.7                           &      6 & -     & 5,6     \\
 J05052286+2531312 & CIDA9A              & 172 &         0.5 &         0.1 & II      &     0.9 & mm          &   1 & 39.52                                & 25.31                             &     21 & 2.3   & 3,4,27,28     \\
 J04333906+2520382 & DLTau               & 159 &         0.9 &         1.5 & II      &     3.7 & mm          &   3 & 46.4,78.1,112.27                     & 14.63,8.6,29.6                    &     20 & -     & 3,4     \\
 J04334871+1810099 & DMTau               & 145 &         0.3 &         0.2 & TD      &     1.6 & mm          &   1 & 24                                   & 16                                &      5 & -     & 7,8,9     \\
 J04352737+2414589 & DNTau               & 128 &         0.5 &         0.7 & II      &     1.2 & mm          &   2 & 15.4,53.4                            & 21.1,7.7                          &     15 & -     & 3,10     \\
 J04474859+2925112 & DSTau               & 159 &         0.5 &         1   & II      &     0.4 & mm          &   1 & 56.8                                 & 17.2                              &     19 & 6.2   & 4,27     \\
 J04233919+2456141 & FTTau               & 128 &         0.3 &         0.4 & II      &     1.3 & mm          &   1 & 31.1                                 & 16.5                              &     15 & -     & 4     \\
 J04551098+3021595 & GMAur               & 160 &         0.8 &         1.2 & TD      &     3.8 & mm          &   3 & 37,87,177                            & 9.6,13,50.4                       &      6 & -     & 9,11,12     \\
 J04430309+2520187 & GOTau               & 145 &         0.4 &         0.3 & II      &     1   & mm          &   2 & 73.0,109.45                          & 9.8,22.2                          &     18 & -     & 3,10     \\
 J04245708+2711565 & IPTau               & 131 &         0.5 &         0.5 & II      &     0.1 & mm          &   1 & 27.1                                 & 10.4                              &     16 & -     & 4,10     \\
 J04295156+2606448 & IQTau               & 131 &         0.4 &         0.8 & II      &     0.9 & mm          &   2 & 48.2,82.8                            & 11.8,24.5                         &     18 & -     & 3,10     \\
 J04154278+2909597 & IRAS04125+2902      & 160 &         0.4 &         0.5 & TD      &     0.4 & mm          &   1 & 55                                   & 30                                &     48 & -     & 13,14     \\
 J04391779+2221034 & LkCa15              & 159 &         0.9 &         1.1 & TD      &     2.7 & mm          &   2 & 69.0,100.11                          & 14.8,34                           &      9 & -     & 15,16,17     \\
  &               &  &          &          &      &      & ir          &   1 & 58                                   & -                                 &      9 & -     & 36     \\
 J04322210+1827426 & MHO6                & 142 &         0.2 &         0.1 & II      &     0.2 & mm          &   1 & 20                                   & 20                                &     14 & -     & 6     \\
 J04584626+2950370 & MWC480/HD31648      & 162 &         2.1 &       21.9   & II      &     5.7 & mm          &   1 & 97.6                                 & 12.6                              &     23 & -     & 3     \\
 J04215740+2826355 & RYTau               & 128 &         2.0 &       12.3 & II      &     2.7 & mm          &   2 & 18.2,49.04                           & 25.6,19.5                         &     16 & -     & 3,4,19      \\
 J04300399+1813493 & UXTauA              & 140 &         1.7 &         3.2 & TD      &     0.9 & mm          &   1 & 37.5                                 & 11                                &     18 & 2.7   & 28,29,30,31     \\
 J04313843+1813576 & HLTau               & 140 &         1   &        11   & I       &    13   & mm          &   7 & 21.4,40.0,49,$\sim$58,               & 12.9,6.4,4.2,$<$15,               &      4 & -     & 20,21     \\
                   &                     &     &             &             &         &         &             &   7 & 72.2,85.4,~102                       & 5.1,12.0,$<$20                    &      4 & -     & 20,21     \\
 J05355845+2444542 & CQTau               & 162 &         1.7 &        10   & II      &     2.3 & mm          &   1 & 45                                   & 23                                &     16 & -     & 22,23,24     \\
 J05380526-0115216 & V1247Ori            & 398 &         1.9 &        15.8 & PTD     &     7.7 & mm          &   1 & 75                                   & 40                                &     16 & -     & 25,26     \\
 J16255615-2420481 & SR4                 & 135 &         0.8 &         1.8 & II      &     1.1 & mm          &   1 & 18                                   & 13.3                              &      5 & -     & 21,32     \\
 J16261033-2420548 & GSS26/ISO-Oph17     & 139 &         0.7 &         4   & II      &     2.9 & mm          &   2 & 25,47                                & 7,20                              &      3 & -     & 33,34     \\
 J16261886-2428196 & Elias2-20           & 138 &         0.9 &         2.6 & II      &     1.6 & mm          &   2 & 29,36                                & 5.2,1.9                           &      4 & -     & 21,32     \\
 J16262367-2443138 & DoAr25              & 138 &         0.8 &         1.5 & II      &     3.9 & mm          &   3 & 86,111,137                           & -,14.3,12.8                       &      4 & -     & 21,32     \\
 J16262407-2416134 & Elias2-24           & 134 &         1.1 &         6.8 & II      &     5.5 & mm          &   2 & 77,123                               & 12.2,-                            &      5 & -     & 21,32     \\
 J16264502-2423077 & GSS39/Elias2-27     & 116 &         0.6 &         1.5 & II      &     3.6 & mm          &   1 & 86                                   & 21                                &      6 & -     & 21,32,33     \\
 J16273718-2430350 & IRS48/WLY2-48       & 121 &         2   &        14.3 & TD      &     0.7 & ir          &   1 & 60                                   & -                                 &     24 & -     & 37     \\
 J16273901-2358187 & DoAr33              & 140 &         1   &         1.6 & II      &     0.6 & mm          &   1 & 17                                   & -                                 &      4 & -     & 21,32     \\
 J16273942-2439155 & ROX27/WSB52         & 137 &         0.5 &         1.1 & II      &     1.1 & mm          &   1 & 25                                   & -                                 &      3 & -     & 21,32    \\
 J16281650-2436579 & ISO-Oph196/WSB60    & 137 &         0.2 &         0.3 & II      &     1.6 & mm          &   2 & 8,34                                 & 8,16                              &      5 & -     & 34     \\
 \enddata
 \normalsize
\end{deluxetable}
 
%\centerwidetable
\begin{deluxetable}{cccccccccccccc}
\rotate
\setcounter{table}{0}
\tabletypesize{\scriptsize}
\tablecaption{Systems with rings}
\renewcommand{\arraystretch}{1.2}
%\tablehead{
%\colhead{2MASS Name} &
%\colhead{Alt. Name} &
%\colhead{d} &
%\colhead{$M_\star$} &
%\colhead{$L_\star$} &
%\colhead{Class} &
%\colhead{$M_d$} &
%\colhead{$\lambda$} &
%\colhead{n} &
%\colhead{Ring location} &
%\colhead{Ring FWHM} &
%\colhead{FWHM} &
%\colhead{sep} & 
%\colhead{ref} \\
% & & \coldhead{(pc)} & \coldhead{($M_\odot$)} & \coldhead{($L_\odot$)} & & \coldhead{($0.01M_\odot$)} &  &  & \coldhead{(au)} & \coldhead{(au)} & \coldhead{(au)} & \coldhead{($\arcsec$)} &
% }
 \tablehead{2MASS Name & Alt. Name & d & $M_\star$ & $L_\star$ & Class & $M_d$ & $\lambda$ & n & Ring location & Ring FWHM & FWHM & binary sep. & ref \\
& & (pc) & ($M_\odot$) & ($L_\odot$) & & ($0.01M_\odot$) & & & (au) & (au) & (au) & ($\arcsec$) & } 
\startdata
 J16313565-2401294 & WLY2-63/IRS63       & 144 &         -   &         -   & I/FS    &     5.9 & mm          &   2 & 27.0,51.1                            & 5.7,13.0                          &      6 & -     & 35     \\
 J16313346-2427372 & DoAr44              & 146 &         1.4 &         1   & II/PT   &     1.5 & mm          &   1 & 47                                   & 13                                &      4 & -     & 32,34     \\
 J16335560-2442049 & RXJ1633.9-2442      & 139 &         1   &         1   & II/TD   &     1.3 & mm          &   1 & 36                                   & 18                                &      3 & -     & 32,34     \\
 J16394544-2402039 & WSB82               & 139 &         1.5 &         5.1 & II      &     3.2 & mm          &   1 & 50                                   & 50                                &      3 & -     & 32,34     \\
 J16491530-1422087 & AS209               & 121 &         0.8 &         1.4 & II      &     3.6 & mm          &   7 & 14.2,27.8,38.7,                      & 8.9,4.7,3.4,                      &      5 & -     & 38     \\
                   &                     &     &             &             &         &         &             &   7 & 74.2,96.7,120.14,141                 & 9.3,8.1,11.2,2.8                  &      5 & -     &      \\
 
 J16265843-2445318 & SR24S               & 114 &         1.6 &         3.8 & II      &     2.3 & mm          &   1 & 42.2                                 & 21.4                              &      4 & 5.2   & 32,34,39     \\
 J16271027–241912  & SR21                & 138 &         2.5 &        12.6 & II      &     2.9 & mm          &   2 & 31,53                                & $<$14,25                          &     14 & -     & 40,41     \\
  &                 &  &         &        &      &      & ir          &   2 & 19,55                                & -,-                               &      7 & -     & 42     \\
 J17562128-2157218 & HD163296            & 101 &         2   &        17   & II      &     6.2 & mm          &   4 & 15.5,67.1,101.2,158.7                & 20.4,15.4,13.6,-                  &      4 & -     & 43     \\
  &             &  &            &           &      &      & ir          &   1 & 64                                   & -                                 &      8 & -     & 44     \\
 J10563044-7711393 & T4/SYCha            & 181 &         0.8 &         1   & II      &     1.3 & mm          &   1 & 90                                   & 45                                &     90 & -     & 45     \\
 J10581677-7717170 & SZCha               & 190 &         1.3 &         1.9 & II/TD   &     3.8 & mm          &   1 & 95                                   & 45                                &     95 & 5.12  & 45     \\
 J11022491-7733357 & CSCha               & 176 &         1.5 &         1.4 & TD      &     2.4 & mm          &   1 & 35.2                                 & -                                 &     10 & -     & 31,45,46     \\
 J11081509-7733531 & HPCha               & 190 &         1.4 &         1.3 & TD      &     2.6 & mm          &   1 & 42                                   & $<$12                             &      8 & -     & 31,46     \\
 J11100010-7634578 & WWCha               & 192 &         1.9 &         2.7 & II      &    17.2 & mm          &   2 & 67,122                               & 8,75                              &     13 & -     & 45     \\
 J11332542-7011412 & HD100546            & 110 &         2.2 &        25.1 & TD      &     4.3 & mm          &   2 & 20.5,29.7                            & 8.3,25                            &      6 & -     & 47     \\
 J11015191-3442170 & TWHya               &  60 &         0.5 &         0.3 & II      &     1.7 & mm          &   5 & 3.0,29.5,33,44.7,$\sim$52            & 2.2,$<6$,$<10$,2.8,-              &      1 & -     & 48, 49     \\
 J14081015-4123525 & PDS70               & 140 &         0.8 &         0.6 & TD      &     1.2 & mm          &   1 & 74                                   & 28                                &      3 & -     & 50,51,52     \\
  &                &  &          &          &      &     & ir          &   1 & 70                                   & -                                 &     10 & -     & 53     \\
 J15560921-3756057 & Sz82,IMLup          & 158 &         1   &         2.6 & II      &     4.4 & mm          &   2 & 117.4,133.5                          & 15.8,18.4                         &      8 & -     & 21,54     \\
 J15564188-4219232 & HD142527            & 157 &         2.1 &        16.2 & II/TD   &    24.9 & mm          &   1 & 205                                  & 70                                &     31 & 0.1   & 55     \\
 J15564230-3749154 & Sz83,RULup          & 160 &         0.7 &         1.5 & II      &     3.6 & mm          &   4 & 17,24,34,50                          & -,$<$8,5.5,-                      &      4 & -     & 21     \\
 J16071159-3903475 & Sz91                & 159 &         0.5 &         0.2 & TD      &     0.2 & mm          &   1 & 110.5                                & 51.7                              &     40 & -     & 54     \\
 J16083070-3828268 & -                   & 156 &         1.5 &         1.8 & TD      &     0.8 & mm          &   1 & 70                                   & 30                                &     47 & -     & 54     \\
 J16083617-3923024 & V1094Sco            & 154 &         0.8 &         1.1 & II      &     3.6 & mm          &   2 & 130,170                              & -,-                               &     14 & -     & 54     \\
 J16085468-3937431 & Sz111               & 158 &         0.5 &         0.2 & TD      &     1.3 & mm          &   1 & 56                                   & 30                                &     14 & -     & 46     \\
 J16090141-3925119 & -                   & 164 &         0.2 &         0.1 & TD      &     0.1 & mm          &   1 & 30                                   & -                                 &     49 & -     & 54     \\
 J16094864-3911169 & Sz118               & 164 &         1   &         0.7 & II      &     0.5 & mm          &   1 & 50                                   & -                                 &     49 & -     & 54     \\
 J16105158-3853137 & Sz123A              & 158 &         0.6 &         0.1 & II      &     0.3 & mm          &   1 & 35                                   & 15                                &      8 & -     & -     \\
 J15154844-3709160 & SAO206462/HD135344B & 135 &         1.6 &         9.8 & TD      &     2.3 & mm          &   1 & 51.3                                 & 19                                &     24 & -     & 56     \\
 J15583692-2257153 & HD143006            & 166 &         1.4 &         3.9 & II      &     0.9 & mm          &   3 & 7.7,40.0,63.6                        & 6,10,22                           &      7 & -     & 21     \\
 J16042165-2130284 & J1604-2130          & 150 &         1   &         0.6 & TD      &     0.9 & mm          &   1 & 70                                   & $<$30                             &     34 & -     & 38    \\
  &           &  &            &        &       &      & ir          &   1 & 60                                   & -                                 &     11 & -     & 57     \\
 J16100501-2132318 & -                   & 145 &         0.7 &         0.5 & II      &     0.3 & mm          &   2 & 28.8,41.1                            & 5.7,7.9                           &      7 & -     &   17   \\
\enddata
\vspace{-0.8cm}
\label{tab:disks_with_rings}
\tablecomments{\scriptsize References: Stellar distances, masses, and luminosities are taken from \cite{testi2022} and references therein. Dust ring properties were taken from: (1) \cite{loomis2017}, (2) \cite{clarke2018}, (3) \cite{long2018}, (4) \cite{long2019}, (5) \cite{pinilla2021}, (6) \cite{kurtovic2021} , (7) \cite{hashimoto2021}, (8) \cite{andrews2011}, (9) \cite{isella2009}, (10) \cite{andrews2018a}, (11) \cite{hughes2009}, (12) \cite{huang2020a}, (13) \cite{luhman2011}, (14) \cite{espaillat2015}, (15) \cite{andrews2011}, (16) \cite{isella2012}, (17) \cite{facchini2020}, (19) \cite{isella2010}, (20) \cite{alma2015}, (21) \cite{Huang2018a}, (22) \cite{Ubeira2019}, (23) \cite{Uyama2020}, (24) \cite{Wolfer2021}, (25) \cite{ohta2016}, (26) \cite{Kraus2017}, (27) \cite{akeson2014}, (28) \cite{Pinilla2018}, (29) \cite{Menard2020}, (30) \cite{Zapata2020}, (31) \cite{Francis+2020}, (32) \cite{cieza2019}, (33) \cite{simon2017}, (34) \cite{cieza2021}, (35) \cite{segura-cox2020}, (36) \cite{thalmann2016}, (37) \cite{follette2015}, (38) \cite{andrews_dsharp}, (39) \cite{pinilla2017}, (40) \cite{perez2014}, (41) \cite{pinilla2015}, (42) \cite{muroarena2020}, (43) \cite{isella2018}, (44) \cite{Guidi2018}, (45) \cite{pascucci2016}, (46) \cite{norfolk2021}, (47) \cite{perez2020}, (48) \cite{andrews2016}, (49) \cite{Tsukagoshi2016}, (50) \cite{keppler2019}, (51) \cite{isella2019}, (52) \cite{benisty2021}, (53) \cite{hashimoto2012},(54) \cite{Ansdell2018}, (55) \cite{fukugawa2013}, (56) \cite{Vandermarel2016}, (57) \cite{mayama2012}}
\normalsize
\end{deluxetable}
\clearpage

%\centerwidetable
\begin{deluxetable}{cccccccccccccc}
%\rotate
%\tabletypesize{\scriptsize}
\tabletypesize{\tiny}
%\tablewidth{0pt}
%\tablewidth{60em}
\tablecaption{Systems with spirals\index{Protoplanetary disk substructure!spirals}}
\renewcommand{\arraystretch}{1.2}
%\tablehead{
%\colhead{2MASS Name} &
%\colhead{Alt. Name} &
%\colhead{d} &
%\colhead{$M_\star$} &
%\colhead{$L_\star$} &
%\colhead{Class} &
%\colhead{$M_d$} &
%\colhead{$\lambda$} &
%\colhead{m} &
%\colhead{$\psi$} &
%\colhead{radial extent} &
%\colhead{FWHM} &
%\colhead{binary sep.} &
%\colhead{ref}\\
% & & (pc) & ($M_\odot)$ & ($L_\odot$) &  & ($0.01M_\odot)$ & & & ($\arcdeg$) & (au) & (au) & ($\arcsec$) & 
%}
 \tablehead{2MASS Name & Alt. Name & d & $M_\star$ & $L_\star$ & Class & $M_d$ & $\lambda$ & m & $\psi$ & radial extent & FWHM & binary sep. & ref \\
& & (pc) & ($M_\odot$) & ($L_\odot$) & & ($0.01M_\odot$) & & & ($\arcdeg$) & (au) & (au) & ($\arcsec$) & } 
\startdata
 J03454828+3224118 & LkHa330             & 309 &             2.95 &           22.91 & II      &        16.97 & ir          &   2 & 12-16       & 60-150     &     46 & -   & 5,9  \\
 J04300399+1813493 & UXTauA              & 140 &             1.67 &            3.24 & TD      &         0.91 & mm line        &   2 & 20-30        & 140-280           &     18 & 2.7  & 11 \\
  &               &  &              &           &       &         & ir          &   2 & -        & -           &     18 & 2.7  & 12 \\
 J04554582+3033043 & ABAur               & 163 &             3.17 &          123.03 & II      &         2.12 & mm line         &   2 & 20       & 30-90       &     18 & -  & 1   \\
  &               &  &              &          &     &          & ir          &   8 & 22       & 30-100      &     10 & -  & 2   \\
 J04555938+3034015 & SUAur               & 158 &             2.18 &           14.45 & II      &         0.58 & ir          &   6 & -        & -           &     11 & -  & 10   \\
 J05194140+0538428 & HD34700A            & 356 &             4.1 &           25.12 & II      &         0.8  & ir          &   6 & 27-55    & 110-320          &     18 & 5.2  & 13,14 \\
 J05302753+2519571 & MWC758/HD36112      & 156 &             1.5  &           10.96 & II      &         1.16 & mm          &   2 & 19       & 30-80       &     31 & -  &  3,4,5  \\
                 &                  &   &                 &                &        &             & ir          &   2 & 19       & 30-80       &      4 & - & 6    \\
 J05355845+2444542 & CQTau               & 162 &             1.67 &           10    & II      &         2.31 & mm line          &   2 & 20-40    & 30-65       &     19 & -   & 7  \\
  &                &  &              &               &      &          & ir          &   2 & 4, 34       & 30-60       &     16 & - & 8    \\
 J05380526-0115216 & V1247Ori            & 398 &             1.9  &           15.81 & PTD     &         7.72 & ir          &   1 & 6.5  & 96-119     &     16 & -   & 5  \\
 J11015191-3442170 & TWHya               &  60 &             0.8 &            0.28 & II      &         1.72 & mm line         &   3 & 3-9      & 70-210      &      1 & -  & 21  \\
 J11100010-7634578 & WWCha               & 192 &             1.9  &            2.69 & II      &        17.18 & ir          &   1 & -        & -           &     13 & -  &  19 \\
 J11330559-5419285 & HD100453            & 103 &             1.5  &           10    & -       &         1.35 & mm          &   2 & 5-7        & 20-30  &      3 & 1.045 & 26 \\
 &            &  &              &             &        &          & mm line         &   2 & 11-25    & 30-100       &      5 & 1.045 & 26 \\
 &            &  &              &             &        &         & ir          &   2 & 14-18    & 20-30       &      3 & 1.045 & 26 \\
 J11332542-7011412 & HD100546            & 110 &             2.2  &           25.12 & TD      &         4.34 & ir          &   6 & -        & -           &      6 & -   & 20  \\
 J11493184-7851011 & DZCha               & 110 &             0.5  &            1    & -       &         0.07 & ir          &   2 & 27     & 5-25        &      6 & -   & 5  \\
  J15154844-3709160 & SAO206462/HD135344B & 135 &             1.6  &            9.77 & TD      &        10.79 & ir          &   2 & 11       & 38-107    &     12 & -  & 5,27   \\
 J15451286-3417305 & Sz68/HTLup          & 154 &             2.15 &            5.37 & II      &         1.35 & mm          &   2 & 17      & 13-39       &      4 & 2.8  &  23 \\
 J15560921-3756057 & Sz82/IMLup          & 158 &             0.95 &            2.57 & II      &         4.37 & mm          &   2 & 10-22   & 30-94   &      8 & -   & 16  \\
 J15564188-4219232 & HD142527            & 157 &             2.1  &           16.22 & II/TD   &        24.85 & mm line         &   3 & 3-17       & 290-670    &     31 & 0.1  &  24 \\
  &             &  &               &          &    &        & ir          &   6 & -        & 80-130      &     31 & 0.1  & 5,25 \\
 J15564230-3749154 & Sz83/RULup          & 160 &             0.67 &            1.48 & II      &         3.65 & mm line        &   5 & 21-31         & 250-1200        &      4 & -  & 22   \\
 J16113134-1838259 & AS205N              & 128 &             0.99 &            2.19 & -       &        12.65 & mm          &   2 & 14       & 19-68       &      6 & 1.3 & 23  \\
  J16264502-2423077 & GSS39/Elias2-27     & 116 &             0.63 &            1.51 & II      &         3.59 & mm          &   2 & 16       & 47-244           &      6 & -   & 15,16  \\
 J16484562-1416359 & WaOph6              & 123 &             0.68 &            2.88 & II      &         2.08 & mm          &   2 & 14.9-18       & 20-70       &      7 & -   & 16  \\
  &               &  &             &            &       &         & ir          &   2 & 14-20       & 20-45       &      7 & -  & 17   \\
   -                 & SR21                & 138 &             2.5  &           12.59 & II      &         2.93 & ir          &   2 & 2-14     & 25-40       &      7 & -   &  18 \\
\enddata
\label{tab:disks_with_spirals}
\tablecomments{References: Stellar distances, masses, and luminosities are taken from \cite{testi2022} and references therein. Spiral properties are taken from: (1) \cite{tang2017}, (2) \cite{hashimoto2011}, (3) \cite{dong2018b}, (4) \cite{Boehler2018}, (5) \cite{yu2019}, (6) \cite{ren2020}, (7) \cite{Wolfer2021}, (8) \cite{Uyama2020b}, (9) \cite{uyama2018}, (10) \cite{Ginski2021}, (11) \cite{Zapata2020}, (12) \cite{Menard2020}, (13) \cite{monnier2019}, (14) \cite{Uyama2020}, (15) \cite{andrews_dsharp}, (16) \cite{huang2018b}, (17) \cite{brown-sevilla2021}, (18) \cite{muroarena2020}, (19) \cite{garufi2020}, (20) \cite{follette2017}, (21) \cite{teague2019}, (22) \cite{Huang2020b}, (23) \cite{Kurtovic2018}, (24) \cite{Christiaens2014}, (25) \cite{Avenhaus2014}, (26) \cite{rosotti2020}, (27) \cite{garufi2013}}
\normalsize
\end{deluxetable}

%\centerwidetable
\begin{deluxetable}{cccccccccccccccc}
%\rotate
\tabletypesize{\tiny}
\tablecaption{Systems with crescents\index{Protoplanetary disk substructure!crescents}}
\renewcommand{\arraystretch}{1.2}
%\tablehead{
%\colhead{2MASS Name} &
%\colhead{Alt. Name} &
%\colhead{d} &
%\colhead{$M_\star$} &
%\colhead{$L_\star$} &
%\colhead{Class} &
%\colhead{$M_d$} &
%\colhead{$\lambda$} &
%\colhead{n} &
%\colhead{rad} &
%\colhead{$w_r$} &
%\colhead{$w_\phi$} &
%\colhead{$\Delta$} &
%\colhead{FWHM} &
%\colhead{binary sep.} &
%\colhead{ref} \\
% & & \coldhead{(pc)} & \coldhead{($M_\odot$)} & \coldhead{($L_\odot$)} & & \coldhead{($0.01 M_\odot$)} &  &  & \coldhead{(au)} & \coldhead{(au)} & \coldhead{(au)} &  & \coldhead{(au)} & \coldhead{($\arcsec$)} & 
% }
  \tablehead{2MASS Name & Alt. Name & d & $M_\star$ & $L_\star$ & Class & $M_d$ & $\lambda$ & n & rad & $w_r$ & $w_\phi$ & $\Delta$ & FWHM & binary sep. & ref \\
& & (pc) & ($M_\odot$) & ($L_\odot$) & & ($0.01M_\odot$) & & & (au) & (au) & (au) &  & (au) & ($\arcsec$) & } 
\startdata
 J03454828+3224118 & LkHa330             & 309 &        2.95 &        22.91 & II      &   16.97 & mm          &   1 & 130   & $<36$        & 116     & 3.3        &     93 & -  & 1   \\
 J04554582+3033043 & ABAur               & 163 &        3.17 &       123.03 & II      &    2.12 & mm          &   1 & 120   & $<50$        & 140     & 3-4        &     23 & -  & 2   \\
 J05194140+0538428 & HD34700A            & 356 &        4.1$^a$  &        25.12 & II      &    0.8  & mm          &   1 & 155   & 72         & 173      & $>$26      &     18 & 5.2 & 3 \\
 J05302753+2519571 & MWC758/HD36112      & 156 &        1.5  &        10.96 & II      &    1.16 & mm          &   2 & 48,83 & 12,25      & 47,78   & 4.4,10     &      6 & -   & 4  \\
 J05380526-0115216 & V1247Ori            & 398 &        1.9  &        15.81 & PTD     &    7.72 & mm          &   1 & 158   & 20         & -       & 5          &     16 & -   & 5  \\
 J11332542-7011412 & HD100546            & 110 &        2.2  &        25.12 & TD      &    4.34 & mm          &   1 & 21.8  & 22         & 40      & 1.5        &      6 & -   & 6  \\
 J15154844-3709160 & SAO206462/HD135344B & 135 &        1.6  &         9.77 & TD      &   10.79 & mm          &   1 & 80.7  & 14.8       & 188     & $>$4       &     24 & -   & 7  \\
 J15404638-4229536 & HD139614            & 135 &        1.6  &         9.23 & -       &    3.02 & ir          &   4 & -     & -          & -       & -          &     68 & 0.04 & 8 \\
 J15564188-4219232 & HD142527            & 157 &        2.1  &        16.22 & II/TD   &   24.85 & mm          &   1 & 185   & 80         & 94      & 20-40      &     31 & 0.1  & 9 \\
 J15583692-2257153 & HD143006            & 166 &        1.4  &         3.89 & II      &    4.12 & mm          &   1 & 74    & 11         & 50      & $>$40      &      7 & -  & 10   \\
  &             &  &         &        &     &    & ir          &   2 & 40,74 & -,-        & -,-     & -,-        &      6 & -   & 11  \\
 J16273718-2430350 & IRS48/WLY2-48       & 121 &        2    &        14.29 & TD      &    0.65 & mm          &   1 & 63    & 42         & 110     & $>$130     &     60 & - & 12 \\
 J17562128-2157218 & HD163296            & 101 &        2    &        16.98 & II      &    6.23 & mm          &   1 & 55    & $\sim$7    & $\sim$48 & $>$20      &      4 & -   & 10  \\
 -                 & SR21                & 138 &        2.5  &        12.59 & II      &    2.93 & mm          &   1 & 46    & 34         & 95      & $\sim$2    &     14 & -  & 13   \\
\enddata
\label{tab:disks_with_crescents}
\tablecomments{$^a$ HD~34700A is a spectroscopic binary system and the stellar mass refers to the total mass of the systems. References: Stellar distances, masses, and luminosities are taken from \cite{testi2022} and references therein. Crescent properties are taken from: (1) \cite{isella2013}, (2) \cite{tang2017}, (3) \cite{benac2020}, (4) \cite{dong2018b}, (5) \cite{Kraus2017}, (6) \cite{perez2020}, (7) \cite{perez2014}, (8) \cite{muroarena2020}, (9) \cite{boehler2021}, (10) \cite{andrews_dsharp},  (11) \cite{benisty2018}, (12) \cite{vandermarel2013}, (13) \cite{pinilla2015}}
\normalsize
\end{deluxetable}

\renewcommand{\arraystretch}{1.5}
%\centerwidetable
\begin{deluxetable}{cccc}
\rotate
\centering
\tabletypesize{\scriptsize}
\tablecaption{Suggested Future Observations to Determine the Origin of Substructures}
%\tablehead{
%\colhead{Origin} & 
%\colhead{Associated substructures} & 
%\colhead{\makecell{Diagnostics}} &
%\colhead{\makecell{Required observations}} 
%}
\tablehead{Origin & Associated substructures & Diagnostics & Required observations}
\startdata
RWI\index{Rossby wave instability} (\S \ref{sec:RWI} ) & spiral/crescent &vorticity across the crescent & high angular + velocity resolution line observations \\
\hline
VSI\index{Vertical shear instability} (\S \ref{sec:vsi}) & ring/crescent & corrugated vertical flows & high angular + velocity resolution line observations \\
\hline
COS\index{Convective overstability} (\S \ref{sec:cos}) & crescent & crescents in the inner few AU & high angular resolution observations with ngVLA \\
\hline
ZVI\index{Zombie vortex instability} (\S \ref{sec:zvi}) & crescent & crescents in the inner few AU & high angular resolution observations with ngVLA \\
\hline
eccentric modes\index{Eccentric modes} (\S \ref{sec:eccentric_modes}) & spiral & small pitch angle ($\lesssim 5^\circ$) and pattern speed ($\ll \Omega_K$) & high angular resolution NIR/molecular line observations + monitoring \\ 
\hline
%infall\index{Infall} (\S \ref{sec:infall}) & spiral/crescent & \makecell{detection of large-scale envelope and streamers \\ shocks} & \makecell{medium/low angular resolution mosaic line observations \\ shock tracers (e.g., SO), chemical tracers}\\
infall\index{Infall} (\S \ref{sec:infall}) & spiral/crescent & detection of large-scale envelope and streamers  &medium/low angular resolution mosaic line observations\\
 & & shocks & shock tracers (e.g., SO), chemical tracers\\
\hline
photoevaporation\index{Photoevaporation} (\S \ref{sec:photoevaporation}) & ring &  detection of outflowing gas & molecular line observations in IR and mm wavelengths \\
\hline
zonal flows\index{Zonal flows} (\S \ref{sec:zonal_flows}, \S \ref{sec:hall}) & ring & direct detection of magnetic fields within rings/gaps & continuum/line polarization observations \\
\hline
MRI turbulence\index{Magnetorotational instability} (\S \ref{sec:mri_turbulence}) & spiral & direct detection of magnetic fields & continuum/line polarization observations \\
\hline
dead-zone\index{Dead-zone} (\S \ref{sec:deadzone}) & ring & variable level of turbulence inside/outside a ring & turbulence measurement using line observations \\
\hline
%companion (\S \ref{sec:planetary_companion}) & ring/spiral/crescent & \makecell{direct detection (IR, H$\alpha$, CPD) \\ kinematic planetary signatures} & \makecell{high angular resolution \& high contrast imaging \\ high angular + velocity resolution line observations} \\
companion (\S \ref{sec:planetary_companion}) & ring/spiral/crescent & direct detection (IR, H$\alpha$, CPD)  & high angular resolution \& high contrast imaging \\
&   & kinematic planetary signatures &   high angular + velocity resolution line observations \\
\hline
stellar flyby\index{Stellar flyby} (\S \ref{sec:flyby}) & spiral & direct detection of flyby stars & large field-of-view imaging, proper motion with Gaia\\
\hline
%GI (\S \ref{sec:GI}) & spiral & \makecell{large disk mass, Toomre $Q$ parameter \\ spiral pattern speed \& pitch angle} & \makecell{measure surface density \& temperature \\ long-term monitoring, high angular resolution imaging}\\
GI (\S \ref{sec:GI}) & spiral & large disk mass, Toomre $Q$ parameter & measure surface density \& temperature \\
 &  &  spiral pattern speed \& pitch angle & long-term monitoring, high angular resolution imaging \\
\hline
%\makecell{iceline\index{Icelines} (\S \ref{sec:icelines})} & ring & \makecell{condensation temperature \\ grain size changes across icelines} & \makecell{direct temperature measurements at the location of rings/gaps \\ multi-wavelength continuum observations}\\
iceline\index{Icelines} (\S \ref{sec:icelines}) & ring & condensation temperature & direct temperature measurements at the location of rings/gaps \\
&  & grain size changes across icelines & multi-wavelength continuum observations \\
\hline
dust-induced instabilities (\S \ref{sec:dust_instability}) & ring & high dust-to-gas mass ratio & measure both gas and dust surface density \enddata
\label{tab:future_prospects}
\end{deluxetable}

\clearpage
\bigskip
\parskip=0pt
{\small
\baselineskip=11pt
\bibliographystyle{pp7}
\bibliography{refs.bib}

\begin{thebibliography}{514}
\parskip=0pt \itemsep=0pt \small \baselineskip=11pt
\providecommand{\natexlab}[1]{#1}

\bibitem[\protect\astroncite{\emph{{Adachi} et~al.}}{1976}]{adachi1976}
{Adachi} I. et~al., 1976 \emph{Progress of Theoretical Physics}, \emph{56},
  1756.

\bibitem[\protect\astroncite{\emph{{Adams}}}{2010}]{adams2010}
{Adams} F.~C., 2010 \emph{\araa}, \emph{48}, 47.

\bibitem[\protect\astroncite{\emph{{Adams} and {Watkins}}}{1995}]{adams1995}
{Adams} F.~C. and {Watkins} R., 1995 \emph{\apj}, \emph{451}, 314.

\bibitem[\protect\astroncite{\emph{{Akeson} and {Jensen}}}{2014}]{akeson2014}
{Akeson} R.~L. and {Jensen} E.~L.~N., 2014 \emph{\apj}, \emph{784}, 1, 62.

\bibitem[\protect\astroncite{\emph{{Akiyama} et~al.}}{2019}]{akiyama2019}
{Akiyama} E. et~al., 2019 \emph{\aj}, \emph{157}, 4, 165.

\bibitem[\protect\astroncite{\emph{{Alexander} et~al.}}{2014}]{alexander2014}
{Alexander} R. et~al., 2014 \emph{Protostars and Planets VI} (H.~{Beuther},
  R.~S. {Klessen}, C.~P. {Dullemond}, and T.~{Henning}), p. 475.

\bibitem[\protect\astroncite{\emph{{ALMA Partnership} et~al.}}{2015}]{alma2015}
{ALMA Partnership} et~al., 2015 \emph{\apjl}, \emph{808}, 1, L3.

\bibitem[\protect\astroncite{\emph{{Alves} et~al.}}{2020}]{alves2020}
{Alves} F.~O. et~al., 2020 \emph{\apjl}, \emph{904}, 1, L6.

\bibitem[\protect\astroncite{\emph{{Aly} et~al.}}{2015}]{aly2015}
{Aly} H. et~al., 2015 \emph{\mnras}, \emph{449}, 65.

\bibitem[\protect\astroncite{\emph{{Andrews} et~al.}}{2009}]{andrews2009}
{Andrews} S.~M. et~al., 2009 \emph{\apj}, \emph{700}, 2, 1502.

\bibitem[\protect\astroncite{\emph{{Andrews}
  et~al.}}{2011{\natexlab{a}}}]{andrews2011b}
{Andrews} S.~M. et~al., 2011{\natexlab{a}} \emph{\apjl}, \emph{742}, 1, L5.

\bibitem[\protect\astroncite{\emph{{Andrews}
  et~al.}}{2011{\natexlab{b}}}]{andrews2011}
{Andrews} S.~M. et~al., 2011{\natexlab{b}} \emph{\apj}, \emph{732}, 1, 42.

\bibitem[\protect\astroncite{\emph{{Andrews} et~al.}}{2013}]{andrews2013}
{Andrews} S.~M. et~al., 2013 \emph{\apj}, \emph{771}, 2, 129.

\bibitem[\protect\astroncite{\emph{{Andrews} et~al.}}{2016}]{andrews2016}
{Andrews} S.~M. et~al., 2016 \emph{\apjl}, \emph{820}, 2, L40.

\bibitem[\protect\astroncite{\emph{{Andrews}
  et~al.}}{2018{\natexlab{a}}}]{andrews2018a}
{Andrews} S.~M. et~al., 2018{\natexlab{a}} \emph{\apj}, \emph{865}, 2, 157.

\bibitem[\protect\astroncite{\emph{{Andrews}
  et~al.}}{2018{\natexlab{b}}}]{andrews_dsharp}
{Andrews} S.~M. et~al., 2018{\natexlab{b}} \emph{\apjl}, \emph{869}, 2, L41.

\bibitem[\protect\astroncite{\emph{{Ansdell} et~al.}}{2018}]{Ansdell2018}
{Ansdell} M. et~al., 2018 \emph{\apj}, \emph{859}, 1, 21.

\bibitem[\protect\astroncite{\emph{{Arlt} and {Urpin}}}{2004}]{arlt2004}
{Arlt} R. and {Urpin} V., 2004 \emph{\aap}, \emph{426}, 755.

\bibitem[\protect\astroncite{\emph{{Artymowicz} and
  {Lubow}}}{1994}]{artymowicz1994}
{Artymowicz} P. and {Lubow} S.~H., 1994 \emph{ApJ}, \emph{421}, 651.

\bibitem[\protect\astroncite{\emph{{Artymowicz} and
  {Lubow}}}{1996}]{artymowicz1996}
{Artymowicz} P. and {Lubow} S.~H., 1996 \emph{\apjl}, \emph{467}, L77.

\bibitem[\protect\astroncite{\emph{{Arzamasskiy}
  et~al.}}{2018}]{Arzamasskiy2018}
{Arzamasskiy} L. et~al., 2018 \emph{\mnras}, \emph{475}, 3, 3201.

\bibitem[\protect\astroncite{\emph{{Asensio-Torres}
  et~al.}}{2021}]{asensio-torres2021}
{Asensio-Torres} R. et~al., 2021 \emph{\aap}, \emph{652}, A101.

\bibitem[\protect\astroncite{\emph{{Avenhaus} et~al.}}{2014}]{Avenhaus2014}
{Avenhaus} H. et~al., 2014 \emph{\apj}, \emph{781}, 2, 87.

\bibitem[\protect\astroncite{\emph{{Avenhaus} et~al.}}{2018}]{avenhaus2018}
{Avenhaus} H. et~al., 2018 \emph{\apj}, \emph{863}, 1, 44.

\bibitem[\protect\astroncite{\emph{{Bae} and
  {Zhu}}}{2018{\natexlab{a}}}]{bae2018a}
{Bae} J. and {Zhu} Z., 2018{\natexlab{a}} \emph{\apj}, \emph{859}, 2, 118.

\bibitem[\protect\astroncite{\emph{{Bae} and
  {Zhu}}}{2018{\natexlab{b}}}]{bae2018b}
{Bae} J. and {Zhu} Z., 2018{\natexlab{b}} \emph{\apj}, \emph{859}, 2, 119.

\bibitem[\protect\astroncite{\emph{{Bae} et~al.}}{2013}]{bae2013}
{Bae} J. et~al., 2013 \emph{\apj}, \emph{774}, 1, 57.

\bibitem[\protect\astroncite{\emph{{Bae} et~al.}}{2014}]{bae2014}
{Bae} J. et~al., 2014 \emph{\apj}, \emph{795}, 1, 61.

\bibitem[\protect\astroncite{\emph{{Bae} et~al.}}{2015}]{bae2015}
{Bae} J. et~al., 2015 \emph{\apj}, \emph{805}, 1, 15.

\bibitem[\protect\astroncite{\emph{{Bae}
  et~al.}}{2016{\natexlab{a}}}]{bae2016a}
{Bae} J. et~al., 2016{\natexlab{a}} \emph{\apj}, \emph{819}, 2, 134.

\bibitem[\protect\astroncite{\emph{{Bae}
  et~al.}}{2016{\natexlab{b}}}]{bae2016b}
{Bae} J. et~al., 2016{\natexlab{b}} \emph{\apj}, \emph{829}, 1, 13.

\bibitem[\protect\astroncite{\emph{{Bae}
  et~al.}}{2016{\natexlab{c}}}]{bae2016c}
{Bae} J. et~al., 2016{\natexlab{c}} \emph{\apj}, \emph{833}, 2, 126.

\bibitem[\protect\astroncite{\emph{{Bae} et~al.}}{2017}]{bae2017}
{Bae} J. et~al., 2017 \emph{\apj}, \emph{850}, 2, 201.

\bibitem[\protect\astroncite{\emph{{Bae} et~al.}}{2019}]{bae2019}
{Bae} J. et~al., 2019 \emph{\apjl}, \emph{884}, 2, L41.

\bibitem[\protect\astroncite{\emph{{Bae} et~al.}}{2021}]{bae2021}
{Bae} J. et~al., 2021 \emph{\apj}, \emph{912}, 1, 56.

\bibitem[\protect\astroncite{\emph{{Bae} et~al.}}{2022}]{bae2022}
{Bae} J. et~al., 2022 \emph{\apjl}, \emph{934}, 2, L20.

\bibitem[\protect\astroncite{\emph{{Baehr} and {Klahr}}}{2015}]{baehr2015}
{Baehr} H. and {Klahr} H., 2015 \emph{\apj}, \emph{814}, 2, 155.

\bibitem[\protect\astroncite{\emph{{Baehr} and
  {Zhu}}}{2021{\natexlab{a}}}]{baehr2021a}
{Baehr} H. and {Zhu} Z., 2021{\natexlab{a}} \emph{\apj}, \emph{909}, 2, 135.

\bibitem[\protect\astroncite{\emph{{Baehr} and
  {Zhu}}}{2021{\natexlab{b}}}]{baehr2021b}
{Baehr} H. and {Zhu} Z., 2021{\natexlab{b}} \emph{\apj}, \emph{909}, 2, 136.

\bibitem[\protect\astroncite{\emph{{Bai}}}{2015}]{bai2015}
{Bai} X.-N., 2015 \emph{\apj}, \emph{798}, 2, 84.

\bibitem[\protect\astroncite{\emph{{Bai} and {Stone}}}{2013}]{bai2013}
{Bai} X.-N. and {Stone} J.~M., 2013 \emph{\apj}, \emph{769}, 1, 76.

\bibitem[\protect\astroncite{\emph{{Bai} and {Stone}}}{2014}]{bai2014}
{Bai} X.-N. and {Stone} J.~M., 2014 \emph{\apj}, \emph{796}, 1, 31.

\bibitem[\protect\astroncite{\emph{{Balbus} and {Hawley}}}{1991}]{balbus1991}
{Balbus} S.~A. and {Hawley} J.~F., 1991 \emph{\apj}, \emph{376}, 214.

\bibitem[\protect\astroncite{\emph{{Bally} and {Scoville}}}{1982}]{bally1982}
{Bally} J. and {Scoville} N.~Z., 1982 \emph{\apj}, \emph{255}, 497.

\bibitem[\protect\astroncite{\emph{{Banzatti} et~al.}}{2015}]{banzatti2015}
{Banzatti} A. et~al., 2015 \emph{\apjl}, \emph{815}, 1, L15.

\bibitem[\protect\astroncite{\emph{{Banzatti} et~al.}}{2022}]{banzatti2022}
{Banzatti} A. et~al., 2022 \emph{arXiv e-prints}, arXiv:2202.03438.

\bibitem[\protect\astroncite{\emph{{Barenfeld} et~al.}}{2017}]{barenfeld2017}
{Barenfeld} S.~A. et~al., 2017 \emph{\apj}, \emph{851}, 2, 85.

\bibitem[\protect\astroncite{\emph{{Barge} and {Sommeria}}}{1995}]{barge1995}
{Barge} P. and {Sommeria} J., 1995 \emph{\aap}, \emph{295}, L1.

\bibitem[\protect\astroncite{\emph{{Barranco} and
  {Marcus}}}{2005}]{barranco2005}
{Barranco} J.~A. and {Marcus} P.~S., 2005 \emph{\apj}, \emph{623}, 2, 1157.

\bibitem[\protect\astroncite{\emph{{Barranco} et~al.}}{2018}]{barranco2018}
{Barranco} J.~A. et~al., 2018 \emph{\apj}, \emph{869}, 2, 127.

\bibitem[\protect\astroncite{\emph{{Barraza-Alfaro}
  et~al.}}{2021}]{barraza2021}
{Barraza-Alfaro} M. et~al., 2021 \emph{\aap}, \emph{653}, A113.

\bibitem[\protect\astroncite{\emph{{Baruteau} et~al.}}{2014}]{Baruteau2014}
{Baruteau} C. et~al., 2014 \emph{Protostars and Planets VI} (H.~{Beuther},
  R.~S. {Klessen}, C.~P. {Dullemond}, and T.~{Henning}), p. 667.

\bibitem[\protect\astroncite{\emph{{Bate} et~al.}}{2000}]{bate2000}
{Bate} M.~R. et~al., 2000 \emph{\mnras}, \emph{317}, 4, 773.

\bibitem[\protect\astroncite{\emph{{Benac} et~al.}}{2020}]{benac2020}
{Benac} P. et~al., 2020 \emph{\apj}, \emph{905}, 2, 120.

\bibitem[\protect\astroncite{\emph{{Benisty} et~al.}}{2015}]{Benisty2015}
{Benisty} M. et~al., 2015 \emph{\aap}, \emph{578}, L6.

\bibitem[\protect\astroncite{\emph{{Benisty} et~al.}}{2017}]{benisty2017}
{Benisty} M. et~al., 2017 \emph{\aap}, \emph{597}, A42.

\bibitem[\protect\astroncite{\emph{{Benisty} et~al.}}{2018}]{benisty2018}
{Benisty} M. et~al., 2018 \emph{\aap}, \emph{619}, A171.

\bibitem[\protect\astroncite{\emph{{Benisty} et~al.}}{2021}]{benisty2021}
{Benisty} M. et~al., 2021 \emph{\apjl}, \emph{916}, 1, L2.

\bibitem[\protect\astroncite{\emph{{B{\'e}thune} et~al.}}{2016}]{bethune2016}
{B{\'e}thune} W. et~al., 2016 \emph{\aap}, \emph{589}, A87.

\bibitem[\protect\astroncite{\emph{{B{\'e}thune} et~al.}}{2017}]{bethune2017}
{B{\'e}thune} W. et~al., 2017 \emph{\aap}, \emph{600}, A75.

\bibitem[\protect\astroncite{\emph{{B{\'e}thune} et~al.}}{2021}]{bethune2021}
{B{\'e}thune} W. et~al., 2021 \emph{\aap}, \emph{650}, A49.

\bibitem[\protect\astroncite{\emph{{Bi} et~al.}}{2020}]{bi2020}
{Bi} J. et~al., 2020 \emph{\apjl}, \emph{895}, 1, L18.

\bibitem[\protect\astroncite{\emph{{Bi} et~al.}}{2021}]{bi2021}
{Bi} J. et~al., 2021 \emph{\apj}, \emph{912}, 2, 107.

\bibitem[\protect\astroncite{\emph{{Binkert} et~al.}}{2021}]{binkert2021}
{Binkert} F. et~al., 2021 \emph{\mnras}, \emph{506}, 4, 5969.

\bibitem[\protect\astroncite{\emph{{Birnstiel} et~al.}}{2010}]{birnstiel2010}
{Birnstiel} T. et~al., 2010 \emph{\aap}, \emph{513}, A79.

\bibitem[\protect\astroncite{\emph{{Birnstiel} et~al.}}{2013}]{birnstiel2013}
{Birnstiel} T. et~al., 2013 \emph{\aap}, \emph{550}, L8.

\bibitem[\protect\astroncite{\emph{{Birnstiel} et~al.}}{2018}]{birnstiel2018}
{Birnstiel} T. et~al., 2018 \emph{\apjl}, \emph{869}, 2, L45.

\bibitem[\protect\astroncite{\emph{{Blanco} et~al.}}{2021}]{blanco2021}
{Blanco} D. et~al., 2021 \emph{arXiv e-prints}, arXiv:2108.00907.

\bibitem[\protect\astroncite{\emph{{Blandford} and
  {Payne}}}{1982}]{blandford1982}
{Blandford} R.~D. and {Payne} D.~G., 1982 \emph{\mnras}, \emph{199}, 883.

\bibitem[\protect\astroncite{\emph{{Blum} and {Wurm}}}{2000}]{blum2000}
{Blum} J. and {Wurm} G., 2000 \emph{\icarus}, \emph{143}, 138.

\bibitem[\protect\astroncite{\emph{{Boehler} et~al.}}{2018}]{Boehler2018}
{Boehler} Y. et~al., 2018 \emph{\apj}, \emph{853}, 2, 162.

\bibitem[\protect\astroncite{\emph{{Boehler} et~al.}}{2021}]{boehler2021}
{Boehler} Y. et~al., 2021 \emph{\aap}, \emph{650}, A59.

\bibitem[\protect\astroncite{\emph{{Bohn} et~al.}}{2021}]{bohn2021}
{Bohn} A.~J. et~al., 2021 \emph{arXiv e-prints}, arXiv:2112.00123.

\bibitem[\protect\astroncite{\emph{{Booth} and {Clarke}}}{2019}]{booth2019}
{Booth} R.~A. and {Clarke} C.~J., 2019 \emph{\mnras}, \emph{483}, 3, 3718.

\bibitem[\protect\astroncite{\emph{{Boss}}}{1997}]{boss1997}
{Boss} A.~P., 1997 \emph{Science}, \emph{276}, 1836.

\bibitem[\protect\astroncite{\emph{{Brandenburg} and
  {Zweibel}}}{1994}]{brandenburg1994}
{Brandenburg} A. and {Zweibel} E.~G., 1994 \emph{\apjl}, \emph{427}, L91.

\bibitem[\protect\astroncite{\emph{{Brown-Sevilla}
  et~al.}}{2021}]{brown-sevilla2021}
{Brown-Sevilla} S.~B. et~al., 2021 \emph{arXiv e-prints}, arXiv:2107.13560.

\bibitem[\protect\astroncite{\emph{{Carpenter}}}{2000}]{carpenter2000}
{Carpenter} J.~M., 2000 \emph{\aj}, \emph{120}, 6, 3139.

\bibitem[\protect\astroncite{\emph{{Carrera} et~al.}}{2021}]{carrera2021}
{Carrera} D. et~al., 2021 \emph{\aj}, \emph{161}, 2, 96.

\bibitem[\protect\astroncite{\emph{{Casassus} and
  {P{\'e}rez}}}{2019}]{casassus2019}
{Casassus} S. and {P{\'e}rez} S., 2019 \emph{\apjl}, \emph{883}, 2, L41.

\bibitem[\protect\astroncite{\emph{{Casassus} et~al.}}{2018}]{casassus2018}
{Casassus} S. et~al., 2018 \emph{\mnras}, \emph{477}, 4, 5104.

\bibitem[\protect\astroncite{\emph{{Cassen} and {Moosman}}}{1981}]{cassen1981}
{Cassen} P. and {Moosman} A., 1981 \emph{\icarus}, \emph{48}, 3, 353.

\bibitem[\protect\astroncite{\emph{{Chambers}}}{2021}]{chambers2021}
{Chambers} J., 2021 \emph{\apj}, \emph{914}, 2, 102.

\bibitem[\protect\astroncite{\emph{{Chen} et~al.}}{2019}]{chen2019}
{Chen} C. et~al., 2019 \emph{\mnras}, \emph{490}, 4, 5634.

\bibitem[\protect\astroncite{\emph{{Chen} et~al.}}{2021}]{chen2021}
{Chen} E. et~al., 2021 \emph{\apj}, \emph{906}, 1, 19.

\bibitem[\protect\astroncite{\emph{{Chiang} and
  {Goldreich}}}{1997}]{chiang1997}
{Chiang} E.~I. and {Goldreich} P., 1997 \emph{\apj}, \emph{490}, 1, 368.

\bibitem[\protect\astroncite{\emph{{Christiaens}
  et~al.}}{2014}]{Christiaens2014}
{Christiaens} V. et~al., 2014 \emph{\apjl}, \emph{785}, 1, L12.

\bibitem[\protect\astroncite{\emph{{Cieza} et~al.}}{2019}]{cieza2019}
{Cieza} L.~A. et~al., 2019 \emph{\mnras}, \emph{482}, 1, 698.

\bibitem[\protect\astroncite{\emph{{Cieza} et~al.}}{2021}]{cieza2021}
{Cieza} L.~A. et~al., 2021 \emph{\mnras}, \emph{501}, 2, 2934.

\bibitem[\protect\astroncite{\emph{{Cimerman} and
  {Rafikov}}}{2021}]{cimerman2021}
{Cimerman} N.~P. and {Rafikov} R.~R., 2021 \emph{\mnras}.

\bibitem[\protect\astroncite{\emph{{Clarke} et~al.}}{2018}]{clarke2018}
{Clarke} C.~J. et~al., 2018 \emph{\apjl}, \emph{866}, 1, L6.

\bibitem[\protect\astroncite{\emph{{Cossins} et~al.}}{2009}]{cossins2009}
{Cossins} P. et~al., 2009 \emph{\mnras}, \emph{393}, 4, 1157.

\bibitem[\protect\astroncite{\emph{{Cossins}
  et~al.}}{2010{\natexlab{a}}}]{cossins2010}
{Cossins} P. et~al., 2010{\natexlab{a}} \emph{\mnras}, \emph{407}, 1, 181.

\bibitem[\protect\astroncite{\emph{{Cossins}
  et~al.}}{2010{\natexlab{b}}}]{Cossins2010a}
{Cossins} P. et~al., 2010{\natexlab{b}} \emph{\mnras}, \emph{401}, 4, 2587.

\bibitem[\protect\astroncite{\emph{{Crnkovic-Rubsamen}
  et~al.}}{2015}]{crnkovic-Rubsamen2015}
{Crnkovic-Rubsamen} I. et~al., 2015 \emph{\mnras}, \emph{450}, 4, 4285.

\bibitem[\protect\astroncite{\emph{{Cuello} et~al.}}{2019}]{cuello2019}
{Cuello} N. et~al., 2019 \emph{\mnras}, \emph{483}, 3, 4114.

\bibitem[\protect\astroncite{\emph{{Cui} and {Bai}}}{2021}]{cui2021}
{Cui} C. and {Bai} X.-N., 2021 \emph{\mnras}, \emph{507}, 1, 1106.

\bibitem[\protect\astroncite{\emph{{Cui} and {Lin}}}{2021}]{cui2021b}
{Cui} C. and {Lin} M.-K., 2021 \emph{\mnras}, \emph{505}, 2, 2983.

\bibitem[\protect\astroncite{\emph{{Currie} et~al.}}{2022}]{currie2022}
{Currie} T. et~al., 2022 \emph{Nature Astronomy}, \emph{6}, 751.

\bibitem[\protect\astroncite{\emph{{Czekala} et~al.}}{2019}]{czekala2019}
{Czekala} I. et~al., 2019 \emph{\apj}, \emph{883}, 1, 22.

\bibitem[\protect\astroncite{\emph{{D'Alessio} et~al.}}{1998}]{dalessio1998}
{D'Alessio} P. et~al., 1998 \emph{\apj}, \emph{500}, 1, 411.

\bibitem[\protect\astroncite{\emph{{de Val-Borro}
  et~al.}}{2007}]{deval-borro2007}
{de Val-Borro} M. et~al., 2007 \emph{\aap}, \emph{471}, 3, 1043.

\bibitem[\protect\astroncite{\emph{{Deng} et~al.}}{2021}]{deng2021}
{Deng} H. et~al., 2021 \emph{Nature Astronomy}, \emph{5}, 440.

\bibitem[\protect\astroncite{\emph{{Doi} and {Kataoka}}}{2021}]{doi2021}
{Doi} K. and {Kataoka} A., 2021 \emph{\apj}, \emph{912}, 2, 164.

\bibitem[\protect\astroncite{\emph{{Dominik} and
  {Tielens}}}{1997}]{dominik1997}
{Dominik} C. and {Tielens} A.~G.~G.~M., 1997 \emph{\apj}, \emph{480}, 2, 647.

\bibitem[\protect\astroncite{\emph{{Dong} and {Fung}}}{2017}]{dong2017}
{Dong} R. and {Fung} J., 2017 \emph{\apj}, \emph{835}, 2, 146.

\bibitem[\protect\astroncite{\emph{{Dong} et~al.}}{2015}]{dong2015c}
{Dong} R. et~al., 2015 \emph{\apjl}, \emph{812}, 2, L32.

\bibitem[\protect\astroncite{\emph{{Dong} et~al.}}{2017}]{dong2017b}
{Dong} R. et~al., 2017 \emph{\apj}, \emph{843}, 2, 127.

\bibitem[\protect\astroncite{\emph{{Dong}
  et~al.}}{2018{\natexlab{a}}}]{dong2018c}
{Dong} R. et~al., 2018{\natexlab{a}} \emph{\apj}, \emph{866}, 2, 110.

\bibitem[\protect\astroncite{\emph{{Dong}
  et~al.}}{2018{\natexlab{b}}}]{dong2018b}
{Dong} R. et~al., 2018{\natexlab{b}} \emph{\apj}, \emph{860}, 2, 124.

\bibitem[\protect\astroncite{\emph{{Doolin} and {Blundell}}}{2011}]{doolin2011}
{Doolin} S. and {Blundell} K.~M., 2011 \emph{\mnras}, \emph{418}, 2656.

\bibitem[\protect\astroncite{\emph{{Draine}}}{2016}]{draine2016}
{Draine} B.~T., 2016 \emph{\apj}, \emph{831}, 1, 109.

\bibitem[\protect\astroncite{\emph{{Drazkowska} and
  {Alibert}}}{2017}]{drazkowska2017}
{Drazkowska} J. and {Alibert} Y., 2017 \emph{\aap}, \emph{608}, A92.

\bibitem[\protect\astroncite{\emph{{Duffell} and
  {MacFadyen}}}{2013}]{duffell2013}
{Duffell} P.~C. and {MacFadyen} A.~I., 2013 \emph{\apj}, \emph{769}, 1, 41.

\bibitem[\protect\astroncite{\emph{{Dullemond}}}{2000}]{dullemond2000}
{Dullemond} C.~P., 2000 \emph{\aap}, \emph{361}, L17.

\bibitem[\protect\astroncite{\emph{{Dullemond} and
  {Penzlin}}}{2018}]{dullemond2018}
{Dullemond} C.~P. and {Penzlin} A.~B.~T., 2018 \emph{\aap}, \emph{609}, A50.

\bibitem[\protect\astroncite{\emph{{Dullemond} et~al.}}{2001}]{dullemond2001}
{Dullemond} C.~P. et~al., 2001 \emph{\apj}, \emph{560}, 2, 957.

\bibitem[\protect\astroncite{\emph{{Dullemond}
  et~al.}}{2018}]{dullemondbirnstiel2018}
{Dullemond} C.~P. et~al., 2018 \emph{\apjl}, \emph{869}, 2, L46.

\bibitem[\protect\astroncite{\emph{{Dullemond} et~al.}}{2020}]{Dullemond2020}
{Dullemond} C.~P. et~al., 2020 \emph{\aap}, \emph{633}, A137.

\bibitem[\protect\astroncite{\emph{{Dutrey} et~al.}}{2014}]{dutrey2014}
{Dutrey} A. et~al., 2014 \emph{Protostars and Planets VI} (H.~{Beuther}, R.~S.
  {Klessen}, C.~P. {Dullemond}, and T.~{Henning}), p. 317.

\bibitem[\protect\astroncite{\emph{{Dzyurkevich}
  et~al.}}{2013}]{dzyurkevich2013}
{Dzyurkevich} N. et~al., 2013 \emph{\apj}, \emph{765}, 2, 114.

\bibitem[\protect\astroncite{\emph{{Epstein}}}{1924}]{epstein1924}
{Epstein} P.~S., 1924 \emph{Physical Review}, \emph{23}, 6, 710.

\bibitem[\protect\astroncite{\emph{{Ercolano} et~al.}}{2021}]{ercolano2021}
{Ercolano} B. et~al., 2021 \emph{\mnras}, \emph{508}, 2, 1675.

\bibitem[\protect\astroncite{\emph{{Espaillat} et~al.}}{2014}]{espaillat2014}
{Espaillat} C. et~al., 2014 \emph{Protostars and Planets VI} (H.~{Beuther},
  R.~S. {Klessen}, C.~P. {Dullemond}, and T.~{Henning}), p. 497.

\bibitem[\protect\astroncite{\emph{{Espaillat} et~al.}}{2015}]{espaillat2015}
{Espaillat} C. et~al., 2015 \emph{\apj}, \emph{807}, 2, 156.

\bibitem[\protect\astroncite{\emph{{Evans} et~al.}}{2009}]{evans2009}
{Evans} Neal~J. I. et~al., 2009 \emph{\apjs}, \emph{181}, 2, 321.

\bibitem[\protect\astroncite{\emph{{Evans} et~al.}}{2017}]{evans2017}
{Evans} M.~G. et~al., 2017 \emph{\mnras}, \emph{470}, 2, 1828.

\bibitem[\protect\astroncite{\emph{{Facchini} et~al.}}{2013}]{facchini2013}
{Facchini} S. et~al., 2013 \emph{MNRAS}, \emph{433}, 2142.

\bibitem[\protect\astroncite{\emph{{Facchini} et~al.}}{2018}]{facchini2018}
{Facchini} S. et~al., 2018 \emph{\mnras}, \emph{473}, 4, 4459.

\bibitem[\protect\astroncite{\emph{{Facchini} et~al.}}{2020}]{facchini2020}
{Facchini} S. et~al., 2020 \emph{\aap}, \emph{639}, A121.

\bibitem[\protect\astroncite{\emph{{Fairbairn} and
  {Rafikov}}}{2022}]{FairbairnRafikov2022}
{Fairbairn} C.~W. and {Rafikov} R.~R., 2022 \emph{\mnras}.

\bibitem[\protect\astroncite{\emph{{Farago} and {Laskar}}}{2010}]{farago2010}
{Farago} F. and {Laskar} J., 2010 \emph{\mnras}, \emph{401}, 1189.

\bibitem[\protect\astroncite{\emph{{Flaherty} et~al.}}{2020}]{flaherty2020}
{Flaherty} K. et~al., 2020 \emph{\apj}, \emph{895}, 2, 109.

\bibitem[\protect\astroncite{\emph{{Flaherty} et~al.}}{2017}]{flaherty2017}
{Flaherty} K.~M. et~al., 2017 \emph{\apj}, \emph{843}, 2, 150.

\bibitem[\protect\astroncite{\emph{{Flaherty} et~al.}}{2018}]{flaherty2018}
{Flaherty} K.~M. et~al., 2018 \emph{\apj}, \emph{856}, 2, 117.

\bibitem[\protect\astroncite{\emph{{Flock} et~al.}}{2011}]{flock2011}
{Flock} M. et~al., 2011 \emph{\apj}, \emph{735}, 2, 122.

\bibitem[\protect\astroncite{\emph{{Flock} et~al.}}{2015}]{flock2015}
{Flock} M. et~al., 2015 \emph{\aap}, \emph{574}, A68.

\bibitem[\protect\astroncite{\emph{{Flock} et~al.}}{2016}]{flock2016}
{Flock} M. et~al., 2016 \emph{\apj}, \emph{827}, 2, 144.

\bibitem[\protect\astroncite{\emph{{Flock}
  et~al.}}{2017{\natexlab{a}}}]{flock2017b}
{Flock} M. et~al., 2017{\natexlab{a}} \emph{\apj}, \emph{835}, 2, 230.

\bibitem[\protect\astroncite{\emph{{Flock}
  et~al.}}{2017{\natexlab{b}}}]{flock2017}
{Flock} M. et~al., 2017{\natexlab{b}} \emph{\apj}, \emph{850}, 2, 131.

\bibitem[\protect\astroncite{\emph{{Flock} et~al.}}{2019}]{flock2019}
{Flock} M. et~al., 2019 \emph{\aap}, \emph{630}, A147.

\bibitem[\protect\astroncite{\emph{{Flock} et~al.}}{2020}]{flock2020}
{Flock} M. et~al., 2020 \emph{\apj}, \emph{897}, 2, 155.

\bibitem[\protect\astroncite{\emph{{Follette} et~al.}}{2015}]{follette2015}
{Follette} K.~B. et~al., 2015 \emph{\apj}, \emph{798}, 2, 132.

\bibitem[\protect\astroncite{\emph{{Follette} et~al.}}{2017}]{follette2017}
{Follette} K.~B. et~al., 2017 \emph{\aj}, \emph{153}, 6, 264.

\bibitem[\protect\astroncite{\emph{{Forgan} et~al.}}{2011}]{forgan2011}
{Forgan} D. et~al., 2011 \emph{\mnras}, \emph{410}, 2, 994.

\bibitem[\protect\astroncite{\emph{{Forgan} et~al.}}{2012}]{forgan2012}
{Forgan} D. et~al., 2012 \emph{\mnras}, \emph{426}, 3, 2419.

\bibitem[\protect\astroncite{\emph{{Forgan} et~al.}}{2018}]{forgan2018}
{Forgan} D.~H. et~al., 2018 \emph{\apjl}, \emph{860}, 1, L5.

\bibitem[\protect\astroncite{\emph{{Foucart} and {Lai}}}{2013}]{foucart2013}
{Foucart} F. and {Lai} D., 2013 \emph{\apj}, \emph{764}, 106.

\bibitem[\protect\astroncite{\emph{{Franchini}
  et~al.}}{2019{\natexlab{a}}}]{franchini2019b}
{Franchini} A. et~al., 2019{\natexlab{a}} \emph{\apjl}, \emph{880}, 2, L18.

\bibitem[\protect\astroncite{\emph{{Franchini}
  et~al.}}{2019{\natexlab{b}}}]{franchini2019}
{Franchini} A. et~al., 2019{\natexlab{b}} \emph{\mnras}, \emph{485}, 315.

\bibitem[\protect\astroncite{\emph{{Franchini} et~al.}}{2020}]{franchini2020}
{Franchini} A. et~al., 2020 \emph{\mnras}, \emph{491}, 4, 5351.

\bibitem[\protect\astroncite{\emph{{Francis} and {van der
  Marel}}}{2020}]{Francis+2020}
{Francis} L. and {van der Marel} N., 2020 \emph{\apj}, \emph{892}, 2, 111.

\bibitem[\protect\astroncite{\emph{{Fu} et~al.}}{2014{\natexlab{a}}}]{fu2014b}
{Fu} W. et~al., 2014{\natexlab{a}} \emph{\apjl}, \emph{795}, 2, L39.

\bibitem[\protect\astroncite{\emph{{Fu} et~al.}}{2014{\natexlab{b}}}]{fu2014a}
{Fu} W. et~al., 2014{\natexlab{b}} \emph{\apjl}, \emph{788}, 2, L41.

\bibitem[\protect\astroncite{\emph{{Fu} et~al.}}{2015}]{fu2015}
{Fu} W. et~al., 2015 \emph{\apj}, \emph{807}, 75.

\bibitem[\protect\astroncite{\emph{{Fukagawa} et~al.}}{2006}]{Fukagawa2006}
{Fukagawa} M. et~al., 2006 \emph{\apjl}, \emph{636}, 2, L153.

\bibitem[\protect\astroncite{\emph{{Fukagawa} et~al.}}{2013}]{fukugawa2013}
{Fukagawa} M. et~al., 2013 \emph{\pasj}, \emph{65}, L14.

\bibitem[\protect\astroncite{\emph{{Fukuhara} et~al.}}{2021}]{fukuhara2021}
{Fukuhara} Y. et~al., 2021 \emph{\apj}, \emph{914}, 2, 132.

\bibitem[\protect\astroncite{\emph{{Fung} et~al.}}{2014}]{fung2014}
{Fung} J. et~al., 2014 \emph{\apj}, \emph{782}, 2, 88.

\bibitem[\protect\astroncite{\emph{{Gammie}}}{1996}]{gammie1996}
{Gammie} C.~F., 1996 \emph{\apj}, \emph{457}, 355.

\bibitem[\protect\astroncite{\emph{{Gammie}}}{2001}]{gammie2001}
{Gammie} C.~F., 2001 \emph{\apj}, \emph{553}, 1, 174.

\bibitem[\protect\astroncite{\emph{{G{\'a}rate} et~al.}}{2021}]{garate2021}
{G{\'a}rate} M. et~al., 2021 \emph{\aap}, \emph{655}, A18.

\bibitem[\protect\astroncite{\emph{{Garufi} et~al.}}{2013}]{garufi2013}
{Garufi} A. et~al., 2013 \emph{\aap}, \emph{560}, A105.

\bibitem[\protect\astroncite{\emph{{Garufi} et~al.}}{2020}]{garufi2020}
{Garufi} A. et~al., 2020 \emph{\aap}, \emph{633}, A82.

\bibitem[\protect\astroncite{\emph{{Garufi} et~al.}}{2021}]{garufi2021}
{Garufi} A. et~al., 2021 \emph{arXiv e-prints}, arXiv:2110.13820.

\bibitem[\protect\astroncite{\emph{{Gibbons} et~al.}}{2014}]{gibbons2014}
{Gibbons} P.~G. et~al., 2014 \emph{\mnras}, \emph{442}, 1, 361.

\bibitem[\protect\astroncite{\emph{{Ginski} et~al.}}{2021}]{Ginski2021}
{Ginski} C. et~al., 2021 \emph{\apjl}, \emph{908}, 2, L25.

\bibitem[\protect\astroncite{\emph{{Ginzburg} and {Sari}}}{2018}]{ginzburg2018}
{Ginzburg} S. and {Sari} R., 2018 \emph{\mnras}, \emph{479}, 2, 1986.

\bibitem[\protect\astroncite{\emph{{Goldreich} and
  {Tremaine}}}{1978}]{goldreich1978}
{Goldreich} P. and {Tremaine} S., 1978 \emph{\apj}, \emph{222}, 850.

\bibitem[\protect\astroncite{\emph{{Goldreich} and
  {Tremaine}}}{1979}]{goldreich1979}
{Goldreich} P. and {Tremaine} S., 1979 \emph{\apj}, \emph{233}, 857.

\bibitem[\protect\astroncite{\emph{{Gonzalez} et~al.}}{2017}]{gonzalez2017}
{Gonzalez} J.~F. et~al., 2017 \emph{\mnras}, \emph{467}, 2, 1984.

\bibitem[\protect\astroncite{\emph{{Goodman} and
  {Rafikov}}}{2001}]{goodman2001}
{Goodman} J. and {Rafikov} R.~R., 2001 \emph{\apj}, \emph{552}, 2, 793.

\bibitem[\protect\astroncite{\emph{{Grady} et~al.}}{1999}]{grady1999}
{Grady} C.~A. et~al., 1999 \emph{\apjl}, \emph{523}, 2, L151.

\bibitem[\protect\astroncite{\emph{{Guidi} et~al.}}{2018}]{Guidi2018}
{Guidi} G. et~al., 2018 \emph{\mnras}, \emph{479}, 2, 1505.

\bibitem[\protect\astroncite{\emph{{Gundlach} and {Blum}}}{2015}]{gundlach2015}
{Gundlach} B. and {Blum} J., 2015 \emph{\apj}, \emph{798}, 1, 34.

\bibitem[\protect\astroncite{\emph{{Gundlach} et~al.}}{2018}]{gundlach2018}
{Gundlach} B. et~al., 2018 \emph{\mnras}, \emph{479}, 1, 1273.

\bibitem[\protect\astroncite{\emph{{Haffert} et~al.}}{2019}]{haffert2019}
{Haffert} S.~Y. et~al., 2019 \emph{Nature Astronomy}, \emph{3}, 749.

\bibitem[\protect\astroncite{\emph{{Hall} et~al.}}{2018}]{hall2018}
{Hall} C. et~al., 2018 \emph{\mnras}, \emph{477}, 1, 1004.

\bibitem[\protect\astroncite{\emph{{Hall} et~al.}}{2019}]{hall2019}
{Hall} C. et~al., 2019 \emph{\apj}, \emph{871}, 2, 228.

\bibitem[\protect\astroncite{\emph{{Hall} et~al.}}{2020}]{hall2020}
{Hall} C. et~al., 2020 \emph{\apj}, \emph{904}, 2, 148.

\bibitem[\protect\astroncite{\emph{{Harrison} et~al.}}{2021}]{harrison2021}
{Harrison} R.~E. et~al., 2021 \emph{\apj}, \emph{908}, 2, 141.

\bibitem[\protect\astroncite{\emph{{Hashimoto} et~al.}}{2011}]{hashimoto2011}
{Hashimoto} J. et~al., 2011 \emph{\apjl}, \emph{729}, 2, L17.

\bibitem[\protect\astroncite{\emph{{Hashimoto} et~al.}}{2012}]{hashimoto2012}
{Hashimoto} J. et~al., 2012 \emph{\apjl}, \emph{758}, 1, L19.

\bibitem[\protect\astroncite{\emph{{Hashimoto} et~al.}}{2021}]{hashimoto2021}
{Hashimoto} J. et~al., 2021 \emph{\apj}, \emph{911}, 1, 5.

\bibitem[\protect\astroncite{\emph{{Hawley}}}{2001}]{hawley2001}
{Hawley} J.~F., 2001 \emph{\apj}, \emph{554}, 1, 534.

\bibitem[\protect\astroncite{\emph{{Hayashi}}}{1981}]{hayashi1981}
{Hayashi} C., 1981 \emph{Progress of Theoretical Physics Supplement},
  \emph{70}, 35.

\bibitem[\protect\astroncite{\emph{{Heinemann} and
  {Papaloizou}}}{2009{\natexlab{a}}}]{heinemann2009a}
{Heinemann} T. and {Papaloizou} J.~C.~B., 2009{\natexlab{a}} \emph{\mnras},
  \emph{397}, 1, 52.

\bibitem[\protect\astroncite{\emph{{Heinemann} and
  {Papaloizou}}}{2009{\natexlab{b}}}]{heinemann2009b}
{Heinemann} T. and {Papaloizou} J.~C.~B., 2009{\natexlab{b}} \emph{\mnras},
  \emph{397}, 1, 64.

\bibitem[\protect\astroncite{\emph{{Helled} et~al.}}{2014}]{helled2014}
{Helled} R. et~al., 2014 \emph{Protostars and Planets VI} (H.~{Beuther}, R.~S.
  {Klessen}, C.~P. {Dullemond}, and T.~{Henning}), p. 643.

\bibitem[\protect\astroncite{\emph{{Hendler} et~al.}}{2020}]{hendler2020}
{Hendler} N. et~al., 2020 \emph{\apj}, \emph{895}, 2, 126.

\bibitem[\protect\astroncite{\emph{{Hennebelle} et~al.}}{2016}]{hennebelle2016}
{Hennebelle} P. et~al., 2016 \emph{\aap}, \emph{590}, A22.

\bibitem[\protect\astroncite{\emph{{Hennebelle} et~al.}}{2017}]{hennebelle2017}
{Hennebelle} P. et~al., 2017 \emph{\aap}, \emph{599}, A86.

\bibitem[\protect\astroncite{\emph{{Hirose} and {Shi}}}{2019}]{hirose2019}
{Hirose} S. and {Shi} J.-M., 2019 \emph{\mnras}, \emph{485}, 1, 266.

\bibitem[\protect\astroncite{\emph{{Hollenbach} et~al.}}{1994}]{hollenbach1994}
{Hollenbach} D. et~al., 1994 \emph{\apj}, \emph{428}, 654.

\bibitem[\protect\astroncite{\emph{{Hu} et~al.}}{2019}]{hu2019}
{Hu} X. et~al., 2019 \emph{\apj}, \emph{885}, 1, 36.

\bibitem[\protect\astroncite{\emph{{Hu} et~al.}}{2021}]{hu2021}
{Hu} X. et~al., 2021 \emph{\apj}, \emph{913}, 2, 133.

\bibitem[\protect\astroncite{\emph{{Hu} et~al.}}{2022}]{Hu2022}
{Hu} X. et~al., 2022 \emph{arXiv e-prints}, arXiv:2203.05629.

\bibitem[\protect\astroncite{\emph{{Huang}
  et~al.}}{2018{\natexlab{a}}}]{huang2018c}
{Huang} J. et~al., 2018{\natexlab{a}} \emph{\apj}, \emph{852}, 2, 122.

\bibitem[\protect\astroncite{\emph{{Huang}
  et~al.}}{2018{\natexlab{b}}}]{Huang2018a}
{Huang} J. et~al., 2018{\natexlab{b}} \emph{\apjl}, \emph{869}, 2, L42.

\bibitem[\protect\astroncite{\emph{{Huang}
  et~al.}}{2018{\natexlab{c}}}]{huang2018b}
{Huang} J. et~al., 2018{\natexlab{c}} \emph{\apjl}, \emph{869}, 2, L43.

\bibitem[\protect\astroncite{\emph{{Huang}
  et~al.}}{2020{\natexlab{a}}}]{huang2020a}
{Huang} J. et~al., 2020{\natexlab{a}} \emph{\apj}, \emph{891}, 1, 48.

\bibitem[\protect\astroncite{\emph{{Huang}
  et~al.}}{2020{\natexlab{b}}}]{Huang2020b}
{Huang} J. et~al., 2020{\natexlab{b}} \emph{\apj}, \emph{898}, 2, 140.

\bibitem[\protect\astroncite{\emph{{Huang} et~al.}}{2019}]{huang2019}
{Huang} P. et~al., 2019 \emph{\apjl}, \emph{883}, 2, L39.

\bibitem[\protect\astroncite{\emph{{Hughes} et~al.}}{2009}]{hughes2009}
{Hughes} A.~M. et~al., 2009 \emph{\apj}, \emph{698}, 1, 131.

\bibitem[\protect\astroncite{\emph{{Hyodo} et~al.}}{2019}]{hyodo2019}
{Hyodo} R. et~al., 2019 \emph{\aap}, \emph{629}, A90.

\bibitem[\protect\astroncite{\emph{{Ida} and {Guillot}}}{2016}]{ida2016}
{Ida} S. and {Guillot} T., 2016 \emph{\aap}, \emph{596}, L3.

\bibitem[\protect\astroncite{\emph{{Inaba} and {Barge}}}{2006}]{inaba2006}
{Inaba} S. and {Barge} P., 2006 \emph{\apj}, \emph{649}, 1, 415.

\bibitem[\protect\astroncite{\emph{{Isella} and
  {Turner}}}{2018}]{isella_turner2018}
{Isella} A. and {Turner} N.~J., 2018 \emph{\apj}, \emph{860}, 1, 27.

\bibitem[\protect\astroncite{\emph{{Isella} et~al.}}{2009}]{isella2009}
{Isella} A. et~al., 2009 \emph{\apj}, \emph{701}, 1, 260.

\bibitem[\protect\astroncite{\emph{{Isella} et~al.}}{2010}]{isella2010}
{Isella} A. et~al., 2010 \emph{\apj}, \emph{714}, 2, 1746.

\bibitem[\protect\astroncite{\emph{{Isella} et~al.}}{2012}]{isella2012}
{Isella} A. et~al., 2012 \emph{\apj}, \emph{747}, 2, 136.

\bibitem[\protect\astroncite{\emph{{Isella} et~al.}}{2013}]{isella2013}
{Isella} A. et~al., 2013 \emph{\apj}, \emph{775}, 1, 30.

\bibitem[\protect\astroncite{\emph{{Isella} et~al.}}{2014}]{Isella2014}
{Isella} A. et~al., 2014 \emph{\apj}, \emph{788}, 2, 129.

\bibitem[\protect\astroncite{\emph{{Isella} et~al.}}{2016}]{isella2016}
{Isella} A. et~al., 2016 \emph{\prl}, \emph{117}, 25, 251101.

\bibitem[\protect\astroncite{\emph{{Isella} et~al.}}{2018}]{isella2018}
{Isella} A. et~al., 2018 \emph{\apjl}, \emph{869}, 2, L49.

\bibitem[\protect\astroncite{\emph{{Isella} et~al.}}{2019}]{isella2019}
{Isella} A. et~al., 2019 \emph{\apjl}, \emph{879}, 2, L25.

\bibitem[\protect\astroncite{\emph{{Jacquemin-Ide}
  et~al.}}{2021}]{Jacquemin-Ide2021}
{Jacquemin-Ide} J. et~al., 2021 \emph{\aap}, \emph{647}, A192.

\bibitem[\protect\astroncite{\emph{{Jin} et~al.}}{2019}]{Jin2019}
{Jin} S. et~al., 2019 \emph{\apj}, \emph{881}, 2, 108.

\bibitem[\protect\astroncite{\emph{{Johansen} and
  {Youdin}}}{2007}]{johansen2007}
{Johansen} A. and {Youdin} A., 2007 \emph{\apj}, \emph{662}, 1, 627.

\bibitem[\protect\astroncite{\emph{{Johansen} et~al.}}{2009}]{johansen2009}
{Johansen} A. et~al., 2009 \emph{\apj}, \emph{697}, 2, 1269.

\bibitem[\protect\astroncite{\emph{{Johansen} et~al.}}{2014}]{johansen2014}
{Johansen} A. et~al., 2014 \emph{Protostars and Planets VI} (H.~{Beuther},
  R.~S. {Klessen}, C.~P. {Dullemond}, and T.~{Henning}), p. 547.

\bibitem[\protect\astroncite{\emph{{Juh{\'a}sz} and
  {Facchini}}}{2017}]{juhasz2017}
{Juh{\'a}sz} A. and {Facchini} S., 2017 \emph{\mnras}, \emph{466}, 4, 4053.

\bibitem[\protect\astroncite{\emph{{Juh{\'a}sz} and
  {Rosotti}}}{2018}]{juhasz2018}
{Juh{\'a}sz} A. and {Rosotti} G.~P., 2018 \emph{\mnras}, \emph{474}, 1, L32.

\bibitem[\protect\astroncite{\emph{{Kanagawa}
  et~al.}}{2015{\natexlab{a}}}]{kanagawa2015a}
{Kanagawa} K.~D. et~al., 2015{\natexlab{a}} \emph{\mnras}, \emph{448}, 1, 994.

\bibitem[\protect\astroncite{\emph{{Kanagawa}
  et~al.}}{2015{\natexlab{b}}}]{kanagawa2015b}
{Kanagawa} K.~D. et~al., 2015{\natexlab{b}} \emph{\apjl}, \emph{806}, 1, L15.

\bibitem[\protect\astroncite{\emph{{Kanagawa} et~al.}}{2016}]{kanagawa2016}
{Kanagawa} K.~D. et~al., 2016 \emph{\pasj}, \emph{68}, 3, 43.

\bibitem[\protect\astroncite{\emph{{Kanagawa} et~al.}}{2017}]{kanagawa2017}
{Kanagawa} K.~D. et~al., 2017 \emph{\pasj}, \emph{69}, 6, 97.

\bibitem[\protect\astroncite{\emph{{Kanagawa} et~al.}}{2020}]{kanagawa2020}
{Kanagawa} K.~D. et~al., 2020 \emph{\apj}, \emph{892}, 2, 83.

\bibitem[\protect\astroncite{\emph{{Kanagawa} et~al.}}{2021}]{kanagawa2021}
{Kanagawa} K.~D. et~al., 2021 \emph{arXiv e-prints}, arXiv:2109.09579.

\bibitem[\protect\astroncite{\emph{{Kataoka} et~al.}}{2015}]{kataoka2015}
{Kataoka} A. et~al., 2015 \emph{\apj}, \emph{809}, 1, 78.

\bibitem[\protect\astroncite{\emph{{Kennedy} et~al.}}{2012}]{kennedy2012}
{Kennedy} G.~M. et~al., 2012 \emph{\mnras}, \emph{421}, 3, 2264.

\bibitem[\protect\astroncite{\emph{{Kennedy} et~al.}}{2019}]{kennedy2019}
{Kennedy} G.~M. et~al., 2019 \emph{Nature Astronomy}, \emph{3}, 230.

\bibitem[\protect\astroncite{\emph{{Keppler} et~al.}}{2018}]{keppler2018}
{Keppler} M. et~al., 2018 \emph{\aap}, \emph{617}, A44.

\bibitem[\protect\astroncite{\emph{{Keppler} et~al.}}{2019}]{keppler2019}
{Keppler} M. et~al., 2019 \emph{\aap}, \emph{625}, A118.

\bibitem[\protect\astroncite{\emph{{Kimura} et~al.}}{2015}]{kimura2015}
{Kimura} H. et~al., 2015 \emph{\apj}, \emph{812}, 1, 67.

\bibitem[\protect\astroncite{\emph{{Klahr} and {Hubbard}}}{2014}]{klahr2014}
{Klahr} H. and {Hubbard} A., 2014 \emph{\apj}, \emph{788}, 1, 21.

\bibitem[\protect\astroncite{\emph{{Kozai}}}{1962}]{kozai1962}
{Kozai} Y., 1962 \emph{AJ}, \emph{67}, 591.

\bibitem[\protect\astroncite{\emph{{Krapp} et~al.}}{2018}]{krapp2018}
{Krapp} L. et~al., 2018 \emph{\apj}, \emph{865}, 2, 105.

\bibitem[\protect\astroncite{\emph{{Kratter} and
  {Lodato}}}{2016}]{kratterlodato2016}
{Kratter} K. and {Lodato} G., 2016 \emph{\araa}, \emph{54}, 271.

\bibitem[\protect\astroncite{\emph{{Kraus} et~al.}}{2017}]{Kraus2017}
{Kraus} S. et~al., 2017 \emph{\apjl}, \emph{848}, 1, L11.

\bibitem[\protect\astroncite{\emph{{Kraus} et~al.}}{2020}]{kraus2020}
{Kraus} S. et~al., 2020 \emph{Science}, \emph{369}, 6508, 1233.

\bibitem[\protect\astroncite{\emph{{Kuffmeier} et~al.}}{2017}]{kuffmeier2017}
{Kuffmeier} M. et~al., 2017 \emph{\apj}, \emph{846}, 1, 7.

\bibitem[\protect\astroncite{\emph{{Kuffmeier} et~al.}}{2018}]{kuffmeier2018}
{Kuffmeier} M. et~al., 2018 \emph{\mnras}, \emph{475}, 2, 2642.

\bibitem[\protect\astroncite{\emph{{Kunitomo} et~al.}}{2020}]{kunitomo2020}
{Kunitomo} M. et~al., 2020 \emph{\mnras}, \emph{492}, 3, 3849.

\bibitem[\protect\astroncite{\emph{{Kunz} and {Lesur}}}{2013}]{kunz2013}
{Kunz} M.~W. and {Lesur} G., 2013 \emph{\mnras}, \emph{434}, 3, 2295.

\bibitem[\protect\astroncite{\emph{{Kurtovic} et~al.}}{2018}]{Kurtovic2018}
{Kurtovic} N.~T. et~al., 2018 \emph{\apjl}, \emph{869}, 2, L44.

\bibitem[\protect\astroncite{\emph{{Kurtovic} et~al.}}{2021}]{kurtovic2021}
{Kurtovic} N.~T. et~al., 2021 \emph{\aap}, \emph{645}, A139.

\bibitem[\protect\astroncite{\emph{{Kuznetsova} et~al.}}{2022}]{kuznetsova2022}
{Kuznetsova} A. et~al., 2022 \emph{arXiv e-prints}, arXiv:2202.05301.

\bibitem[\protect\astroncite{\emph{{Lada} and {Lada}}}{2003}]{lada2003}
{Lada} C.~J. and {Lada} E.~A., 2003 \emph{\araa}, \emph{41}, 57.

\bibitem[\protect\astroncite{\emph{{Larwood} et~al.}}{1996}]{larwood1996}
{Larwood} J.~D. et~al., 1996 \emph{MNRAS}, \emph{282}, 597.

\bibitem[\protect\astroncite{\emph{{Latter} and
  {Papaloizou}}}{2018}]{latter2018}
{Latter} H.~N. and {Papaloizou} J., 2018 \emph{\mnras}, \emph{474}, 3, 3110.

\bibitem[\protect\astroncite{\emph{{Law} et~al.}}{2021}]{law2021a}
{Law} C.~J. et~al., 2021 \emph{\apjs}, \emph{257}, 1, 3.

\bibitem[\protect\astroncite{\emph{{Lee} et~al.}}{2019}]{lee2019}
{Lee} W.-K. et~al., 2019 \emph{\apj}, \emph{872}, 2, 184.

\bibitem[\protect\astroncite{\emph{{Lesur} and {Papaloizou}}}{2010}]{lesur2010}
{Lesur} G. and {Papaloizou} J.~C.~B., 2010 \emph{\aap}, \emph{513}, A60.

\bibitem[\protect\astroncite{\emph{{Lesur} et~al.}}{2015}]{lesur2015}
{Lesur} G. et~al., 2015 \emph{\aap}, \emph{582}, L9.

\bibitem[\protect\astroncite{\emph{{Lesur} and {Latter}}}{2016}]{lesur2016}
{Lesur} G. R.~J. and {Latter} H., 2016 \emph{\mnras}, \emph{462}, 4, 4549.

\bibitem[\protect\astroncite{\emph{{Li} et~al.}}{2000}]{li2000}
{Li} H. et~al., 2000 \emph{\apj}, \emph{533}, 2, 1023.

\bibitem[\protect\astroncite{\emph{{Li} et~al.}}{2021}]{li2021}
{Li} J. et~al., 2021 \emph{\apj}, \emph{910}, 1, 79.

\bibitem[\protect\astroncite{\emph{{Li} and {Youdin}}}{2021}]{li_youdin2021}
{Li} R. and {Youdin} A.~N., 2021 \emph{\apj}, \emph{919}, 2, 107.

\bibitem[\protect\astroncite{\emph{{Li} et~al.}}{2018{\natexlab{a}}}]{li2018}
{Li} R. et~al., 2018{\natexlab{a}} \emph{\apj}, \emph{862}, 1, 14.

\bibitem[\protect\astroncite{\emph{{Li}
  et~al.}}{2018{\natexlab{b}}}]{LiYoudinSimon2018}
{Li} R. et~al., 2018{\natexlab{b}} \emph{\apj}, \emph{862}, 1, 14.

\bibitem[\protect\astroncite{\emph{{Lidov}}}{1962}]{lidov1962}
{Lidov} M.~L., 1962 \emph{Planet. Space Sci.}, \emph{9}, 719.

\bibitem[\protect\astroncite{\emph{{Lin}}}{2014}]{lin2014}
{Lin} M.-K., 2014 \emph{\mnras}, \emph{437}, 1, 575.

\bibitem[\protect\astroncite{\emph{{Lin}}}{2015}]{lin2015}
{Lin} M.-K., 2015 \emph{\mnras}, \emph{448}, 4, 3806.

\bibitem[\protect\astroncite{\emph{{Lin}}}{2019}]{lin2019}
{Lin} M.-K., 2019 \emph{\mnras}, \emph{485}, 4, 5221.

\bibitem[\protect\astroncite{\emph{{Lin} and
  {Papaloizou}}}{2011}]{LinPapaloizou2011}
{Lin} M.-K. and {Papaloizou} J. C.~B., 2011 \emph{\mnras}, \emph{415}, 2, 1426.

\bibitem[\protect\astroncite{\emph{{Lin} and {Youdin}}}{2015}]{linyoudin2015}
{Lin} M.-K. and {Youdin} A.~N., 2015 \emph{\apj}, \emph{811}, 1, 17.

\bibitem[\protect\astroncite{\emph{{Lodato} and {Clarke}}}{2011}]{lodato2011}
{Lodato} G. and {Clarke} C.~J., 2011 \emph{\mnras}, \emph{413}, 4, 2735.

\bibitem[\protect\astroncite{\emph{{Lodato} and {Facchini}}}{2013}]{lodato2013}
{Lodato} G. and {Facchini} S., 2013 \emph{\mnras}, \emph{433}, 2157.

\bibitem[\protect\astroncite{\emph{{Lodato} et~al.}}{2019}]{lodato2019}
{Lodato} G. et~al., 2019 \emph{\mnras}, \emph{486}, 1, 453.

\bibitem[\protect\astroncite{\emph{{Long} et~al.}}{2018}]{long2018}
{Long} F. et~al., 2018 \emph{\apj}, \emph{869}, 1, 17.

\bibitem[\protect\astroncite{\emph{{Long} et~al.}}{2019}]{long2019}
{Long} F. et~al., 2019 \emph{\apj}, \emph{882}, 1, 49.

\bibitem[\protect\astroncite{\emph{{Long} et~al.}}{2022}]{long2022}
{Long} F. et~al., 2022 \emph{\apjl}, \emph{937}, 1, L1.

\bibitem[\protect\astroncite{\emph{{Loomis} et~al.}}{2017}]{loomis2017}
{Loomis} R.~A. et~al., 2017 \emph{\apj}, \emph{840}, 1, 23.

\bibitem[\protect\astroncite{\emph{{Lovelace} and
  {Hohlfeld}}}{2013}]{lovelace2013}
{Lovelace} R.~V.~E. and {Hohlfeld} R.~G., 2013 \emph{\mnras}, \emph{429}, 1,
  529.

\bibitem[\protect\astroncite{\emph{{Lovelace} et~al.}}{1999}]{lovelace1999}
{Lovelace} R.~V.~E. et~al., 1999 \emph{\apj}, \emph{513}, 2, 805.

\bibitem[\protect\astroncite{\emph{{Lubow}}}{1991}]{lubow1991}
{Lubow} S.~H., 1991 \emph{\apj}, \emph{381}, 259.

\bibitem[\protect\astroncite{\emph{{Lubow} and {Martin}}}{2016}]{lubow2016}
{Lubow} S.~H. and {Martin} R.~G., 2016 \emph{\apj}, \emph{817}, 30.

\bibitem[\protect\astroncite{\emph{{Lubow} and {Martin}}}{2018}]{lubow2018}
{Lubow} S.~H. and {Martin} R.~G., 2018 \emph{\mnras}, \emph{473}, 3, 3733.

\bibitem[\protect\astroncite{\emph{{Lubow} and {Ogilvie}}}{2000}]{lubow2000}
{Lubow} S.~H. and {Ogilvie} G.~I., 2000 \emph{\apj}, \emph{538}, 326.

\bibitem[\protect\astroncite{\emph{{Lubow} and {Zhu}}}{2014}]{lubow2014}
{Lubow} S.~H. and {Zhu} Z., 2014 \emph{\apj}, \emph{785}, 1, 32.

\bibitem[\protect\astroncite{\emph{{Lubow} et~al.}}{2015}]{lubow2015}
{Lubow} S.~H. et~al., 2015 \emph{ApJ}, \emph{800}, 96.

\bibitem[\protect\astroncite{\emph{{Luhman} et~al.}}{2011}]{luhman2011}
{Luhman} K.~L. et~al., 2011 \emph{\apjl}, \emph{730}, 1, L9.

\bibitem[\protect\astroncite{\emph{{Lyra}}}{2014}]{lyra2014}
{Lyra} W., 2014 \emph{\apj}, \emph{789}, 1, 77.

\bibitem[\protect\astroncite{\emph{{Lyra} and {Lin}}}{2013}]{lyra2013}
{Lyra} W. and {Lin} M.-K., 2013 \emph{\apj}, \emph{775}, 1, 17.

\bibitem[\protect\astroncite{\emph{{Lyra} and {Mac Low}}}{2012}]{lyra2012}
{Lyra} W. and {Mac Low} M.-M., 2012 \emph{\apj}, \emph{756}, 1, 62.

\bibitem[\protect\astroncite{\emph{{Lyra} and {Umurhan}}}{2019}]{lyra2019}
{Lyra} W. and {Umurhan} O.~M., 2019 \emph{\pasp}, \emph{131}, 1001, 072001.

\bibitem[\protect\astroncite{\emph{{Lyra} et~al.}}{2008}]{lyra2008}
{Lyra} W. et~al., 2008 \emph{\aap}, \emph{491}, 3, L41.

\bibitem[\protect\astroncite{\emph{{Lyra} et~al.}}{2009}]{lyra2009}
{Lyra} W. et~al., 2009 \emph{\aap}, \emph{493}, 3, 1125.

\bibitem[\protect\astroncite{\emph{{Lyra} et~al.}}{2015}]{lyra2015}
{Lyra} W. et~al., 2015 \emph{\aap}, \emph{574}, A10.

\bibitem[\protect\astroncite{\emph{{Malygin} et~al.}}{2017}]{malygin2017}
{Malygin} M.~G. et~al., 2017 \emph{\aap}, \emph{605}, A30.

\bibitem[\protect\astroncite{\emph{{Manger} et~al.}}{2020}]{manger2020}
{Manger} N. et~al., 2020 \emph{\mnras}, \emph{499}, 2, 1841.

\bibitem[\protect\astroncite{\emph{{Marcus} et~al.}}{2013}]{marcus2013}
{Marcus} P.~S. et~al., 2013 \emph{\prl}, \emph{111}, 8, 084501.

\bibitem[\protect\astroncite{\emph{{Marcus} et~al.}}{2015}]{marcus2015}
{Marcus} P.~S. et~al., 2015 \emph{\apj}, \emph{808}, 1, 87.

\bibitem[\protect\astroncite{\emph{{Marcus} et~al.}}{2016}]{marcus2016}
{Marcus} P.~S. et~al., 2016 \emph{\apj}, \emph{833}, 2, 148.

\bibitem[\protect\astroncite{\emph{{Marino} et~al.}}{2015}]{marino2015}
{Marino} S. et~al., 2015 \emph{\apjl}, \emph{798}, 2, L44.

\bibitem[\protect\astroncite{\emph{{Marley} et~al.}}{2007}]{marley2007}
{Marley} M.~S. et~al., 2007 \emph{\apj}, \emph{655}, 1, 541.

\bibitem[\protect\astroncite{\emph{{Marr} and {Dong}}}{2022}]{metea2022}
{Marr} M. and {Dong} R., 2022 \emph{\apj}, \emph{930}, 1, 80.

\bibitem[\protect\astroncite{\emph{{Martin} and {Livio}}}{2012}]{martin2012}
{Martin} R.~G. and {Livio} M., 2012 \emph{\mnras}, \emph{425}, 1, L6.

\bibitem[\protect\astroncite{\emph{{Martin} and {Lubow}}}{2017}]{martin2017}
{Martin} R.~G. and {Lubow} S.~H., 2017 \emph{\apjl}, \emph{835}, L28.

\bibitem[\protect\astroncite{\emph{{Martin} and {Lubow}}}{2019}]{martin2019}
{Martin} R.~G. and {Lubow} S.~H., 2019 \emph{\mnras}, \emph{490}, 1, 1332.

\bibitem[\protect\astroncite{\emph{{Martin} et~al.}}{2014}]{martin2014}
{Martin} R.~G. et~al., 2014 \emph{ApJL}, \emph{792}, L33.

\bibitem[\protect\astroncite{\emph{{Martin} et~al.}}{2016}]{martin2016}
{Martin} R.~G. et~al., 2016 \emph{\mnras}, \emph{458}, 4345.

\bibitem[\protect\astroncite{\emph{{Mayama} et~al.}}{2012}]{mayama2012}
{Mayama} S. et~al., 2012 \emph{\apjl}, \emph{760}, 2, L26.

\bibitem[\protect\astroncite{\emph{{Mayama} et~al.}}{2018}]{mayama2018}
{Mayama} S. et~al., 2018 \emph{\apjl}, \emph{868}, 1, L3.

\bibitem[\protect\astroncite{\emph{{McNally} et~al.}}{2020}]{mcnally2020}
{McNally} C.~P. et~al., 2020 \emph{\mnras}, \emph{493}, 3, 4382.

\bibitem[\protect\astroncite{\emph{{Meheut} et~al.}}{2010}]{meheut2010}
{Meheut} H. et~al., 2010 \emph{\aap}, \emph{516}, A31.

\bibitem[\protect\astroncite{\emph{{M{\'e}nard} et~al.}}{2020}]{Menard2020}
{M{\'e}nard} F. et~al., 2020 \emph{\aap}, \emph{639}, L1.

\bibitem[\protect\astroncite{\emph{{Meru} and {Bate}}}{2011}]{meru2011}
{Meru} F. and {Bate} M.~R., 2011 \emph{\mnras}, \emph{411}, 1, L1.

\bibitem[\protect\astroncite{\emph{{Meru} et~al.}}{2019}]{meru2019}
{Meru} F. et~al., 2019 \emph{\mnras}, \emph{482}, 3, 3678.

\bibitem[\protect\astroncite{\emph{{Mesa} et~al.}}{2019}]{mesa2019}
{Mesa} D. et~al., 2019 \emph{\aap}, \emph{632}, A25.

\bibitem[\protect\astroncite{\emph{{Michael} et~al.}}{2012}]{michael2012}
{Michael} S. et~al., 2012 \emph{\apj}, \emph{746}, 1, 98.

\bibitem[\protect\astroncite{\emph{{Min} et~al.}}{2017}]{min2017}
{Min} M. et~al., 2017 \emph{\aap}, \emph{604}, L10.

\bibitem[\protect\astroncite{\emph{{Miranda} and {Lai}}}{2015}]{miranda2015}
{Miranda} R. and {Lai} D., 2015 \emph{\mnras}, \emph{452}, 2396.

\bibitem[\protect\astroncite{\emph{{Miranda} and
  {Rafikov}}}{2019}]{miranda2019}
{Miranda} R. and {Rafikov} R.~R., 2019 \emph{\apj}, \emph{875}, 1, 37.

\bibitem[\protect\astroncite{\emph{{Miranda} and
  {Rafikov}}}{2020}]{miranda2020}
{Miranda} R. and {Rafikov} R.~R., 2020 \emph{\apj}, \emph{892}, 1, 65.

\bibitem[\protect\astroncite{\emph{{Miranda} et~al.}}{2017}]{miranda2017}
{Miranda} R. et~al., 2017 \emph{\apj}, \emph{835}, 2, 118.

\bibitem[\protect\astroncite{\emph{{Moll}}}{2012}]{moll2012}
{Moll} R., 2012 \emph{\aap}, \emph{548}, A76.

\bibitem[\protect\astroncite{\emph{{Monnier} et~al.}}{2019}]{monnier2019}
{Monnier} J.~D. et~al., 2019 \emph{\apj}, \emph{872}, 2, 122.

\bibitem[\protect\astroncite{\emph{{Mordasini} et~al.}}{2017}]{mordasini2017}
{Mordasini} C. et~al., 2017 \emph{\aap}, \emph{608}, A72.

\bibitem[\protect\astroncite{\emph{{Morishima}}}{2012}]{morishima2012}
{Morishima} R., 2012 \emph{\mnras}, \emph{420}, 4, 2851.

\bibitem[\protect\astroncite{\emph{{Mu{\~n}oz} and
  {Lithwick}}}{2020}]{munoz2020}
{Mu{\~n}oz} D.~J. and {Lithwick} Y., 2020 \emph{\apj}, \emph{905}, 2, 106.

\bibitem[\protect\astroncite{\emph{{Muro-Arena} et~al.}}{2020}]{muroarena2020}
{Muro-Arena} G.~A. et~al., 2020 \emph{\aap}, \emph{635}, A121.

\bibitem[\protect\astroncite{\emph{{Musiolik} and {Wurm}}}{2019}]{musiolik2019}
{Musiolik} G. and {Wurm} G., 2019 \emph{\apj}, \emph{873}, 1, 58.

\bibitem[\protect\astroncite{\emph{{Musiolik}
  et~al.}}{2016{\natexlab{a}}}]{musiolik2016a}
{Musiolik} G. et~al., 2016{\natexlab{a}} \emph{\apj}, \emph{818}, 1, 16.

\bibitem[\protect\astroncite{\emph{{Musiolik}
  et~al.}}{2016{\natexlab{b}}}]{musiolik2016b}
{Musiolik} G. et~al., 2016{\natexlab{b}} \emph{\apj}, \emph{827}, 1, 63.

\bibitem[\protect\astroncite{\emph{{Muto} et~al.}}{2012}]{muto2012}
{Muto} T. et~al., 2012 \emph{\apjl}, \emph{748}, 2, L22.

\bibitem[\protect\astroncite{\emph{{Nakagawa} et~al.}}{1986}]{nakagawa1986}
{Nakagawa} Y. et~al., 1986 \emph{\icarus}, \emph{67}, 3, 375.

\bibitem[\protect\astroncite{\emph{{Nazari} et~al.}}{2019}]{nazari2019}
{Nazari} P. et~al., 2019 \emph{\mnras}, \emph{485}, 4, 5914.

\bibitem[\protect\astroncite{\emph{{Nealon} et~al.}}{2019}]{nealon2019}
{Nealon} R. et~al., 2019 \emph{\mnras}, \emph{484}, 4, 4951.

\bibitem[\protect\astroncite{\emph{{Nelson} et~al.}}{2013}]{nelson2013}
{Nelson} R.~P. et~al., 2013 \emph{\mnras}, \emph{435}, 3, 2610.

\bibitem[\protect\astroncite{\emph{{Nixon} et~al.}}{2013}]{nixon2013}
{Nixon} C. et~al., 2013 \emph{MNRAS}, \emph{434}, 1946.

\bibitem[\protect\astroncite{\emph{{Nixon} et~al.}}{2011}]{nixon2011}
{Nixon} C.~J. et~al., 2011 \emph{MNRAS}, \emph{417}, L66.

\bibitem[\protect\astroncite{\emph{{Norfolk} et~al.}}{2021}]{norfolk2021}
{Norfolk} B.~J. et~al., 2021 \emph{\mnras}, \emph{502}, 4, 5779.

\bibitem[\protect\astroncite{\emph{{{\"O}berg} et~al.}}{2021}]{oberg2021}
{{\"O}berg} K.~I. et~al., 2021 \emph{\apjs}, \emph{257}, 1, 1.

\bibitem[\protect\astroncite{\emph{{Ogilvie} and {Lubow}}}{2002}]{ogilvie2002}
{Ogilvie} G.~I. and {Lubow} S.~H., 2002 \emph{\mnras}, \emph{330}, 4, 950.

\bibitem[\protect\astroncite{\emph{{Ohashi} and {Kataoka}}}{2019}]{ohashi2019}
{Ohashi} S. and {Kataoka} A., 2019 \emph{\apj}, \emph{886}, 2, 103.

\bibitem[\protect\astroncite{\emph{{Ohta} et~al.}}{2016}]{ohta2016}
{Ohta} Y. et~al., 2016 \emph{\pasj}, \emph{68}, 4, 53.

\bibitem[\protect\astroncite{\emph{{Okuzumi}}}{2009}]{okuzumi2009}
{Okuzumi} S., 2009 \emph{\apj}, \emph{698}, 2, 1122.

\bibitem[\protect\astroncite{\emph{{Okuzumi} and {Tazaki}}}{2019}]{okuzumi2019}
{Okuzumi} S. and {Tazaki} R., 2019 \emph{\apj}, \emph{878}, 2, 132.

\bibitem[\protect\astroncite{\emph{{Okuzumi} et~al.}}{2016}]{okuzumi2016}
{Okuzumi} S. et~al., 2016 \emph{\apj}, \emph{821}, 2, 82.

\bibitem[\protect\astroncite{\emph{{Okuzumi} et~al.}}{2022}]{okuzumi2022}
{Okuzumi} S. et~al., 2022 \emph{arXiv e-prints}, arXiv:2201.09241.

\bibitem[\protect\astroncite{\emph{{Ono} et~al.}}{2016}]{ono2016}
{Ono} T. et~al., 2016 \emph{\apj}, \emph{823}, 2, 84.

\bibitem[\protect\astroncite{\emph{{Ono} et~al.}}{2018}]{ono2018}
{Ono} T. et~al., 2018 \emph{\apj}, \emph{864}, 1, 70.

\bibitem[\protect\astroncite{\emph{{Owen}}}{2020}]{owen2020}
{Owen} J.~E., 2020 \emph{\mnras}, \emph{495}, 3, 3160.

\bibitem[\protect\astroncite{\emph{{Owen} et~al.}}{2012}]{owen2012}
{Owen} J.~E. et~al., 2012 \emph{\mnras}, \emph{422}, 3, 1880.

\bibitem[\protect\astroncite{\emph{{Paardekooper}}}{2012}]{paardekooper2012}
{Paardekooper} S.-J., 2012 \emph{\mnras}, \emph{421}, 4, 3286.

\bibitem[\protect\astroncite{\emph{{Paardekooper}
  et~al.}}{2010}]{paardekooper2010}
{Paardekooper} S.-J. et~al., 2010 \emph{\apj}, \emph{725}, 1, 146.

\bibitem[\protect\astroncite{\emph{{Paardekooper}
  et~al.}}{2020}]{Paardekooper2020}
{Paardekooper} S.-J. et~al., 2020 \emph{\mnras}, \emph{499}, 3, 4223.

\bibitem[\protect\astroncite{\emph{{Paneque-Carre{\~n}o}
  et~al.}}{2021}]{paneque2021}
{Paneque-Carre{\~n}o} T. et~al., 2021 \emph{\apj}, \emph{914}, 2, 88.

\bibitem[\protect\astroncite{\emph{{Papaloizou} and
  {Lin}}}{1995}]{papaloizou1995}
{Papaloizou} J.~C.~B. and {Lin} D.~N.~C., 1995 \emph{ApJ}, \emph{438}, 841.

\bibitem[\protect\astroncite{\emph{{Pascucci} et~al.}}{2016}]{pascucci2016}
{Pascucci} I. et~al., 2016 \emph{\apj}, \emph{831}, 2, 125.

\bibitem[\protect\astroncite{\emph{{P{\'e}rez} et~al.}}{2014}]{perez2014}
{P{\'e}rez} L.~M. et~al., 2014 \emph{\apjl}, \emph{783}, 1, L13.

\bibitem[\protect\astroncite{\emph{{P{\'e}rez} et~al.}}{2018}]{perez2018}
{P{\'e}rez} L.~M. et~al., 2018 \emph{\apjl}, \emph{869}, 2, L50.

\bibitem[\protect\astroncite{\emph{{P{\'e}rez} et~al.}}{2020}]{perez2020}
{P{\'e}rez} S. et~al., 2020 \emph{\apjl}, \emph{889}, 1, L24.

\bibitem[\protect\astroncite{\emph{{Pfeil} and {Klahr}}}{2019}]{pfeil2019}
{Pfeil} T. and {Klahr} H., 2019 \emph{\apj}, \emph{871}, 2, 150.

\bibitem[\protect\astroncite{\emph{{Picogna} and
  {Marzari}}}{2015}]{picogna2015}
{Picogna} G. and {Marzari} F., 2015 \emph{\aap}, \emph{583}, A133.

\bibitem[\protect\astroncite{\emph{{Pierens} and {Lin}}}{2018}]{pierens2018}
{Pierens} A. and {Lin} M.-K., 2018 \emph{\mnras}, \emph{479}, 4, 4878.

\bibitem[\protect\astroncite{\emph{{Pineda} et~al.}}{2020}]{pineda2020}
{Pineda} J.~E. et~al., 2020 \emph{Nature Astronomy}, \emph{4}, 1158.

\bibitem[\protect\astroncite{\emph{{Pinilla} et~al.}}{2012}]{pinilla2012}
{Pinilla} P. et~al., 2012 \emph{\aap}, \emph{538}, A114.

\bibitem[\protect\astroncite{\emph{{Pinilla} et~al.}}{2015}]{pinilla2015}
{Pinilla} P. et~al., 2015 \emph{\aap}, \emph{584}, A16.

\bibitem[\protect\astroncite{\emph{{Pinilla}
  et~al.}}{2017{\natexlab{a}}}]{pinilla2017_icelines}
{Pinilla} P. et~al., 2017{\natexlab{a}} \emph{\apj}, \emph{845}, 1, 68.

\bibitem[\protect\astroncite{\emph{{Pinilla}
  et~al.}}{2017{\natexlab{b}}}]{pinilla2017}
{Pinilla} P. et~al., 2017{\natexlab{b}} \emph{\apj}, \emph{845}, 1, 68.

\bibitem[\protect\astroncite{\emph{{Pinilla} et~al.}}{2018}]{Pinilla2018}
{Pinilla} P. et~al., 2018 \emph{\apj}, \emph{859}, 1, 32.

\bibitem[\protect\astroncite{\emph{{Pinilla} et~al.}}{2021}]{pinilla2021}
{Pinilla} P. et~al., 2021 \emph{\aap}, \emph{649}, A122.

\bibitem[\protect\astroncite{\emph{{Pinte} et~al.}}{2016}]{pinte2016}
{Pinte} C. et~al., 2016 \emph{\apj}, \emph{816}, 1, 25.

\bibitem[\protect\astroncite{\emph{{Pinte} et~al.}}{2018}]{pinte2018}
{Pinte} C. et~al., 2018 \emph{\apjl}, \emph{860}, 1, L13.

\bibitem[\protect\astroncite{\emph{{Pinte} et~al.}}{2019}]{pinte2019}
{Pinte} C. et~al., 2019 \emph{Nature Astronomy}, \emph{3}, 1109.

\bibitem[\protect\astroncite{\emph{{Pohl} et~al.}}{2015}]{Pohl2015}
{Pohl} A. et~al., 2015 \emph{\mnras}, \emph{453}, 2, 1768.

\bibitem[\protect\astroncite{\emph{{Pollack} et~al.}}{1996}]{pollack1996}
{Pollack} J.~B. et~al., 1996 \emph{\icarus}, \emph{124}, 1, 62.

\bibitem[\protect\astroncite{\emph{{Price}
  et~al.}}{2018{\natexlab{a}}}]{price2018b}
{Price} D.~J. et~al., 2018{\natexlab{a}} \emph{\mnras}, \emph{477}, 1, 1270.

\bibitem[\protect\astroncite{\emph{{Price}
  et~al.}}{2018{\natexlab{b}}}]{price2018}
{Price} D.~J. et~al., 2018{\natexlab{b}} \emph{\pasa}, \emph{35}, e031.

\bibitem[\protect\astroncite{\emph{{Raettig} et~al.}}{2021}]{raettig2021}
{Raettig} N. et~al., 2021 \emph{\apj}, \emph{913}, 2, 92.

\bibitem[\protect\astroncite{\emph{{Rafikov}}}{2002}]{rafikov2002}
{Rafikov} R.~R., 2002 \emph{\apj}, \emph{569}, 2, 997.

\bibitem[\protect\astroncite{\emph{{Rafikov}}}{2005}]{Rafikov2005}
{Rafikov} R.~R., 2005 \emph{\apjl}, \emph{621}, 1, L69.

\bibitem[\protect\astroncite{\emph{{Ragusa} et~al.}}{2017}]{Ragusa2017}
{Ragusa} E. et~al., 2017 \emph{\mnras}, \emph{464}, 2, 1449.

\bibitem[\protect\astroncite{\emph{{Ragusa} et~al.}}{2020}]{ragusa2020}
{Ragusa} E. et~al., 2020 \emph{\mnras}, \emph{499}, 3, 3362.

\bibitem[\protect\astroncite{\emph{{Railton} and
  {Papaloizou}}}{2014}]{railton2014}
{Railton} A.~D. and {Papaloizou} J.~C.~B., 2014 \emph{\mnras}, \emph{445}, 4,
  4409.

\bibitem[\protect\astroncite{\emph{{Rayleigh}}}{1917}]{rayleigh}
{Rayleigh} L., 1917 \emph{Proceedings of the Royal Society of London Series A},
  \emph{93}, 648, 148.

\bibitem[\protect\astroncite{\emph{{Raymond} et~al.}}{2014}]{raymond2014}
{Raymond} S.~N. et~al., 2014 \emph{Protostars and Planets VI} (H.~{Beuther},
  R.~S. {Klessen}, C.~P. {Dullemond}, and T.~{Henning}), p. 595.

\bibitem[\protect\astroncite{\emph{{Ren} et~al.}}{2020}]{ren2020}
{Ren} B. et~al., 2020 \emph{\apjl}, \emph{898}, 2, L38.

\bibitem[\protect\astroncite{\emph{{Rice} et~al.}}{2003}]{rice2003}
{Rice} W.~K.~M. et~al., 2003 \emph{\mnras}, \emph{339}, 4, 1025.

\bibitem[\protect\astroncite{\emph{{Rice} et~al.}}{2004}]{rice2004}
{Rice} W.~K.~M. et~al., 2004 \emph{\mnras}, \emph{355}, 2, 543.

\bibitem[\protect\astroncite{\emph{{Rice} et~al.}}{2005}]{rice2005}
{Rice} W.~K.~M. et~al., 2005 \emph{\mnras}, \emph{364}, 1, L56.

\bibitem[\protect\astroncite{\emph{{Rice} et~al.}}{2006}]{rice2006b}
{Rice} W.~K.~M. et~al., 2006 \emph{\mnras}, \emph{372}, 1, L9.

\bibitem[\protect\astroncite{\emph{{Rice} et~al.}}{2012}]{rice2012}
{Rice} W.~K.~M. et~al., 2012 \emph{\mnras}, \emph{420}, 2, 1640.

\bibitem[\protect\astroncite{\emph{{Rice} et~al.}}{2014}]{rice2014}
{Rice} W.~K.~M. et~al., 2014 \emph{\mnras}, \emph{438}, 2, 1593.

\bibitem[\protect\astroncite{\emph{{Rich} et~al.}}{2021}]{rich2021}
{Rich} E.~A. et~al., 2021 \emph{\apj}, \emph{913}, 2, 138.

\bibitem[\protect\astroncite{\emph{{Richard} et~al.}}{2016}]{richard2016}
{Richard} S. et~al., 2016 \emph{\mnras}, \emph{456}, 4, 3571.

\bibitem[\protect\astroncite{\emph{{Riols} and {Lesur}}}{2018}]{riols2018}
{Riols} A. and {Lesur} G., 2018 \emph{\aap}, \emph{617}, A117.

\bibitem[\protect\astroncite{\emph{{Riols} and {Lesur}}}{2019}]{riols2019}
{Riols} A. and {Lesur} G., 2019 \emph{\aap}, \emph{625}, A108.

\bibitem[\protect\astroncite{\emph{{Riols} et~al.}}{2020}]{riols2020}
{Riols} A. et~al., 2020 \emph{\aap}, \emph{639}, A95.

\bibitem[\protect\astroncite{\emph{{Rogers} and {Wadsley}}}{2012}]{rogers2012}
{Rogers} P.~D. and {Wadsley} J., 2012 \emph{\mnras}, \emph{423}, 2, 1896.

\bibitem[\protect\astroncite{\emph{{Ros} and {Johansen}}}{2013}]{ros2013}
{Ros} K. and {Johansen} A., 2013 \emph{\aap}, \emph{552}, A137.

\bibitem[\protect\astroncite{\emph{{Rosotti}
  et~al.}}{2020{\natexlab{a}}}]{rosotti2020}
{Rosotti} G.~P. et~al., 2020{\natexlab{a}} \emph{\mnras}, \emph{491}, 1, 1335.

\bibitem[\protect\astroncite{\emph{{Rosotti}
  et~al.}}{2020{\natexlab{b}}}]{rosotti2020b}
{Rosotti} G.~P. et~al., 2020{\natexlab{b}} \emph{\mnras}, \emph{495}, 1, 173.

\bibitem[\protect\astroncite{\emph{{Ruge} et~al.}}{2016}]{ruge2016}
{Ruge} J.~P. et~al., 2016 \emph{\aap}, \emph{590}, A17.

\bibitem[\protect\astroncite{\emph{{Safronov}}}{1972}]{safronov1972}
{Safronov} V.~S., 1972 \emph{{Evolution of the protoplanetary cloud and
  formation of the earth and planets.}}

\bibitem[\protect\astroncite{\emph{{Saito} and {Sirono}}}{2011}]{saito2011}
{Saito} E. and {Sirono} S.-i., 2011 \emph{\apj}, \emph{728}, 1, 20.

\bibitem[\protect\astroncite{\emph{{Salyk} et~al.}}{2014}]{salyk2014}
{Salyk} C. et~al., 2014 \emph{\apj}, \emph{792}, 1, 68.

\bibitem[\protect\astroncite{\emph{{Sanchis} et~al.}}{2020}]{sanchis2020}
{Sanchis} E. et~al., 2020 \emph{\mnras}, \emph{492}, 3, 3440.

\bibitem[\protect\astroncite{\emph{{Sano} et~al.}}{2000}]{sano2000}
{Sano} T. et~al., 2000 \emph{\apj}, \emph{543}, 1, 486.

\bibitem[\protect\astroncite{\emph{{Schoonenberg} and
  {Ormel}}}{2017}]{schoonenberg2017}
{Schoonenberg} D. and {Ormel} C.~W., 2017 \emph{\aap}, \emph{602}, A21.

\bibitem[\protect\astroncite{\emph{{Segura-Cox} et~al.}}{2020}]{segura-cox2020}
{Segura-Cox} D.~M. et~al., 2020 \emph{\nat}, \emph{586}, 7828, 228.

\bibitem[\protect\astroncite{\emph{{Shakura} and
  {Sunyaev}}}{1973}]{shakura1973}
{Shakura} N.~I. and {Sunyaev} R.~A., 1973 \emph{\aap}, \emph{500}, 33.

\bibitem[\protect\astroncite{\emph{{Shi} and {Chiang}}}{2014}]{shi2014}
{Shi} J.-M. and {Chiang} E., 2014 \emph{\apj}, \emph{789}, 1, 34.

\bibitem[\protect\astroncite{\emph{{Shi} et~al.}}{2012}]{shi2012}
{Shi} J.-M. et~al., 2012 \emph{\apj}, \emph{749}, 2, 118.

\bibitem[\protect\astroncite{\emph{{Shu}}}{1977}]{shu1977}
{Shu} F.~H., 1977 \emph{\apj}, \emph{214}, 488.

\bibitem[\protect\astroncite{\emph{{Shu}}}{2016}]{shu2016}
{Shu} F.~H., 2016 \emph{\araa}, \emph{54}, 667.

\bibitem[\protect\astroncite{\emph{{Shu} et~al.}}{1993}]{shu1993}
{Shu} F.~H. et~al., 1993 \emph{\icarus}, \emph{106}, 1, 92.

\bibitem[\protect\astroncite{\emph{{Sierra} et~al.}}{2019}]{sierra2019}
{Sierra} A. et~al., 2019 \emph{\apj}, \emph{876}, 1, 7.

\bibitem[\protect\astroncite{\emph{{Simon} and {Armitage}}}{2014}]{simon2014}
{Simon} J.~B. and {Armitage} P.~J., 2014 \emph{\apj}, \emph{784}, 1, 15.

\bibitem[\protect\astroncite{\emph{{Simon} et~al.}}{2012}]{simon2012}
{Simon} J.~B. et~al., 2012 \emph{\mnras}, \emph{422}, 3, 2685.

\bibitem[\protect\astroncite{\emph{{Simon} et~al.}}{2017}]{simon2017}
{Simon} M. et~al., 2017 \emph{\apj}, \emph{844}, 2, 158.

\bibitem[\protect\astroncite{\emph{{Sirono}}}{2011}]{sirono2011}
{Sirono} S.-i., 2011 \emph{\apj}, \emph{735}, 2, 131.

\bibitem[\protect\astroncite{\emph{{Sirono} and {Ueno}}}{2017}]{sirono2017}
{Sirono} S.-i. and {Ueno} H., 2017 \emph{\apj}, \emph{841}, 1, 36.

\bibitem[\protect\astroncite{\emph{{Smallwood} et~al.}}{2020}]{smallwood2020}
{Smallwood} J.~L. et~al., 2020 \emph{\mnras}, \emph{494}, 1, 487.

\bibitem[\protect\astroncite{\emph{{Smallwood} et~al.}}{2021}]{smallwood2021}
{Smallwood} J.~L. et~al., 2021 \emph{\apjl}, \emph{907}, 1, L14.

\bibitem[\protect\astroncite{\emph{{Speedie} et~al.}}{2022}]{Speedie2022}
{Speedie} J. et~al., 2022 \emph{arXiv e-prints}, arXiv:2203.00692.

\bibitem[\protect\astroncite{\emph{{Spiegel} and
  {Burrows}}}{2012}]{spiegel2012}
{Spiegel} D.~S. and {Burrows} A., 2012 \emph{\apj}, \emph{745}, 2, 174.

\bibitem[\protect\astroncite{\emph{{Stammler} et~al.}}{2017}]{stammler2017}
{Stammler} S.~M. et~al., 2017 \emph{\aap}, \emph{600}, A140.

\bibitem[\protect\astroncite{\emph{{Steinpilz} et~al.}}{2019}]{steinpilz2019}
{Steinpilz} T. et~al., 2019 \emph{\apj}, \emph{874}, 1, 60.

\bibitem[\protect\astroncite{\emph{{Stoll} et~al.}}{2017}]{stoll2017}
{Stoll} M. H.~R. et~al., 2017 \emph{\aap}, \emph{599}, L6.

\bibitem[\protect\astroncite{\emph{{Strom} et~al.}}{1989}]{strom1989}
{Strom} K.~M. et~al., 1989 \emph{\aj}, \emph{97}, 1451.

\bibitem[\protect\astroncite{\emph{{Sturm} et~al.}}{2020}]{Sturm2020}
{Sturm} J.~A. et~al., 2020 \emph{\aap}, \emph{643}, A92.

\bibitem[\protect\astroncite{\emph{{Suriano} et~al.}}{2018}]{suriano2018}
{Suriano} S.~S. et~al., 2018 \emph{\mnras}, \emph{477}, 1, 1239.

\bibitem[\protect\astroncite{\emph{{Suriano} et~al.}}{2019}]{suriano2019}
{Suriano} S.~S. et~al., 2019 \emph{\mnras}, \emph{484}, 1, 107.

\bibitem[\protect\astroncite{\emph{{Surville} and
  {Barge}}}{2015}]{surville2015}
{Surville} C. and {Barge} P., 2015 \emph{\aap}, \emph{579}, A100.

\bibitem[\protect\astroncite{\emph{{Suzuki} and {Inutsuka}}}{2014}]{suzuki2014}
{Suzuki} T.~K. and {Inutsuka} S.-i., 2014 \emph{\apj}, \emph{784}, 2, 121.

\bibitem[\protect\astroncite{\emph{{Suzuki} et~al.}}{2016}]{suzuki2016}
{Suzuki} T.~K. et~al., 2016 \emph{\aap}, \emph{596}, A74.

\bibitem[\protect\astroncite{\emph{{Szul{\'a}gyi} and
  {Ercolano}}}{2020}]{Szuagyi2020}
{Szul{\'a}gyi} J. and {Ercolano} B., 2020 \emph{\apj}, \emph{902}, 2, 126.

\bibitem[\protect\astroncite{\emph{{Szul{\'a}gyi} et~al.}}{2019}]{Szulagyi2019}
{Szul{\'a}gyi} J. et~al., 2019 \emph{\mnras}, \emph{487}, 1, 1248.

\bibitem[\protect\astroncite{\emph{{Takahashi} and
  {Inutsuka}}}{2014}]{takahashi2014}
{Takahashi} S.~Z. and {Inutsuka} S.-i., 2014 \emph{\apj}, \emph{794}, 1, 55.

\bibitem[\protect\astroncite{\emph{{Takahashi} and
  {Inutsuka}}}{2016}]{takahashiinutsuka2016}
{Takahashi} S.~Z. and {Inutsuka} S.-i., 2016 \emph{\aj}, \emph{152}, 6, 184.

\bibitem[\protect\astroncite{\emph{{Takahashi} and
  {Muto}}}{2018}]{takahashi2018}
{Takahashi} S.~Z. and {Muto} T., 2018 \emph{\apj}, \emph{865}, 2, 102.

\bibitem[\protect\astroncite{\emph{{Takahashi} et~al.}}{2016}]{takahashi2016}
{Takahashi} S.~Z. et~al., 2016 \emph{\mnras}, \emph{458}, 4, 3597.

\bibitem[\protect\astroncite{\emph{{Takeuchi} and {Lin}}}{2002}]{takeuchi02}
{Takeuchi} T. and {Lin} D.~N.~C., 2002 \emph{\apj}, \emph{581}, 2, 1344.

\bibitem[\protect\astroncite{\emph{{Tang} et~al.}}{2017}]{tang2017}
{Tang} Y.-W. et~al., 2017 \emph{\apj}, \emph{840}, 1, 32.

\bibitem[\protect\astroncite{\emph{{Tanga} et~al.}}{1996}]{tanga1996}
{Tanga} P. et~al., 1996 \emph{\icarus}, \emph{121}, 1, 158.

\bibitem[\protect\astroncite{\emph{{Teague}
  et~al.}}{2018{\natexlab{a}}}]{teague2018}
{Teague} R. et~al., 2018{\natexlab{a}} \emph{\apjl}, \emph{860}, 1, L12.

\bibitem[\protect\astroncite{\emph{{Teague}
  et~al.}}{2018{\natexlab{b}}}]{teague2018b}
{Teague} R. et~al., 2018{\natexlab{b}} \emph{\apj}, \emph{868}, 2, 113.

\bibitem[\protect\astroncite{\emph{{Teague}
  et~al.}}{2018{\natexlab{c}}}]{teague2018c}
{Teague} R. et~al., 2018{\natexlab{c}} \emph{\apj}, \emph{864}, 2, 133.

\bibitem[\protect\astroncite{\emph{{Teague} et~al.}}{2019}]{teague2019}
{Teague} R. et~al., 2019 \emph{\apjl}, \emph{884}, 2, L56.

\bibitem[\protect\astroncite{\emph{{Teague} et~al.}}{2021}]{teague2021b}
{Teague} R. et~al., 2021 \emph{\apj}, \emph{922}, 2, 139.

\bibitem[\protect\astroncite{\emph{{Testi} et~al.}}{2014}]{testi2014}
{Testi} L. et~al., 2014 \emph{Protostars and Planets VI} (H.~{Beuther}, R.~S.
  {Klessen}, C.~P. {Dullemond}, and T.~{Henning}), p. 339.

\bibitem[\protect\astroncite{\emph{{Testi} et~al.}}{2022}]{testi2022}
{Testi} L. et~al., 2022 \emph{arXiv e-prints}, arXiv:2201.04079.

\bibitem[\protect\astroncite{\emph{{Thalmann} et~al.}}{2016}]{thalmann2016}
{Thalmann} C. et~al., 2016 \emph{\apjl}, \emph{828}, 2, L17.

\bibitem[\protect\astroncite{\emph{{Tominaga} et~al.}}{2018}]{tominaga2018}
{Tominaga} R.~T. et~al., 2018 \emph{\pasj}, \emph{70}, 1, 3.

\bibitem[\protect\astroncite{\emph{{Tominaga} et~al.}}{2019}]{tominaga2019}
{Tominaga} R.~T. et~al., 2019 \emph{\apj}, \emph{881}, 1, 53.

\bibitem[\protect\astroncite{\emph{{Tominaga} et~al.}}{2021}]{tominaga2021}
{Tominaga} R.~T. et~al., 2021 \emph{\apj}, \emph{923}, 1, 34.

\bibitem[\protect\astroncite{\emph{{Toomre}}}{1964}]{toomre1964}
{Toomre} A., 1964 \emph{\apj}, \emph{139}, 1217.

\bibitem[\protect\astroncite{\emph{{Tremaine}}}{2001}]{tremaine2001}
{Tremaine} S., 2001 \emph{\aj}, \emph{121}, 3, 1776.

\bibitem[\protect\astroncite{\emph{{Tsukagoshi} et~al.}}{2016}]{Tsukagoshi2016}
{Tsukagoshi} T. et~al., 2016 \emph{\apjl}, \emph{829}, 2, L35.

\bibitem[\protect\astroncite{\emph{{Turner} et~al.}}{2014}]{turner2014}
{Turner} N.~J. et~al., 2014 \emph{Protostars and Planets VI} (H.~{Beuther},
  R.~S. {Klessen}, C.~P. {Dullemond}, and T.~{Henning}), p. 411.

\bibitem[\protect\astroncite{\emph{{Ubeira Gabellini}
  et~al.}}{2019}]{Ubeira2019}
{Ubeira Gabellini} M.~G. et~al., 2019 \emph{\mnras}, \emph{486}, 4, 4638.

\bibitem[\protect\astroncite{\emph{{Ueda} et~al.}}{2021}]{ueda2021}
{Ueda} T. et~al., 2021 \emph{\apjl}, \emph{914}, 2, L38.

\bibitem[\protect\astroncite{\emph{{Ulrich}}}{1976}]{ulrich1976}
{Ulrich} R.~K., 1976 \emph{\apj}, \emph{210}, 377.

\bibitem[\protect\astroncite{\emph{{Urpin} and
  {Brandenburg}}}{1998}]{urpin1998}
{Urpin} V. and {Brandenburg} A., 1998 \emph{\mnras}, \emph{294}, 3, 399.

\bibitem[\protect\astroncite{\emph{{Uyama} et~al.}}{2018}]{uyama2018}
{Uyama} T. et~al., 2018 \emph{\aj}, \emph{156}, 2, 63.

\bibitem[\protect\astroncite{\emph{{Uyama}
  et~al.}}{2020{\natexlab{a}}}]{Uyama2020b}
{Uyama} T. et~al., 2020{\natexlab{a}} \emph{\aj}, \emph{159}, 3, 118.

\bibitem[\protect\astroncite{\emph{{Uyama}
  et~al.}}{2020{\natexlab{b}}}]{Uyama2020}
{Uyama} T. et~al., 2020{\natexlab{b}} \emph{\apj}, \emph{900}, 2, 135.

\bibitem[\protect\astroncite{\emph{{van der Marel}
  et~al.}}{2013}]{vandermarel2013}
{van der Marel} N. et~al., 2013 \emph{Science}, \emph{340}, 6137, 1199.

\bibitem[\protect\astroncite{\emph{{van der Marel}
  et~al.}}{2016}]{Vandermarel2016}
{van der Marel} N. et~al., 2016 \emph{\apj}, \emph{832}, 2, 178.

\bibitem[\protect\astroncite{\emph{{van der Marel}
  et~al.}}{2019}]{vandermarel2019}
{van der Marel} N. et~al., 2019 \emph{\apj}, \emph{872}, 1, 112.

\bibitem[\protect\astroncite{\emph{{Varni{\`e}re} and
  {Tagger}}}{2006}]{varniere2006}
{Varni{\`e}re} P. and {Tagger} M., 2006 \emph{\aap}, \emph{446}, 2, L13.

\bibitem[\protect\astroncite{\emph{{Verrier} and {Evans}}}{2009}]{verrier2009}
{Verrier} P.~E. and {Evans} N.~W., 2009 \emph{\mnras}, \emph{394}, 1721.

\bibitem[\protect\astroncite{\emph{{Villenave} et~al.}}{2020}]{villenave2020}
{Villenave} M. et~al., 2020 \emph{\aap}, \emph{642}, A164.

\bibitem[\protect\astroncite{\emph{{Vorobyov} and {Basu}}}{2005}]{vorobyov2005}
{Vorobyov} E.~I. and {Basu} S., 2005 \emph{\apjl}, \emph{633}, 2, L137.

\bibitem[\protect\astroncite{\emph{{Vorobyov} et~al.}}{2020}]{vorobyov2020}
{Vorobyov} E.~I. et~al., 2020 \emph{\aap}, \emph{638}, A102.

\bibitem[\protect\astroncite{\emph{{Wada} et~al.}}{2009}]{wada2009}
{Wada} K. et~al., 2009 \emph{\apj}, \emph{702}, 2, 1490.

\bibitem[\protect\astroncite{\emph{{Wafflard-Fernandez} and
  {Baruteau}}}{2020}]{wafflard-fernandez2020}
{Wafflard-Fernandez} G. and {Baruteau} C., 2020 \emph{\mnras}, \emph{493}, 4,
  5892.

\bibitem[\protect\astroncite{\emph{{Wagner} et~al.}}{2015}]{wagner2015}
{Wagner} K. et~al., 2015 \emph{\apjl}, \emph{813}, 1, L2.

\bibitem[\protect\astroncite{\emph{{Wagner} et~al.}}{2018}]{wagner2018}
{Wagner} K. et~al., 2018 \emph{\apjl}, \emph{863}, 1, L8.

\bibitem[\protect\astroncite{\emph{{Walsh} et~al.}}{2017}]{walsh2017}
{Walsh} C. et~al., 2017 \emph{\aap}, \emph{607}, A114.

\bibitem[\protect\astroncite{\emph{{Wardle}}}{2007}]{wardle2007}
{Wardle} M., 2007 \emph{\apss}, \emph{311}, 1-3, 35.

\bibitem[\protect\astroncite{\emph{{Watanabe} and {Lin}}}{2008}]{watanabe2008}
{Watanabe} S.-i. and {Lin} D.~N.~C., 2008 \emph{\apj}, \emph{672}, 2, 1183.

\bibitem[\protect\astroncite{\emph{{Weber} et~al.}}{2019}]{weber2019}
{Weber} P. et~al., 2019 \emph{\apj}, \emph{884}, 2, 178.

\bibitem[\protect\astroncite{\emph{{Weidenschilling}}}{1977{\natexlab{a}}}]{weidenschilling1977}
{Weidenschilling} S.~J., 1977{\natexlab{a}} \emph{\mnras}, \emph{180}, 57.

\bibitem[\protect\astroncite{\emph{{Weidenschilling}}}{1977{\natexlab{b}}}]{weidenschilling1977_MMSN}
{Weidenschilling} S.~J., 1977{\natexlab{b}} \emph{\apss}, \emph{51}, 1, 153.

\bibitem[\protect\astroncite{\emph{{Whipple}}}{1972}]{whipple1972}
{Whipple} F.~L., 1972 \emph{From Plasma to Planet} (A.~{Elvius}), p. 211.

\bibitem[\protect\astroncite{\emph{{W{\"o}lfer} et~al.}}{2021}]{Wolfer2021}
{W{\"o}lfer} L. et~al., 2021 \emph{\aap}, \emph{648}, A19.

\bibitem[\protect\astroncite{\emph{{Wu} and {Lithwick}}}{2021}]{wu2021}
{Wu} Y. and {Lithwick} Y., 2021 \emph{\apj}, \emph{923}, 1, 123.

\bibitem[\protect\astroncite{\emph{{Xiang-Gruess} and
  {Papaloizou}}}{2013}]{Xiang-Gruess2013}
{Xiang-Gruess} M. and {Papaloizou} J.~C.~B., 2013 \emph{\mnras}, \emph{431}, 2,
  1320.

\bibitem[\protect\astroncite{\emph{{Xie} et~al.}}{2021}]{xie2021}
{Xie} C. et~al., 2021 \emph{\apjl}, \emph{906}, 2, L9.

\bibitem[\protect\astroncite{\emph{{Xu} and
  {Kunz}}}{2021{\natexlab{a}}}]{XuKunz2021a}
{Xu} W. and {Kunz} M.~W., 2021{\natexlab{a}} \emph{\mnras}, \emph{502}, 4,
  4911.

\bibitem[\protect\astroncite{\emph{{Xu} and
  {Kunz}}}{2021{\natexlab{b}}}]{XuKunz2021b}
{Xu} W. and {Kunz} M.~W., 2021{\natexlab{b}} \emph{\mnras}, \emph{508}, 2,
  2142.

\bibitem[\protect\astroncite{\emph{{Yang} and {Johansen}}}{2014}]{yang2014}
{Yang} C.-C. and {Johansen} A., 2014 \emph{\apj}, \emph{792}, 2, 86.

\bibitem[\protect\astroncite{\emph{{Yang} and {Zhu}}}{2021}]{yang2021}
{Yang} C.-C. and {Zhu} Z., 2021 \emph{\mnras}, \emph{508}, 4, 5538.

\bibitem[\protect\astroncite{\emph{{Yang} et~al.}}{2017}]{yang2017}
{Yang} C.~C. et~al., 2017 \emph{\aap}, \emph{606}, A80.

\bibitem[\protect\astroncite{\emph{{Yen} et~al.}}{2019}]{yen2019}
{Yen} H.-W. et~al., 2019 \emph{\apj}, \emph{880}, 2, 69.

\bibitem[\protect\astroncite{\emph{{Youdin}}}{2011}]{youdin2011}
{Youdin} A.~N., 2011 \emph{\apj}, \emph{731}, 2, 99.

\bibitem[\protect\astroncite{\emph{{Youdin} and {Goodman}}}{2005}]{youdin2005}
{Youdin} A.~N. and {Goodman} J., 2005 \emph{\apj}, \emph{620}, 1, 459.

\bibitem[\protect\astroncite{\emph{{Youdin} and {Lithwick}}}{2007}]{youdin2007}
{Youdin} A.~N. and {Lithwick} Y., 2007 \emph{\icarus}, \emph{192}, 2, 588.

\bibitem[\protect\astroncite{\emph{{Yu} et~al.}}{2019}]{yu2019}
{Yu} S.-Y. et~al., 2019 \emph{\apj}, \emph{877}, 2, 100.

\bibitem[\protect\astroncite{\emph{{Yun} et~al.}}{2019}]{yun2019}
{Yun} H.~G. et~al., 2019 \emph{\apj}, \emph{884}, 2, 142.

\bibitem[\protect\astroncite{\emph{{Yun} et~al.}}{2022}]{yun2022}
{Yun} H.-G. et~al., 2022 \emph{arXiv e-prints}, arXiv:2209.05417.

\bibitem[\protect\astroncite{\emph{{Zanazzi} and {Lai}}}{2018}]{zanazzi2018}
{Zanazzi} J.~J. and {Lai} D., 2018 \emph{\mnras}, \emph{473}, 1, 603.

\bibitem[\protect\astroncite{\emph{{Zapata} et~al.}}{2020}]{Zapata2020}
{Zapata} L.~A. et~al., 2020 \emph{\apj}, \emph{896}, 2, 132.

\bibitem[\protect\astroncite{\emph{{Zhang} et~al.}}{2015}]{zhang2015}
{Zhang} K. et~al., 2015 \emph{\apjl}, \emph{806}, 1, L7.

\bibitem[\protect\astroncite{\emph{{Zhang} et~al.}}{2021}]{zhang2021}
{Zhang} K. et~al., 2021 \emph{\apjs}, \emph{257}, 1, 5.

\bibitem[\protect\astroncite{\emph{{Zhang} and {Zhu}}}{2020}]{zhang2020}
{Zhang} S. and {Zhu} Z., 2020 \emph{\mnras}, \emph{493}, 2, 2287.

\bibitem[\protect\astroncite{\emph{{Zhang} et~al.}}{2018}]{zhang2018}
{Zhang} S. et~al., 2018 \emph{\apjl}, \emph{869}, 2, L47.

\bibitem[\protect\astroncite{\emph{{Zhu}}}{2019}]{zhu2019}
{Zhu} Z., 2019 \emph{\mnras}, \emph{483}, 3, 4221.

\bibitem[\protect\astroncite{\emph{{Zhu} and {Baruteau}}}{2016}]{zhu2016}
{Zhu} Z. and {Baruteau} C., 2016 \emph{\mnras}, \emph{458}, 4, 3918.

\bibitem[\protect\astroncite{\emph{{Zhu} and {Stone}}}{2014}]{zhu2014b}
{Zhu} Z. and {Stone} J.~M., 2014 \emph{\apj}, \emph{795}, 1, 53.

\bibitem[\protect\astroncite{\emph{{Zhu} and {Stone}}}{2018}]{ZhuStone2018}
{Zhu} Z. and {Stone} J.~M., 2018 \emph{\apj}, \emph{857}, 1, 34.

\bibitem[\protect\astroncite{\emph{{Zhu} and
  {Zhang}}}{2022{\natexlab{a}}}]{Zhu2022}
{Zhu} Z. and {Zhang} R.~M., 2022{\natexlab{a}} \emph{\mnras}, \emph{510}, 3,
  3986.

\bibitem[\protect\astroncite{\emph{{Zhu} and
  {Zhang}}}{2022{\natexlab{b}}}]{zhuzhang2022}
{Zhu} Z. and {Zhang} R.~M., 2022{\natexlab{b}} \emph{\mnras}, \emph{510}, 3,
  3986.

\bibitem[\protect\astroncite{\emph{{Zhu}
  et~al.}}{2012{\natexlab{a}}}]{zhu2012c}
{Zhu} Z. et~al., 2012{\natexlab{a}} \emph{\apj}, \emph{746}, 1, 110.

\bibitem[\protect\astroncite{\emph{{Zhu} et~al.}}{2012{\natexlab{b}}}]{zhu2012}
{Zhu} Z. et~al., 2012{\natexlab{b}} \emph{\apjl}, \emph{758}, 2, L42.

\bibitem[\protect\astroncite{\emph{{Zhu} et~al.}}{2015{\natexlab{a}}}]{Zhu2015}
{Zhu} Z. et~al., 2015{\natexlab{a}} \emph{\apj}, \emph{801}, 2, 81.

\bibitem[\protect\astroncite{\emph{{Zhu}
  et~al.}}{2015{\natexlab{b}}}]{zhu2015b}
{Zhu} Z. et~al., 2015{\natexlab{b}} \emph{\apj}, \emph{813}, 2, 88.

\bibitem[\protect\astroncite{\emph{{Zhu} et~al.}}{2019}]{zhu2019b}
{Zhu} Z. et~al., 2019 \emph{\apjl}, \emph{877}, 2, L18.

\bibitem[\protect\astroncite{\emph{{Zurlo} et~al.}}{2020}]{zurlo2020}
{Zurlo} A. et~al., 2020 \emph{\aap}, \emph{633}, A119.

\end{thebibliography}
}

%\printindex
%\renewcommand{\indexname}{Object Index}
%\printindex[obj]

\end{document}